\documentclass[pra,aps,nofootinbib,notitlepage,twocolumn]{revtex4-1}

\usepackage{amsmath}
\usepackage{amssymb}
\usepackage{graphicx}

\usepackage{txfonts}

\usepackage[colorlinks=true,linkcolor=blue,citecolor=blue,urlcolor=blue]{hyperref}

\begin{document}

\title{Statistical speed of quantum states:\\Generalized quantum Fisher information and Schatten speed}
\author{Manuel Gessner}
\email{manuel.gessner@ino.it}
\affiliation{QSTAR, CNR-INO and LENS, Largo Enrico Fermi 2, I-50125 Firenze, Italy}
\author{Augusto Smerzi}
\affiliation{QSTAR, CNR-INO and LENS, Largo Enrico Fermi 2, I-50125 Firenze, Italy}
\date{\today}

\begin{abstract}
We analyze families of measures for the quantum statistical speed which include as special cases the quantum Fisher information, the trace speed, i.e., the quantum statistical speed obtained from the trace distance, and more general quantifiers obtained from the family of Schatten norms. These measures quantify the statistical speed under generic quantum evolutions and are obtained by maximizing classical measures over all possible quantum measurements. We discuss general properties, optimal measurements and upper bounds on the speed of separable states. We further provide a physical interpretation for the trace speed by linking it to an analog of the quantum Cram\'{e}r-Rao bound for median-unbiased quantum phase estimation.
\end{abstract}

\maketitle

\section{Introduction}\label{sec:intro}
Measures of quantum statistical speed quantify the sensitivity of an initial state with respect to changes of the parameter of a dynamical evolution. A higher sensitivity implies that this parameter, which could be an unknown phase shift of interest, can be estimated with higher precision \cite{PhysRevLett.72.3439,PhysRevLett.96.010401}. Thus, quantifiers of statistical speed in Hilbert space can be linked to technological tasks such as quantum phase estimation \cite{Varenna}. Fast evolutions that lead to highly sensitive phase estimation can only be achieved through the opportune usage of entangled states, as limits for the quantum statistical speed exist for separable states \cite{PhysRevLett.102.100401}. This, in turn, renders the statistical speed of a quantum state an observable witness for entanglement \cite{LucaRMP,Strobel2014,PezzePNAS2016,Zhang2016}.

Quantum states are generally distinguished by measurement. The probabilistic nature of the quantum measurement process connects the statistical distance of quantum states to the classical statistical distance between probability distributions. Since the probability distribution that is generated by a quantum state depends strongly on the choice of measurement, a quantum statistical distance must be defined as the maximum classical distance over all possible quantum measurements \cite{Fuchs1999,NielsenChuang}. This immediately ensures that the obtained measure is experimentally attainable by an optimal measurement.

\begin{figure}
\centering
\includegraphics[width=.35\textwidth]{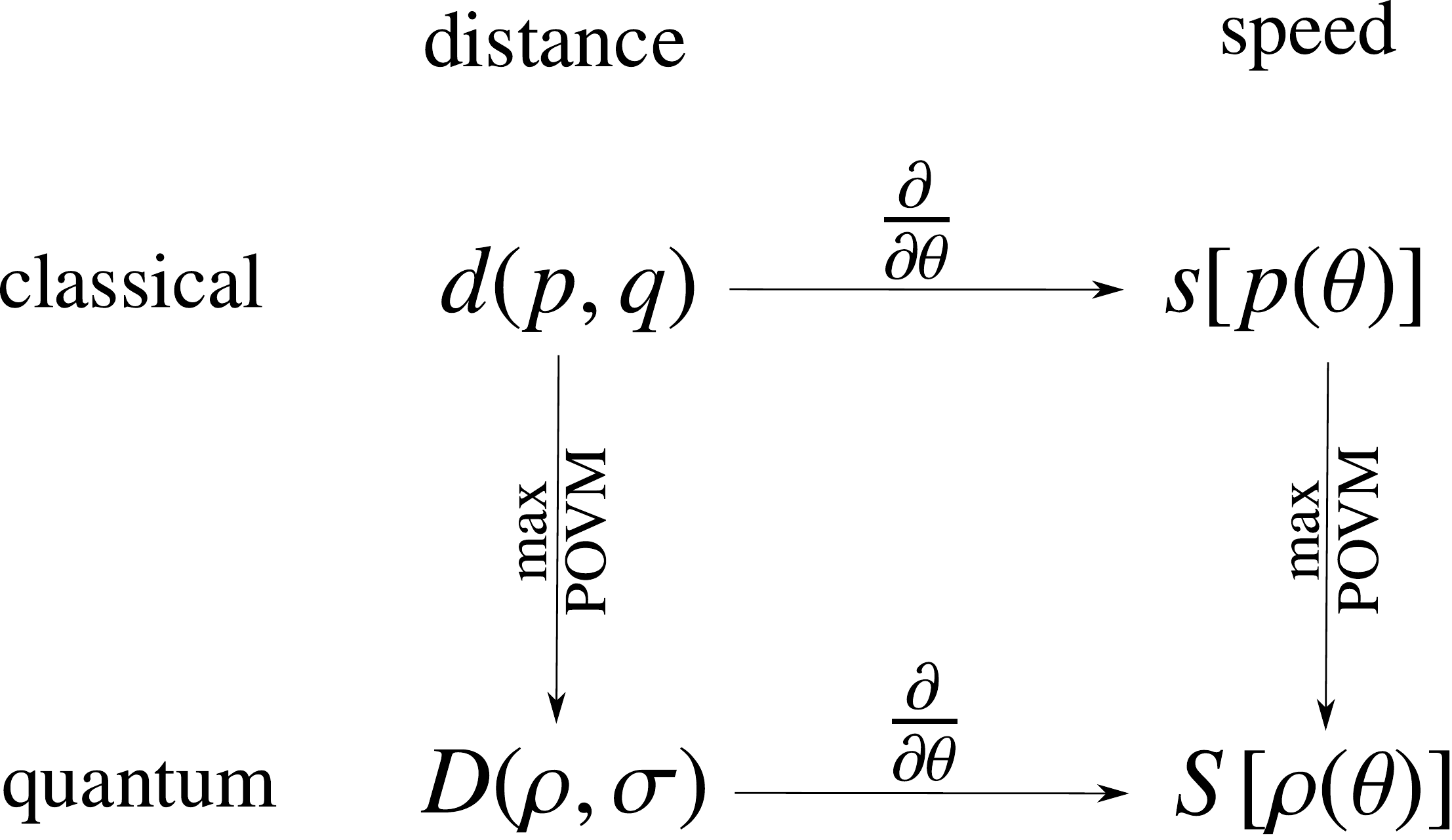}
\caption{Correspondence of classical and quantum statistical distance and speed. The statistical speed is derived from infinitesimal changes of the statistical distance for a parametric evolution. Measures for quantum states are obtained by maximizing the classical measures over all possible measurements.}
\label{fig:diagram}
\end{figure}

Every measure of statistical distance naturally induces a statistical speed for parametric evolutions of probability distributions or quantum states. This statistical speed is given by the change in distance induced by a small change of this parameter (i.e., the derivative of the distance). A maximization over the classical statistical speed over all quantum measurements yields the quantum statistical speed, as visualized in the diagram in Fig.~\ref{fig:diagram}.

Quantifiers of the statistical distance and speed of quantum states play an integral role in quantum information theory \cite{Hayashi2006,NielsenChuang,Zyczkowski2006}. They are used to quantify the distinguishability of quantum states \cite{Hayashi2006,NielsenChuang,Spehner2014}, to understand geometrical aspects of quantum mechanics and quantum phase transitions \cite{WignerYanase,Petz1996,PhysRevD.23.357,Gibilisco2003,PhysRevLett.99.100603,Zyczkowski2006,Petz2011,PhysRevLett.109.230405,PhysRevA.87.032324,Hauke2016,PhysRevLett.119.250401}, to quantify initial correlations, information flow and non-Markovian effects in quantum evolutions \cite{Laine2010,RevModPhys.88.021002,GessnerDiss,PhysRevA.96.012105}, to determine the precision of phase estimation \cite{PhysRevLett.72.3439,Toth2014,Varenna,LucaRMP}, to derive limits on the evolution time \cite{PhysRevA.67.052109,PhysRevLett.110.050402,PhysRevLett.110.050403,PhysRevX.6.021031} or on the occurrence of quantum Zeno dynamics \cite{PhysRevLett.109.150410}, and to quantify and detect quantum properties such as coherence \cite{Streltsov2016}, quantum correlations and discord \cite{Horodecki2009,Modi2012,Adesso2016,Gessner2017}, uncertainty \cite{Luo2003}, asymmetry \cite{Marvian2014,Zhang2016}, or purity \cite{PhysRevA.67.062104,Streltsov2016a}.

An important quantifier of quantum statistical speed is the quantum Fisher information. It can be derived as the quantum statistical speed associated with the (classical) Hellinger distance \cite{Hellinger1909,Bhattacharyya1946,Jeffreys1946,Matusita1967}
\begin{align}\label{eq:Hellinger}
\left(d_{2}(p,q)\right)^{2}=\frac{1}{2}\sum_x\left(\sqrt{p_x}-\sqrt{q_x}\right)^{2},
\end{align}
where $p=\{p_x\}_x$ and $q=\{q_x\}_x$ are probability distributions. Here we assumed the random variable $x$ to take on discrete values; in the case of a continuum of values, the sum must be replaced by an integral. The statistical speed associated with this distance is the (classical) Fisher information. Formally, we obtain the statistical speed from any statistical distance by quantifying the distance between infinitesimally close distributions taken from a one-parameter family $p_x(\theta)$ with the parameter $\theta$. Performing a Taylor expansion at $\theta_0$ for small values of $\theta$ yields the classical statistical speed
\begin{align}
s_2[p(\theta_0)]=\frac{d}{d \theta}d_{2}(p(\theta_0+\theta),p(\theta_0))=\sqrt{\frac{f_{2}[p(\theta_0)]}{8}},
\end{align}
where
\begin{align}\label{eq:FI}
f_2[p(\theta)]=\sum_xp_x(\theta)\left(\frac{\partial \log p_x(\theta)}{\partial \theta}\right)^2
\end{align}
is the Fisher information \cite{PhysRevD.23.357}. 

To connect these classical notions to the quantum case, we write $p_x=\mathrm{Tr}\{E_x\rho\}$ and $q_x=\mathrm{Tr}\{E_x\sigma\}$ as the measurement probabilities associated with the positive-operator-valued measure (POVM) defined by the $E_x\geq 0$, which satisfy $\sum_xE_x=\mathbb{I}$. For a given pair of quantum states $\rho$ and $\sigma$, the classical distance~(\ref{eq:Hellinger}) can be maximized over all possible choices of POVMs \cite{Fuchs1995}, leading to the Bures distance \cite{Bures1969,Huebner1992}
\begin{align}\label{eq:Bures}
D_2(\rho,\sigma):=\max_{\{E_x\}}d_{2}(p,q)=\sqrt{1-\mathcal{F}(\rho,\sigma)},
\end{align}
where $\mathcal{F}(\rho,\sigma)=\mathrm{Tr}\sqrt{\sqrt{\rho}\sigma\sqrt{\rho}}$ is the fidelity \cite{Uhlmann1976,Josza1994}. By expanding the quantum statistical distance for a one-parameter family of quantum states $\rho(\theta)$, we obtain the quantum statistical speed \cite{PhysRevLett.72.3439}
\begin{align}\label{eq:V2}
S_2[\rho(\theta_0)]=\frac{d}{d \theta}D_2(\rho(\theta_0+\theta),\rho(\theta_0))=\sqrt{\frac{F_{2}[\rho(\theta_0)]}{8}},
\end{align}
where $F_2[\rho(\theta)]=\mathrm{Tr}\{\rho(\theta)L_{\theta}^2\}$ is the quantum Fisher information \cite{PhysRevLett.72.3439} with the symmetric logarithmic derivative $L_{\theta}$, defined by 
\begin{align}\label{eq:SLD}
\frac{d\rho(\theta)}{d \theta}=\frac{1}{2}(L_{\theta}\rho(\theta)+\rho(\theta)L_{\theta}).
\end{align}
We obtain the same quantum statistical speed by maximizing the classical statistical speed over all possible POVMs \cite{PhysRevLett.72.3439}
\begin{align}\label{eq:QFI}
F_{2}[\rho(\theta)]:=\max_{\{E_x\}}f_2[p(\theta)],
\end{align}
where $p(\theta)=\{p_x(\theta)\}_x$ and $p_x(\theta)=\mathrm{Tr}\{E_x\rho(\theta)\}$. An optimal POVM is given by projectors onto the eigenstates of $L_{\theta}$.

The fundamental relevance of the statistical speed defined by the Fisher information and its corresponding quantum bound is expressed by the Cram\'{e}r-Rao bound, which states that \cite{Frechet1943,Rao1945,Cramer1946}
\begin{align}\label{eq:CR}
\Delta\theta_{\mathrm{est}}\geq\Delta\theta_{\mathrm{CR}}=\frac{1}{\sqrt{f_2[p(\theta)]}},
\end{align}
where $\theta_{\mathrm{est}}$ is an arbitrary mean-unbiased (i.e., the mean value of $\theta_{\mathrm{est}}$ yields $\theta$) estimator for $\theta$ with variance $(\Delta\theta_{\mathrm{est}})^2$. The Fisher information therefore defines the precision limit for mean-unbiased phase estimation procedures. By maximizing over all POVMs, we can further identify the quantum Cram\'{e}r-Rao bound \cite{PhysRevLett.72.3439}
\begin{align}\label{eq:QCR}
\Delta\theta_{\mathrm{CR}}\geq \Delta\theta_{\mathrm{QCR}}=\frac{1}{\sqrt{F_2[\rho(\theta)]}},
\end{align}
which sets the precision limit for quantum phase estimation \cite{paris2009,Giovannetti2011,Varenna,Toth2014,LucaRMP}.

The quantum statistical speed as given by the Fisher information furthermore exhibits a strict bound for separable states \cite{PhysRevLett.102.100401}. For instance, the evolution of separable $N$-qubit states generated by local Hamiltonians $J_{\mathbf{n}}=\frac{1}{2}\sum_i\mathbf{n}\cdot\boldsymbol{\sigma}_i$ cannot exceed a quantum statistical speed of $F_2[\rho_{\mathrm{sep}}]\leq N$, whereas with the aid of entangled states values up to $F_2[\rho]\leq N^2$ can be reached.

It should be emphasized that the existence of a quantum bound which can be saturated by an optimal measurement as in Eqs.~(\ref{eq:Bures}) and~(\ref{eq:QFI}) is highly nontrivial and has been found only for few other quantifiers of classical distances or speed \cite{Fuchs1995,Fuchs1999,NielsenChuang,PhysRevLett.72.3439,PhysRevLett.98.160501,Temme2015}. Most popular (pseudo)distance measures for quantum states are derived from classical expressions using heuristic arguments, such as replacing the probability distributions with density matrices and the sum with the trace operation \cite{footnotePseudoDistance}. While the resulting measures often have interesting properties, these procedures have no clear physical interpretation and their ambiguity often leads to several competing quantum versions for the same classical distance. Furthermore, such quantities can often only be accessed experimentally through full state tomography. The existence of an achievable quantum bound renders a quantity, in principle, observable without the need to construct the quantum state entirely. The realization of the optimal quantum measurement may however also be challenging in practice. Nevertheless, any other (suboptimal) measurement observable produces an accessible lower bound. Another distance with a well-defined quantum limit is the trace distance, which provides an operational interpretation of the distinguishability of two quantum states and plays an important role for quantum information theory \cite{Helstrom1976,Hayashi2006,NielsenChuang}.

The derivation of the Fisher information as classical and quantum statistical speed from the Hellinger distance suggests that similarly interesting quantifiers may be derived from other classical distances, following an analogous procedure. While the entanglement recognized by the quantum Fisher information is useful for quantum metrology, other distance measures may identify useful entanglement for different tasks. 

In this article, we first discuss a family of classical distance measures which continuously interpolates between the Hellinger and the trace distances (Sec.~\ref{sec:dalpha}). This leads to a family of statistical speed measures which contain the two important special cases of the trace speed and the quantum Fisher information. We analyze the properties of the trace speed, i.e., the quantum statistical speed associated with the trace distance (Sec.~\ref{sec:trace}). We further present the optimal measurement for pure states and some bounds for the general case in Sec.~\ref{sec:Falpha}. In Sec.~\ref{sec:Schatten} we introduce another one-parameter family of classical distance measures which, after maximizing over all quantum measurements, yields the Schatten-norm quantifiers of quantum statistical distance and speed. These quantities are experimentally accessible and again include the trace speed as a special case. We finally discuss the role of the trace speed for quantum technologies. In particular, we derive an analogue of the quantum Cram\'{e}r-Rao bound for median-unbiased quantum phase estimation in Sec.~\ref{sec:Applications}.

\section{Classical statistical distance and speed}\label{sec:dalpha}
In the following we will discuss a family of classical distance measures and their associated statistical speeds, which generalize the Fisher information~(\ref{eq:FI}).

\subsection{Classical statistical distance}
We begin by considering the family of distance measures \cite{Jeffreys1946,Boekee1977}
\begin{align}\label{eq:dalpha}
\left(d_{\alpha}(p,q)\right)^{\alpha}&=\frac{1}{2}\sum_x\left|p_x^{\frac{1}{\alpha}}-q_x^{\frac{1}{\alpha}}\right|^{\alpha}\notag\\&=\frac{1}{2}\sum_xp_x\left|1-\left(\frac{q_x}{p_x}\right)^{\frac{1}{\alpha}}\right|^{\alpha}
\end{align}
for any $\alpha\geq 1$. All these distances satisfy the following basic properties
\begin{itemize}
\item Non-negativity and normalization: $0\leq d_{\alpha}(p,q)\leq 1$,\\ where $d_{\alpha}(p,q)=0\quad\Leftrightarrow\quad p\equiv q$,
\item Symmetry: $d_{\alpha}(p,q)=d_{\alpha}(q,p)$,
\item Triangle inequality:\\ $d_{\alpha}(p_1,p_3)\leq d_{\alpha}(p_1,p_2)+d_{\alpha}(p_2,p_3)$.
\end{itemize}
Furthermore, they are ordered according to the following hierarchy:
\begin{itemize}
\item Ordering:\begin{align}(d_{\alpha}(p,q))^{\alpha}\leq (d_{\beta}(p,q))^{\beta}\:\mathrm{for}\:\alpha\geq \beta.\label{eq:ordering}\end{align}
\end{itemize}
The triangle inequality is satisfied as a consequence of the Minkowski inequality. The ordering relation~(\ref{eq:ordering}) is proven in Appendix~\ref{sec:ordering}. Two special cases are of particular interest. For $\alpha=2$, we recover the Hellinger distance~(\ref{eq:Hellinger}) and all the results entailed by it as summarized in Sec.~\ref{sec:intro}. For $\alpha=1$, this distance reduces to the Kolmogorov distance \cite{Fuchs1999,NielsenChuang}
\begin{align}\label{eq:Kolmogorov}
d_1(p,q)=\frac{1}{2}\sum_x|p_x-q_x|.
\end{align}
This distance is associated with a well-defined quantum distance, the trace distance. This will be discussed in further detail in Sec.~\ref{sec:trace}.

\subsection{Classical statistical speed: The generalized Fisher information}\label{sec:CLspeed}
To obtain the associated statistical speed, we parametrize the probability distribution $p(\theta)$ and expand it to first order in $\theta$ at $\theta_0$, i.e.,
\begin{align}\label{eq:expansionP}
p_x(\theta_0+\theta)=p_x(\theta_0)+\left.\frac{\partial p_x(\theta)}{\partial\theta}\right|_{\theta=\theta_0}\theta+\mathcal{O}(\theta^2).
\end{align}
The distance between $p(\theta_0)$ and $p(\theta_0+\theta)$ reads
\begin{align}
&\quad\left[d_{\alpha}(p(\theta_0+\theta),p(\theta_0))\right]^{\alpha}\notag\\&=\frac{1}{2}\sum_xp_x(\theta_0)\left|1-\left(\frac{p_x(\theta_0)+p_x'(\theta_0)\theta+\mathcal{O}(\theta^2)}{p_x(\theta_0)}\right)^{\frac{1}{\alpha}}\right|^{\alpha}\notag\\
&=\frac{1}{2}\sum_xp_x(\theta_0)\left|1-\left(1+\frac{p'_x(\theta_0)\theta+\mathcal{O}(\theta^2)}{p_x(\theta_0)}\right)^{\frac{1}{\alpha}}\right|^{\alpha}\notag\\
&=\frac{1}{2}\sum_xp_x(\theta_0)\left|\frac{p'_x(\theta_0)\theta+\mathcal{O}(\theta^2)}{\alpha p_x(\theta_0)}\right|^{\alpha},
\end{align}
where $p_x'(\theta_0)=\left.\frac{\partial p_x(\theta)}{\partial\theta}\right|_{\theta=\theta_0}$ and we used $(1+x)^{\alpha}=1+\alpha x+\mathcal{O}(x^2)$. Retaining only terms up to first order in $\theta$ in the numerator, we obtain
\begin{align}
\left[d_{\alpha}(p(\theta_0+\theta),p(\theta_0))\right]^{\alpha}=\frac{1}{2\alpha^{\alpha}}f_{\alpha}[p(\theta_0)]\theta^{\alpha}+\mathcal{O}(\theta^{2\alpha}),
\end{align}
with the generalized Fisher information
\begin{align}\label{eq:generalizedFisher}
f_{\alpha}[p(\theta)]=\sum_x p_x(\theta)\left|\frac{\partial}{\partial\theta}\log p_x(\theta)\right|^{\alpha}.
\end{align}
Consequently, the statistical speed is determined by
\begin{align}
s_{\alpha}[p(\theta_0)]=\frac{d}{d\theta}d_{\alpha}(p(\theta_0+\theta),p(\theta_0))=\frac{1}{\alpha}\left(\frac{f_{\alpha}(\theta_0)}{2}\right)^{\frac{1}{\alpha}}.
\end{align}

The generalized Fisher information has been studied in the literature on statistical inference and information theory \cite{Barankin1949,Boekee1977,Stangenhaus1977,Ferentinos1981,Lutwak2005,Bercher2012}. In Ref.~\cite{PhysRevA.87.034101} it was applied in the context of quantum estimation problems to a special class of probe states. In Ref.~\cite{Ferentinos1981}, parametric expansions of a series of classical distance and pseudodistance measures were studied. Interestingly, the statistical speed of the R\'{e}nyi divergence of order $\alpha$ \cite{Renyi1961} is given by $f_{2}$, scaled by a prefactor proportional to $\alpha$.

Next, following the literature, we briefly review basic properties of the $f_{\alpha}[p(\theta)]$, specifically, convexity, subadditivity, and ordering, and their role in the Barankin bounds. Then we will introduce a family of lower bounds which can be obtained from the moments of quantum observables in Sec.~\ref{sec:lbounds}.

\subsubsection{Convexity and subadditivity}
The generalized Fisher information is a convex function of $p(\theta)$ for $\alpha\geq 1$ \cite{Boekee1977a,Bercher2013}. Hence, for $p_x(\theta)=\sum_iq_ip^{(i)}_x(\theta)$, we have $f_{\alpha}[p_{x}(\theta)]\leq\sum_iq_if_{\alpha}[p_x^{(i)}(\theta)]$. Furthermore, also the quantity $\left(f_{\alpha}[p(\theta)]\right)^{\frac{1}{\alpha-1}}$ is convex in $p(\theta)$ for $1<\alpha\leq 2$ \cite{Boekee1977}.

For multiple measurements, i.e., probability distributions of the form $p_{\mathbf{x}}(\theta)=p_{x_1}(\theta)\cdots p_{x_m}(\theta)$, we have a subadditivity property: $\left(f_{\alpha}[p_{\mathbf{x}}(\theta)]\right)^{\frac{1}{\alpha}}\leq \sum_{i=1}^m\left(f_{\alpha}[p_{x_i}(\theta)]\right)^{\frac{1}{\alpha}}$ \cite{Boekee1977}. A strict additivity property holds only for the standard Fisher information $(\alpha=2$), specifically, $f_{2}[p_{\mathbf{x}}(\theta)]= \sum_{i=1}^mf_{2}[p_{x_i}(\theta)]$ \cite{Varenna}. 

\subsubsection{Ordering}
The generalized Fisher information obeys the following order: For $\beta\geq \alpha\geq 1$ we have \cite{Boekee1977}
\begin{align}\label{eq:monotonicity}
\left(f_\beta[p(\theta)]\right)^{\frac{1}{\beta}}\geq\left(f_\alpha[p(\theta)]\right)^{\frac{1}{\alpha}}.
\end{align}
A proof is provided in Appendix~\ref{app:monotonicity}. 

\subsubsection{Barankin bounds}\label{sec:Barankin}
Barankin \cite{Barankin1949} derived a family of generalized Cram\'{e}r-Rao bounds, introducing the generalized Fisher information. In general, the Barankin bounds are obtained by optimizing over a family of bounds for the $\beta$th absolute central moment. As a special case one obtains, for $\beta>1$ and an unbiased estimator $\theta_{\mathrm{est}}$,
\begin{align}\label{eq:Barankin}
\left(\sum_xp_x(\theta)|\theta_{\mathrm{est}}(x)-\theta|^{\beta}\right)^{\frac{1}{\beta}}\geq\frac{1}{(f_{\alpha}[p(\theta)])^{1/\alpha}},
\end{align}
where $\alpha=\beta/(\beta-1)$. These bounds can be derived in a way similar to Eq.~(\ref{eq:moments}), where the observable $M$ is replaced by the unbiased estimator and $g(\theta)=\theta$. Barankin further discusses conditions for the existence and uniqueness of estimators that minimize the $\beta$th central moment. These ideas were generalized even further in Refs.~\cite{Lutwak2005,Bercher2012,Bercher2013,Cianchi2014}. 

\subsubsection{Lower bounds}\label{sec:lbounds}
The generalized Fisher information can be bounded in terms of the absolute moments of an observable. Let $p(\theta)=\{p_x(\theta)\}_x$ be the probability distribution obtained from a quantum state $\rho(\theta)$ when measuring an observable $M$ with spectral decomposition $M=\sum_xm_x|x\rangle\langle x|$, i.e., $p_x(\theta)=\langle x|\rho(\theta)|x\rangle$. The expectation value is given by $\langle M\rangle_{\rho(\theta)}=\mathrm{Tr}\{M\rho(\theta)\}=\sum_xp_x(\theta)m_x$. For $\alpha,\beta\geq 1$ with $1/\alpha+1/\beta=1$ and an arbitrary function $g(\theta)$, the generalized Fisher information is bounded by
\begin{align}\label{eq:moments}
\left(f_{\alpha}[p(\theta)]\right)^{\frac{1}{\alpha}}\geq \frac{\left\vert\frac{d\langle M\rangle_{\rho(\theta)}}{d\theta}\right\vert}{\left(\sum_xp_x(\theta)|m_x-g(\theta)|^{\beta}\right)^{\frac{1}{\beta}}}.
\end{align}
This is proven in Appendix~\ref{app:moments}. The special case $\alpha=\beta=2$ is well known \cite{Varenna} and is used to bound the Fisher information with spin squeezing coefficients \cite{PhysRevLett.102.100401,ResolutionEnhanced}. For unitary evolution we obtain $\frac{d\langle M\rangle_{\rho(\theta)}}{d\theta}=-i\langle[M,H]\rangle_{\rho(\theta)}$ and with $g(\theta)=\langle M\rangle_{\rho(\theta)}$ we obtain the bound
\begin{align}
\left(f_{\alpha}[p(\theta)]\right)^{\frac{1}{\alpha}}\geq \frac{\left\vert\langle[M,H]\rangle_{\rho(\theta)}\right\vert}{\langle|M-\langle M\rangle_{\rho(\theta)}\mathbb{I}|^{\beta}\rangle_{\rho(\theta)}^{\frac{1}{\beta}}},
\end{align}
where the denominator contains the $\beta$th absolute central moment of $M$.

\section{Trace distance and trace speed}\label{sec:trace}
Let us now turn to the discussion of the associated quantum distance and speed. The quantum bounds can be determined for arbitrary states when $\alpha=1$ or $\alpha=2$. The case $\alpha=2$ was discussed in the Introduction. In this section we discuss the case $\alpha=1$. Some results on other values of $\alpha$ will be presented in Sec.~\ref{sec:Falpha}.

\subsection{Optimal quantum measurement}\label{sec:V1}
A well known result from quantum information theory states that the quantum distance associated to the classical Kolmogorov distance~(\ref{eq:Kolmogorov}) is the trace distance \cite{Fuchs1999,NielsenChuang,Hayashi2006}. Specifically, under the conditions that lead to Eq.~(\ref{eq:Bures}), we obtain
\begin{align}\label{eq:Trace}
D_1(\rho,\sigma)&:=\max_{\{E_x\}}d_{1}(p,q)\notag\\
&=\frac{1}{2}\|\rho-\sigma\|_1,
\end{align}
where $\|X\|_1=\mathrm{Tr}|X|$ is the trace norm with $|X|=\sqrt{X^{\dagger}X}$. By considering a family of quantum states parametrized by $\theta$ and expanding
\begin{align}
\rho(\theta_0+\theta)=\rho(\theta_0)+\left.\frac{d \rho(\theta)}{d\theta}\right|_{\theta=\theta_0}\theta+\mathcal{O}(\theta^2),
\end{align}
we obtain the corresponding quantum statistical speed
\begin{align}\label{eq:V1}
S_1[\rho(\theta_0)]:=\:&\frac{d}{d\theta}D_1(\rho(\theta_0+\theta),\rho(\theta_0))\notag\\=\:&\frac{1}{2}\mathrm{Tr}\left|\frac{d\rho(\theta_0)}{d\theta_0}\right|=\frac{1}{2}F_1[\rho(\theta_0)].
\end{align}
The quantity $F_1[\rho(\theta)]$ will be called the trace speed and can be interpreted as the generalized quantum Fisher information for $\alpha=1$.

Alternatively, we can derive the quantum statistical speed by maximizing the classical speed over all quantum measurements. Notice that the generalized Fisher information reduces for $\alpha=1$ to the expression
\begin{align}\label{eq:clFisher1}
f_{1}[p(\theta)]=\sum_x \left|\frac{\partial p_x(\theta)}{\partial\theta}\right|.
\end{align}
As is detailed in Appendix~\ref{app:F1}, we find indeed
\begin{align}\label{eq:F1}
F_{1}[\rho(\theta)]&:=\max_{\{E_x\}}f_1[p(\theta)]=\mathrm{Tr}\left|\frac{d\rho(\theta)}{d\theta}\right|.
\end{align}
An optimal measurement is achieved by the projectors on the eigenstates of $\frac{d\rho(\theta)}{d\theta}$.

\subsection{Properties of the trace speed}\label{sec:F1}
The trace speed can be expressed in terms of the trace norm as $F_{1}[\rho(\theta)]=\left\|\frac{d\rho(\theta)}{d\theta}\right\|_1$. This ensures basic properties of the trace speed such as convexity in $\rho(\theta)$ and subadditivity under product states of the form $\rho(\theta)=\rho_1(\theta)\otimes\cdots\otimes\rho_m(\theta)$ by virtue of the triangle inequality.

The trace speed was previously considered in Refs.~\cite{Marvian2014,PhysRevX.6.041044} and some basic properties were studied, focusing primarily on the case of unitary evolution. We will see that these results can be extended to nonunitary cases. Furthermore, the relationship of the trace speed with the quantum Fisher information can be understood by considering their common origin provided by the generalized Fisher information.

\subsubsection{Unitary evolution}
For unitary evolution generated by the Hamiltonian $H$, i.e., quantum states of the form
\begin{align}
\rho(\theta)=e^{-iH\theta}\rho e^{iH\theta},
\end{align}
the trace speed only depends on $\rho$ and $H$ and assumes the simple form
\begin{align}
F_1[\rho,H]=\mathrm{Tr}\left|\left[H,\rho\right]\right|.
\end{align}
This quantity has been proposed as a measure for asymmetry in Ref.~\cite{Marvian2014}.

\subsubsection{Relation to the quantum Fisher information}
The trace speed represents a lower bound for the quantum Fisher information: $F_{1}[\rho(\theta)]\leq \sqrt{F_2[\rho(\theta)]}$ \cite{PhysRevX.6.041044}. We can understand this relation as a direct consequence of the ordering~(\ref{eq:monotonicity}) in the special case of $\beta=2$ and $\alpha=1$. Let $\{E_x^{\max}\}_x$ denote the POVM that achieves the maximum $F_{1}[\rho(\theta)]=\max_{\{E_x\}}f_{1}[\{\mathrm{Tr}\{E_x\rho(\theta)\}\}_x]=f_{1}[\{\mathrm{Tr}\{E^{\max}_x\rho(\theta)\}\}_x]$. Then we have
\begin{align}\label{eq:F1F2bound}
F_{1}[\rho(\theta)] &\:= f_{1}[\{\mathrm{Tr}\{E^{\max}_x\rho(\theta)\}\}_x]\notag\\
&\stackrel{(\ref{eq:monotonicity})}{\leq} \sqrt{f_{2}[\{\mathrm{Tr}\{E^{\max}_x\rho(\theta)\}\}_x]} \notag\\
&\:\leq \sqrt{\max_{\{E_x\}}f_{2}[\{\mathrm{Tr}\{E_x\rho(\theta)\}\}_x]}=\sqrt{F_2[\rho(\theta)]}.
\end{align}
The same argument applies to the generalized quantum Fisher information with arbitrary $\alpha,\beta$, as will be discussed further in Sec.~\ref{sec:Falpha}.

\subsubsection{Pure states}\label{sec:purestates}
Remarkably, the bound~(\ref{eq:F1F2bound}) is saturated by any pure state, i.e., we have
\begin{align}\label{eq:F1F2pure}
F_1[\Psi(\theta)]=\sqrt{F_{2}[\Psi(\theta)]},
\end{align}
for arbitrary pure states $\Psi(\theta)$. This was shown for unitary evolution in Ref.~\cite{PhysRevX.6.041044} but holds for any dependence on $\theta$. In this case also the optimal quantum measurements coincide, i.e., the same POVM achieves the maximum in Eqs.~(\ref{eq:QFI}) and~(\ref{eq:F1}), respectively. This is proven in Appendix~\ref{app:F1F2pure}.

Let us return to the special case of unitary evolution, for which it is known that \cite{PhysRevLett.72.3439}
\begin{align}\label{eq:F2pure}
F_2[\Psi,H]=4(\Delta H)^2_{|\Psi\rangle},
\end{align}
where $(\Delta H)^2_{|\Psi\rangle}=\langle\Psi|H^2|\Psi\rangle-\langle\Psi|H|\Psi\rangle^2$ denotes the variance. With Eq.~(\ref{eq:F1F2pure}), this implies that \cite{Marvian2014}
\begin{align}\label{eq:F1pure}
F_1[\Psi,H]=2(\Delta H)_{|\Psi\rangle}.
\end{align}

\subsubsection{Non-Hermitian Hamiltonians}
Let us consider a time evolution described by a non-Hermitian Hamiltonian
\begin{align}
H_{\mathrm{eff}}=H-i\Gamma,
\end{align}
with $H=H^{\dagger}$ and $\Gamma=\Gamma^{\dagger}$. Effective Hamiltonians of this type are not self-adjoint and therefore do not represent observables. Nevertheless, they can be useful for the description of dissipative processes \cite{Moiseyev2011} and they appear in stochastic unravelings of open-system evolutions \cite{BreuerPetruccione2006}. The evolution is given by
\begin{align}\label{eq:nonHevo}
\frac{\partial\rho}{\partial \theta}=-i(H_{\mathrm{eff}}\rho-\rho H_{\mathrm{eff}}^{\dagger}).
\end{align}
As we show in Appendix~\ref{app:F1nonH}, the trace speed of a pure state under such a time evolution is given by
\begin{align}\label{eq:F1nonH}
F_{1}[\Psi,H-i\Gamma]
&=2\sqrt{\langle\Psi|H_{\mathrm{eff}}^{\dagger}H_{\mathrm{eff}}|\Psi\rangle-\langle\Psi|H|\Psi\rangle^2}.
\end{align}
This quantity can be interpreted as a generalization of the variance for a non-Hermitian operator $H_{\mathrm{eff}}$ with real part $H$. This follows from
\begin{align}\label{eq:ineq1}
&\quad\langle\Psi|H_{\mathrm{eff}}^{\dagger}H_{\mathrm{eff}}|\Psi\rangle-\langle\Psi|H|\Psi\rangle^2\notag\\&=\langle\Psi|(H_{\mathrm{eff}}-\langle\Psi|H|\Psi\rangle\mathbb{I})^{\dagger}(H_{\mathrm{eff}}-\langle\Psi|H|\Psi\rangle\mathbb{I})|\Psi\rangle\notag\\&\leq \langle\Psi|(H_{\mathrm{eff}}-r\mathbb{I})^{\dagger}(H_{\mathrm{eff}}-r\mathbb{I})|\Psi\rangle
\end{align}
for all $r\in\mathbb{R}$. This is in analogy to the fact that the mean-square deviation of a Hermitian operator is minimized when taking the mean value as central point, in which case it yields the variance. The inequality~(\ref{eq:ineq1}) is obtained immediately upon observing that the last line in~(\ref{eq:ineq1}) as a function of $r$ has a minimum at
\begin{align}
r=\langle\Psi|\frac{H_{\mathrm{eff}}+H^{\dagger}_{\mathrm{eff}}}{2}|\Psi\rangle.
\end{align}
For $H_{\mathrm{eff}}=H-i\Gamma$ this yields $r=\langle\Psi|H|\Psi\rangle$.

For $\Gamma=0$, Eq.~(\ref{eq:F1nonH}) reduces to the result~(\ref{eq:F1pure}) for standard unitary evolutions. Furthermore, by means of Eq.~(\ref{eq:F1F2pure}) we obtain the following result for the quantum Fisher information of pure states:
\begin{align}
F_2[\Psi,H-i\Gamma]=4\left(\langle\Psi|H_{\mathrm{eff}}^{\dagger}H_{\mathrm{eff}}|\Psi\rangle-\langle\Psi|H|\Psi\rangle^2\right).
\end{align}

\subsubsection{Temperature sensitivity}
As an example, we consider the temperature sensitivity of a thermal state $\rho(\beta)=e^{-\beta H}/\mathrm{Tr}e^{-\beta H}$, where $\beta=1/kT$ is the inverse temperature. We obtain
\begin{align}
F_1[\rho(\beta)]=\langle|H-\langle H\rangle\mathbb{I}|\rangle,
\end{align}
where $\langle X\rangle=\mathrm{Tr}\{X\rho(\beta)\}$, i.e., the right-hand side is given by the mean absolute deviation of the energy. Similarly, the quantum Fisher information yields the second central moment, $F_2[\rho(\beta)]=\langle(H-\langle H\rangle\mathbb{I})^2\rangle=(\Delta H)^2_{\rho(\beta)}$ \cite{PhysRevLett.99.100603}. Both expressions coincide with the classical expression for measurements in the energy eigenbasis, which shows that this is the optimal POVM in both cases. This is due to the fact that the eigenvectors do not change under variations of the parameter $\beta$ and thus the quantum expressions reduce to the classical ones. One can further directly confirm that the optimal POVMs in these cases coincide for the trace speed and quantum Fisher information, by observing that $\rho(\beta)$, $d\rho(\beta)/d\beta$, and the symmetric logarithmic derivative $L_{\beta}=\langle H\rangle\mathbb{I}-H$ [recall Eq.~(\ref{eq:SLD})] commute and thus share the same eigenbasis. The optimal measurements for Eqs.~(\ref{eq:QFI}) and~(\ref{eq:F1}) are given by projectors onto the eigenstates of $L_{\beta}$ and $d\rho(\beta)/d\beta$, respectively, and thus coincide. For this choice of measurement, the classical generalized Fisher information~(\ref{eq:generalizedFisher}) yields the central moment of order $\alpha$, i.e.,
\begin{align}\label{eq:energymean}
f_{\alpha}[p(\beta)]=\sum_mp_m(\beta)|\epsilon_m-\langle H\rangle|^{\alpha},
\end{align}
where $p_m(\beta)=e^{-\beta \epsilon_m}/Z(\beta)$, $Z(\beta)=\sum_me^{-\beta \epsilon_m}$, and $\epsilon_m$ are the eigenvalues of $H$.

\subsection{General upper bounds for the quantum statistical speed:\\The Heisenberg limit}\label{sec:HL}
A wide class of open-system quantum evolutions can be modeled as
\begin{align}\label{eq:Levo}
\frac{d\rho(\theta)}{d\theta}=\mathcal{L}[\rho(\theta)],
\end{align}
where $\mathcal{L}$ is a linear superoperator. If $\mathcal{L}$ is independent of $\theta$, it generates a dynamical semigroup, described in terms of a Lindblad master equation, whereas more general, non-Markovian evolutions are described if $\mathcal{L}$ is itself a function of $\theta$ \cite{BreuerPetruccione2006}. In the following, we consider the simple case of $\theta$-independent $\mathcal{L}$, but the results can be generalized straightforwardly. For unitary evolutions, the superoperator $\mathcal{L}$ is given by the commutator with the Hamiltonian $H$.

\subsubsection{General quantum evolution}
For any given superoperator, there exists an upper bound for the quantum statistical speed, given by
\begin{align}
F_1[\rho(\theta)]&\leq \sup_{\rho(\theta)}F_1[\rho(\theta)]=\|\mathcal{L}\|_{1}.
\end{align}
The quantity
\begin{align}\label{eq:SOPnorm1}
\|\mathcal{L}\|_{1}:=\sup_{\substack{\rho\geq 0\\\mathrm{Tr}\rho=1}}\|\mathcal{L}[\rho]\|_1
\end{align}
is a superoperator norm induced by density operators via the trace norm; see Refs.~\cite{Kitaev1997,Amosov2000,Watrous2005,Devetak2006} for similar constructions and applications in the context of completely positive maps. From convexity it follows that the supremum is attained by a pure state \cite{Watrous2005}, i.e.,
\begin{align}
\|\mathcal{L}\|_{1}=\sup_{|\Psi\rangle}\|\mathcal{L}[|\Psi\rangle\langle\Psi|]\|_1,
\end{align}
where $|\Psi\rangle$ is some unit vector. This quantity denotes the ultimate bound on the quantum statistical speed for a fixed generator $\mathcal{L}$. We remark that the superoperator norm~(\ref{eq:SOPnorm1}) is not stable under enlargements of the Hilbert space and extensions of $\mathcal{L}$ via a tensor product with an identity operator \cite{Kitaev1997}. The statistical speed can indeed increase if the local system is entangled with another, noninteracting subsystem. Clearly this operation would require additional resources as the full quantum state of both systems needs to be manipulated and measured to observe the enhanced statistical speed. Effectively, we are no longer quantifying the statistical speed of the original system, but that of a larger system. Therefore, unlike for the discrimination of quantum channels \cite{PhysRevA.71.062310}, the stability property is not a natural requirement in our context.

The bounds obtained for the trace speed immediately also define the corresponding bounds for the quantum Fisher information. This follows from the convexity of the quantum Fisher information, which leads to
\begin{align}\label{eq:HL}
F_2[\rho(\theta)]&\leq \max_{|\Psi(\theta)\rangle}F_2[|\Psi(\theta)\rangle\langle\Psi(\theta)|]\notag\\
&=\left(\max_{|\Psi(\theta)\rangle}F_1[|\Psi(\theta)\rangle\langle\Psi(\theta)|]\right)^2\notag\\
&=\|\mathcal{L}\|_{1}^2.
\end{align}
In the second step we used the equivalence of the two statistical speed measures~(\ref{eq:F1F2pure}) for pure states. This number can be interpreted as the most general form of the Heisenberg limit. According to Eqs.~(\ref{eq:CR}) and~(\ref{eq:QCR}), the value~(\ref{eq:HL}) defines the maximal achievable precision in a phase estimation procedure where the phase is imprinted through an evolution described by $\mathcal{L}$.

For Hamiltonian evolutions $\mathcal{L}_H[\rho]=-i[H,\rho]$, the quantum Fisher information of arbitrary states is bounded from above by the variance~\cite{PhysRevLett.72.3439,Varenna}, i.e., $F_2[\rho,H]\leq 4(\Delta H)^2_{\rho}$, where pure states lead to equality [Eq.~(\ref{eq:F2pure})]. Using an upper bound for the variance by Bhatia and Davis \cite{Bhatia2000}, we find
\begin{align}\label{eq:BD}
F_2[\rho,H]&\leq 4(\Delta H)^2_{\rho}\notag\\&\leq 4[\lambda_{\max}(H)-\langle H\rangle_{\rho}][\langle H\rangle_{\rho} -\lambda_{\min}(H)],
\end{align}
where $\langle H\rangle_{\rho}=\mathrm{Tr}\{\rho H\}$ and $\lambda_{\min,\max}(H)$ denote the smallest and largest eigenvalue of $H$, respectively. The upper bound reaches its maximum value for $\langle H\rangle_{\rho}=[\lambda_{\max}(H)+\lambda_{\min}(H)]/2$, leading to \cite{PhysRevLett.96.010401}
\begin{align}\label{eq:HLH}
F_2[\rho,H]&\leq \max_{|\Psi\rangle}F_2[\Psi,H]=\max_{|\Psi\rangle}4(\Delta H)^2_{|\Psi\rangle}\notag\\
&=(\lambda_{\max}(H)-\lambda_{\min}(H))^2.
\end{align}

For a collective spin $J_{\mathbf{n}}$, we have $\lambda_{\max}(J_{\mathbf{n}})=-\lambda_{\min}(J_{\mathbf{n}})=N/2$. From Eq.~(\ref{eq:BD}) we obtain the upper bound
\begin{align}
F_2[\rho,J_{\mathbf{n}}]&\leq N^2-4\langle J_{\mathbf{n}}\rangle_{\rho}^2,
\end{align}
which is saturated by states of the form $|\Psi\rangle=\alpha|N,0\rangle+\beta|0,N\rangle$, where $|n,m\rangle$ are Dicke states with $n$($m$) particles in the highest (lowest) eigenstate of $J_{\mathbf{n}}$ and complex amplitudes satisfying $|\alpha|^2+|\beta|^2=1$. A nonzero expectation value $\langle J_{\mathbf{n}}\rangle_{\rho}$ lowers the achievable sensitivity and the maximum value $N^2$ is reached by the above states when $|\alpha|^2=|\beta|^2=1/2$ \cite{PhysRevLett.96.010401}.

With Eq.~(\ref{eq:F1F2pure}), these bounds, which are attained by pure states, define upper bounds for the trace speed. From Eqs.~(\ref{eq:BD}) and~(\ref{eq:HLH}), we obtain
\begin{align}\label{eq:HLHF1}
F_1[\rho,H]&\leq \max_{|\Psi\rangle}F_1[\Psi,H]\notag\\&\leq 2\sqrt{[\lambda_{\max}(H)-\langle H\rangle_{\rho}][\langle H\rangle_{\rho} -\lambda_{\min}(H)]}\notag\\
&\leq\lambda_{\max}(H)-\lambda_{\min}(H).
\end{align}

Determining upper bounds for the quantum Fisher information beyond Hamiltonian evolutions is a difficult task \cite{Escher2011,PhysRevLett.112.120405,Rafal2012,PhysRevLett.116.120801,Toth2014}. Most existing results are obtained from a Kraus map representation of the solution of the dynamical equation, which is not unique and therefore demands an additional optimization. In contrast, the bound~(\ref{eq:HL}) is a function of the unique generator only.

\subsubsection{Non-Hermitian Hamiltonians} 
The expressions~(\ref{eq:HLH}) and~(\ref{eq:HLHF1}) for the maximal achievable quantum statistical speed can be generalized for non-Hermitian Hamiltonian evolutions using Eq.~(\ref{eq:F1nonH}). We obtain
\begin{align}\label{eq:nonHboundF1}
F_{1}[\rho,H_{\mathrm{eff}}]\leq 2\min_{r\in\mathbb{R}}\|H_{\mathrm{eff}}-r\mathbb{I} \|_{\infty},
\end{align}
where $\|X\|_{\infty}=\sigma_{\max}(X)$ is the operator norm, given by the largest singular value. 
The details are given in Appendix~\ref{app:nonHboundF1}. With Eq.~(\ref{eq:F1F2pure}) we have further identified the upper bound for the quantum Fisher information for non-Hermitian Hamiltonians, i.e., $F_2[\rho,H_{\mathrm{eff}}]\leq 4\min_{r\in\mathbb{R}}\|H_{\mathrm{eff}}-r\mathbb{I} \|_{\infty}^2$. Despite its simple appearance, the general solution of the minimum in Eq.~(\ref{eq:nonHboundF1}) is nontrivial. An illustrative example can be found in Appendix~\ref{app:nonHboundF1}.

\section{Towards a generalized quantum Fisher information}\label{sec:Falpha}
The preceding section focused on the trace speed, i.e., the generalized Fisher information for $\alpha=1$. We now consider the generalized quantum Fisher information~(\ref{eq:generalizedFisher}) for different values of $\alpha$, by maximizing over all possible quantum measurements, i.e.,
\begin{align}\label{eq:Falpha}
F_{\alpha}[\rho(\theta)]:=\max_{\{E_x\}}f_{\alpha}[p(\theta)].
\end{align}

\subsection{An optimal quantum measurement for pure states}\label{sec:optmeas}
Let us first consider the evolution of a pure state $|\Psi(\theta)\rangle=e^{-iH\theta}|\Psi_0\rangle$ under some Hamiltonian $H$. In this case we know that the optimal measurement coincides for both $\alpha=1$ and $\alpha=2$ and that it is given by the projectors onto the eigenvectors of $\frac{d\Psi(\theta)}{d\theta}$ with $\Psi(\theta)=|\Psi(\theta)\rangle\langle\Psi(\theta)|$ (recall Sec.~\ref{sec:purestates} and Appendix~\ref{app:F1F2pure}). Notice that the hierarchy~(\ref{eq:monotonicity}) extends to the quantum realm as a direct generalization of Eq.~(\ref{eq:F1F2bound}), i.e., in general we have for $1\leq\alpha\leq\beta$ that $F_{\beta}[\rho(\theta)]^{\frac{1}{\beta}}\geq F_{\alpha}[\rho(\theta)]^{\frac{1}{\alpha}}$. 

Let us now focus on the cases $1\leq\alpha\leq 2$. We obtain from the hierarchy
\begin{align}\label{eq:FalphaF1F2}
F_1[\rho(\theta)]\leq F_{\alpha}[\rho(\theta)]^{\frac{1}{\alpha}}\leq \sqrt{F_{2}[\rho(\theta)]}.
\end{align}
Among the generalized quantum Fisher information $F_{\alpha}[\rho(\theta)]^{\frac{1}{\alpha}}$ with $1\leq\alpha\leq 2$, this identifies the standard version $\sqrt{F_{2}[\rho(\theta)]}$ as the maximal one.

In the case of a pure state, the equality~(\ref{eq:F1F2pure}) implies that both inequalities in~(\ref{eq:FalphaF1F2}) are saturated with
\begin{align}\label{eq:FalphaVar}
F_{\alpha}[\Psi,H]^{\frac{1}{\alpha}}=2(\Delta H)_{|\Psi\rangle}.
\end{align}
The results~(\ref{eq:FalphaF1F2}) and~(\ref{eq:FalphaVar}) may be compared to Refs.~\cite{PhysRevA.87.032324,Yu2013}, where it was shown that the quantum Fisher information $F_{2}[\rho(\theta)]$ is the largest among all convex functions which reduce to the variance for pure states.

Since the inequalities~(\ref{eq:FalphaF1F2}) can only be saturated if the optimal measurements coincide, it follows that for pure states under unitary evolution, the maximum in Eq.~(\ref{eq:Falpha}) is reached by the same measurement that maximizes Eqs.~(\ref{eq:QFI}) and~(\ref{eq:F1}). This measurement is given by the projectors onto $\frac{d\Psi(\theta)}{d\theta}$. As is shown in Appendix~\ref{app:optmeas}, an optimal measurement for a pure state under unitary evolution can always be achieved with a projective measurement involving only two projectors. Explicitly, the optimal projectors are given as the projectors onto the states
\begin{align}\label{eq:optstates}
|\varphi_{\pm}\rangle=\frac{1}{\sqrt{2}}(|\Psi\rangle\pm i|\widetilde{\Psi}\rangle),
\end{align}
where
\begin{align}\label{eq:psitilde}
|\widetilde{\Psi}\rangle=\frac{(H-\langle\Psi|H|\Psi\rangle\mathbb{I})}{(\Delta H)_{|\Psi\rangle}}|\Psi\rangle
\end{align}
is a normalized state, orthogonal to $|\Psi\rangle$. This choice of measurement yields $f_{\alpha}[p(\theta)]^{\frac{1}{\alpha}}=2(\Delta H)_{|\Psi\rangle}$ for \textit{arbitrary} values of $\alpha\geq 1$. By Eq.~(\ref{eq:FalphaVar}) this constitutes an optimal measurement for $1\leq \alpha\leq 2$.

The identification of the optimal pure-state measurement recipe~(\ref{eq:optstates}) for arbitrary $\alpha$ improves and extends an important result for the quantum Fisher information and, thus, may lead to useful applications in the field of quantum metrology. To be precise, in the case of the quantum Fisher information ($\alpha=2$) it is known that a projective measurement involving the projector onto the initial state $|\Psi\rangle$ and some other orthogonal projector is optimal in the limit $\theta\rightarrow 0$ \cite{Varenna}. The projection~(\ref{eq:optstates}) involves two orthogonal states that are ``halfway'' between the initial state $|\Psi\rangle$ and the orthogonal state $|\widetilde{\Psi}\rangle$. This choice is always optimal, independently of the value of $\theta$.

\subsection{Upper and lower bounds}
The generalized quantum Fisher information with $1\leq \alpha\leq 2$ of arbitrary mixed states $\rho$ under unitary evolutions is bounded from above by
\begin{align}\label{eq:Falphaupperbound}
F_{\alpha}[\rho,H]^{\frac{1}{\alpha}}\leq 2(\Delta H)_{\rho}.
\end{align}
This bound is known for $\alpha=2$ \cite{PhysRevLett.72.3439}. For other values of $\alpha$ it follows from inequality~(\ref{eq:FalphaF1F2}).

The lower bounds~(\ref{eq:moments}) also hold for the generalized quantum Fisher information. Let us consider the case of a rotation of a collective spin, generated by $H=J_{\mathbf{n}_2}$, leading to $p_x(\theta)=\mathrm{Tr}\{E_x e^{-iJ_{\mathbf{n}_2}\theta}\rho e^{iJ_{\mathbf{n}_2}\theta}\}$. Using $[J_{\mathbf{n}_1},J_{\mathbf{n}_2}]=iJ_{\mathbf{n}_3}$ for three orthogonal directions $\mathbf{n}_1,\mathbf{n}_2,\mathbf{n}_3\in\mathbb{R}^3$, we obtain
\begin{align}\label{eq:crookedSpinSqueez}
F_{\alpha}[\rho,J_{\mathbf{n}_2}]^{\frac{1}{\alpha}}\geq f_{\alpha}[p(\theta)]^{\frac{1}{\alpha}}\geq \frac{\left\vert\langle J_{\mathbf{n}_3}\rangle_{\rho(\theta)}\right\vert}{\langle|J_{\mathbf{n}_1}-\langle J_{\mathbf{n}_1}\rangle_{\rho(\theta)}\mathbb{I}|^{\beta}\rangle_{\rho(\theta)}^{\frac{1}{\beta}}}.
\end{align}
For $1\leq \alpha\leq 2$ (which implies $\beta\geq 2$) the left-hand side of~(\ref{eq:crookedSpinSqueez}) can further be bounded by the variance $2(\Delta J_{\mathbf{n}_2})_{\rho}$ using Eq.~(\ref{eq:Falphaupperbound}). By employing bounds on the variance for separable states this allows us to generalize spin-squeezing coefficients \cite{SpinSqueezing} using moments of order $\beta$. For example, a fully separable state of $N$ qubits obeys $2(\Delta J_{\mathbf{n}_2})_{\rho_{\mathrm{sep}}}\leq \sqrt{N}$ (a more detailed account and other separability bounds will be provided in Sec.~\ref{sec:separable}), which leads to the condition $\xi_{\beta}[\rho_{\mathrm{sep}}]\geq 1$ for all separable states where
\begin{align}
\xi_{\beta}[\rho]:=\frac{\sqrt{N}\langle|J_{\mathbf{n}_1}-\langle J_{\mathbf{n}_1}\rangle_{\rho}\mathbb{I}|^{\beta}\rangle_{\rho}^{\frac{1}{\beta}}}{\left\vert\langle J_{\mathbf{n}_3}\rangle_{\rho}\right\vert},
\end{align}
and the directions $\mathbf{n}_1,\mathbf{n}_2,\mathbf{n}_3\in\mathbb{R}^3$ can be chosen in an optimal way. Notice that, in general [since $\beta\geq 2$, this follows from Eq.~(\ref{eq:generalmeanorder})],
\begin{align}
\xi_{\beta}[\rho]\geq \xi_{2}[\rho],
\end{align}
where $\xi_2$ coincides with the spin-squeezing coefficient introduced in Ref.~\cite{PhysRevA.46.R6797}.

For values of $\alpha>2$ the measurement discussed in Sec.~\ref{sec:optmeas} provides a lower bound to the generalized quantum Fisher information of pure states
\begin{align}
F_{\alpha}[\Psi,H]^{\frac{1}{\alpha}}\geq 2(\Delta H)_{|\Psi\rangle}.
\end{align}

In summary, the analytic expressions for the generalized quantum Fisher information~(\ref{eq:Falpha}) and the optimal measurement are available for arbitrary quantum states only for $\alpha=1$ and $\alpha=2$. We were further able to provide analytical results for pure states for all values of $1\leq \alpha\leq 2$. In the future, it may be possible to extend these results to arbitrary states and to a larger range of $\alpha$.

\subsection{Bounds for the quantum statistical distance}
The generalized quantum statistical distance associated with the classical distance~(\ref{eq:dalpha}) can be bounded with methods similar to those employed above for the generalized Fisher information. By combining the ordering relation~(\ref{eq:ordering}) with a maximization over all POVMs [in analogy to Eq.~(\ref{eq:F1F2bound})], we obtain $D_{\alpha}(\rho,\sigma)^{\alpha}\leq D_{\beta}(\rho,\sigma)^{\beta}$ for $\alpha\geq \beta$, and, in particular,
\begin{align}\label{eq:D1D2ineq}
D_2(\rho,\sigma)^2\leq D_{\alpha}(\rho,\sigma)^{\alpha} \leq D_{1}(\rho,\sigma)
\end{align}
for $1\leq \alpha\leq 2$, where $D_{\alpha}(\rho,\sigma)=\max_{\{E_x\}}d_{\alpha}(p,q)$ with $p_x=\mathrm{Tr}\{E_x\rho\}$ and $q_x=\mathrm{Tr}\{E_x\sigma\}$. Recall also the definitions~(\ref{eq:Bures}) and~(\ref{eq:Trace}). Unlike Eq.~(\ref{eq:FalphaF1F2}), the inequality~(\ref{eq:D1D2ineq}) is not saturated by pure states $\Psi =|\Psi\rangle\langle\Psi|$ and $\Phi=|\Phi\rangle\langle \Phi|$, but instead reads
\begin{align}
1-|\langle\Psi|\Phi\rangle|\leq D_{\alpha}(\Psi,\Phi)^{\alpha} \leq \sqrt{1-|\langle\Psi|\Phi\rangle|^2}.
\end{align}

\section{Schatten norm distance and speed}\label{sec:Schatten}
The family of classical distance measures~(\ref{eq:dalpha}) led us in special cases to the quantum Fisher information and the trace speed, respectively. These two are the only cases in which the corresponding quantum distance could be obtained for arbitrary quantum states by maximizing explicitly over all quantum measurements. In this section we discuss another family of classical distance measures whose quantum bounds can be found analytically for the full set of parameters and which includes the trace distance as a special case.

\subsection{Classical statistical distance and speed}
Consider the family of distance measures
\begin{align}\label{eq:newdalpha}
(\mathsf{d}_{\alpha}(p,q))^{\alpha}=\frac{1}{2}\sum_x|p_x-q_x|^{\alpha},
\end{align}
with $\alpha\geq 1$. For $\alpha=1$, we recover the Kolmogorov distance~(\ref{eq:Kolmogorov}). These distances satisfy all of the basic properties listed below Eq.~(\ref{eq:dalpha}). An ordering relation is inherited from the monotonicity of Schatten norms if the normalizing prefactor is removed, i.e., $2^{\frac{1}{\alpha}}\mathsf{d}_{\alpha}(p,q)\leq 2^{\frac{1}{\beta}}\mathsf{d}_{\beta}(p,q)$ for $\alpha\geq\beta$. Employing the perturbative expansion~(\ref{eq:expansionP}), we obtain
\begin{align}
\mathsf{d}_{\alpha}(p(\theta_0+\theta),p(\theta_0))=\left(\frac{1}{2}\sum_x\left|p'_x(\theta_0)\right|^{\alpha}\right)^{\frac{1}{\alpha}}\theta+\mathcal{O}(\theta^2)
\end{align}
and the classical statistical speed is given by
\begin{align}\label{eq:newvalpha}
\mathsf{s}_{\alpha}[p(\theta_0)]=\frac{d}{d\theta}\mathsf{d}_{\alpha}(p(\theta_0+\theta),p(\theta_0))=2^{-\frac{1}{\alpha}}\mathsf{f}_{\alpha}[p(\theta_0)],
\end{align}
where $\mathsf{f}_{\alpha}[p(\theta)]=\left(\sum_x\left|p'_x(\theta)\right|^{\alpha}\right)^{\frac{1}{\alpha}}$.

\subsection{Quantum statistical distance and speed}
The associated quantum distance is defined as
\begin{align}
\mathsf{D}_{\alpha}(\rho,\sigma):=\max_{\{E_x\}}\mathsf{d}_{\alpha}(p,q),
\end{align}
where $p_x=\mathrm{Tr}\{\rho E_x\}$ and $q_x=\mathrm{Tr}\{\sigma E_x\}$. In Appendix~\ref{app:Dalpha} we show that the bound is achieved by the Schatten norm distance
\begin{align}\label{eq:Dalpha}
(\mathsf{D}_{\alpha}(\rho,\sigma))^{\alpha}=\frac{1}{2}\mathrm{Tr}|\rho-\sigma|^{\alpha}.
\end{align}

Notice that the quantum distance $(\mathsf{D}_{\alpha}(\rho,\sigma))^{\alpha}$ is contractive under positive operations only if $\alpha=1$ (trace distance) \cite{Spehner2014}. For $\alpha=2$ we obtain the Hilbert-Schmidt distance, which allows for a simple evaluation as it does not require a diagonalization of the argument operator.

Following the methods of Sec.~\ref{sec:V1} (see also Appendix~\ref{app:Dalpha}), the associated quantum statistical speed is now straightforward to obtain and reads
\begin{align}\label{eq:Valpha}
\mathsf{S}_{\alpha}[\rho(\theta)]&=\max_{\{E_x\}}\mathsf{s}_{\alpha}[p(\theta)]\notag\\&=\left(\frac{1}{2}\mathrm{Tr}\left|\frac{d\rho(\theta)}{d\theta}\right|^{\alpha}\right)^{\frac{1}{\alpha}}\notag\\
&=2^{-\frac{1}{\alpha}}\mathsf{F}_{\alpha}[\rho(\theta)],
\end{align}
where we introduced the Schatten speed
\begin{align}\label{eq:SchattenSpeed}
\mathsf{F}_{\alpha}[\rho(\theta)]=\left\|\frac{d\rho(\theta)}{d\theta}\right\|_{\alpha}
\end{align}
and $\|X\|_{\alpha}=(\sum_i\sigma_i(X)^{\alpha})^{\frac{1}{\alpha}}$ is the Schatten $\alpha$-norm, which can be expressed as a function of the singular values $\sigma_i(X)$ of $X$. These are operator norms, i.e., they respect basic properties such as unitary invariance and the triangle inequality. This implies that the properties convexity and subadditivity hold for the entire family of statistical speed measures; recall the discussion of the special case $\alpha=1$ in Sec.~\ref{sec:F1}. The optimal measurement is again given by projections on the eigenstates of $\frac{d\rho(\theta)}{d\theta}$ (see also Appendix~\ref{app:F1}).

The natural ordering of Schatten norms implies the sequence of bounds for $1\leq \alpha\leq \beta\leq \infty$,
\begin{align}\label{eq:SchattenHierarchy}
F_{1}[\rho(\theta)]=\mathsf{F}_{1}[\rho(\theta)]\geq \mathsf{F}_{\alpha}[\rho(\theta)]\geq \mathsf{F}_{\beta}[\rho(\theta)]\geq \mathsf{F}_{\infty}[\rho(\theta)],
\end{align}
where $F_{1}[\rho(\theta)]$ was introduced in Eq.~(\ref{eq:F1}). With Eq.~(\ref{eq:F1F2bound}) this further leads to a sequence of lower bounds on the quantum Fisher information [and its generalizations; see Eq.~(\ref{eq:FalphaF1F2})], as was pointed out for the Hilbert-Schmidt case $\alpha=2$ in Refs.~\cite{PhysRevX.6.041044,Girolami2017}.

The results of Appendix~\ref{app:F1nonH} imply that for pure states under non-Hermitian evolutions~(\ref{eq:nonHevo}), one obtains
\begin{align}
&\quad\left(\mathsf{F}_{\alpha}[\Psi,H-i\Gamma]\right)^{\alpha}\notag\\&=\left(\sqrt{\langle\Psi|H_{\mathrm{eff}}^{\dagger}H_{\mathrm{eff}}|\Psi\rangle-\langle\Psi|H|\Psi\rangle^2} +\langle\Psi|\Gamma|\Psi\rangle\right)^{\alpha}\notag\\&\quad+ \left(\sqrt{\langle\Psi|H_{\mathrm{eff}}^{\dagger}H_{\mathrm{eff}}|\Psi\rangle-\langle\Psi|H|\Psi\rangle^2} -\langle\Psi|\Gamma|\Psi\rangle\right)^{\alpha}.
\end{align}
For $\Gamma=0$ this implies, with Eq.~(\ref{eq:Valpha}),
\begin{align}\label{eq:Valphapure}
\mathsf{S}_{\alpha}[\Psi,H]&=(\Delta H)_{|\Psi\rangle}
\end{align}
for all $\alpha\geq 1$.

\subsection{The Hilbert-Schmidt speed}
For $\alpha=2$ the quantum statistical speed is given by the particularly simple expression
\begin{align}
\mathsf{S}_{2}[\rho(\theta)]=\sqrt{\frac{1}{2}\mathrm{Tr}\left(\frac{d\rho(\theta)}{d\theta}\right)^2},
\end{align}
which does not require diagonalization of $\frac{d\rho(\theta)}{d\theta}$. For instance, in the case of a Hamiltonian evolution $\frac{d\rho(\theta)}{d\theta}=-i[H,\rho(\theta)]$, we find
\begin{align}
\mathsf{S}_{2}[\rho,H]&=\sqrt{-\frac{1}{2}\mathrm{Tr}\{[H,\rho][H,\rho]\}}\notag\\
&=\sqrt{\mathrm{Tr}\{\rho^2H^2\}-\mathrm{Tr}\{(H\rho)^2\}},
\end{align}
which reduces to Eq.~(\ref{eq:Valphapure}) for a pure state. 

A closely related quantity $S_{\mathrm{HS}}[\rho,H,\theta]=(\mathsf{D}_{2}(\rho,e^{-iH\theta}\rho e^{iH\theta})/\theta)^{2}$ was considered in Ref.~\cite{Zhang2016} to assess asymmetry \cite{Marvian2014}. It holds that $\lim_{\theta\rightarrow 0}S_{\mathrm{HS}}[\rho,H,\theta]=\mathsf{S}_2[\rho,H]^2$. The lower bound $4S_{\mathrm{HS}}[\rho,H,\theta]\leq F_2[\rho,H]$ was shown for all $\rho$, $H$, and $\theta$ \cite{Zhang2016}. Furthermore, in  Ref.~\cite{PhysRevLett.110.050403} an upper bound for the rate of change (quantified by the relative purity, which is closely related to $\mathsf{D}_{2}$) was derived, which coincides with the Hilbert-Schmidt speed.

\subsection{Lower bounding the quantum statistical speed from experimental data}
To measure the functions~(\ref{eq:SchattenSpeed}) experimentally, we recall their definitions in terms of an optimal POVM over the classical statistical speed, obtained by a parametric expansion of Eq.~(\ref{eq:newdalpha}). A lower bound for the quantum statistical speed can be extracted following a protocol which was implemented experimentally for the Fisher information in a system of cold atoms in Ref.~\cite{Strobel2014}. In the first step, the probability distribution $p_x(\theta)$ for the possible measurement outcomes of an observable is determined by repeated measurements on many copies of the state $\rho(\theta)$. Next the system is allowed to evolve for a short time $\delta\theta$, producing the state $\rho(\theta+\delta\theta)$. Again one determines the full probability distribution $p_x(\theta+\delta\theta)$ for the same observable. Repeating this for a number of small evolution steps leads to a family of probability distributions $\{p_x(\theta),p_x(\theta+\delta\theta),p_x(\theta+2\delta\theta),p_x(\theta+3\delta\theta),\dots\}$ from which the classical distance~(\ref{eq:newdalpha}) can be determined. By fitting the obtained statistical distance as a function of $\theta$ to the function~(\ref{eq:newvalpha}), the classical statistical speed $\mathsf{s}[p(\theta)]$ can be obtained. This quantity provides a lower bound on~(\ref{eq:Valpha}) and hence on (\ref{eq:SchattenSpeed}) for any observable. In the case of an optimally chosen observable the quantities coincide.

\subsection{Lower bounds from the evolution of mean values}
A lower bound for the classical statistical speed~(\ref{eq:newvalpha}) can be obtained from the changes of the mean value of an arbitrary observable $M$ with $\theta$. We show in Appendix~\ref{app:SchattenBound} using the H\"older inequality that the classical Schatten speed is bounded from below by
\begin{align}\label{eq:SchattenBound}
\mathsf{f}_{\alpha}[p(\theta)]\geq \frac{\left\vert\frac{d\langle M\rangle_{\rho(\theta)}}{d\theta}\right\vert}{\|M-c\mathbb{I}\|_{\beta}},
\end{align}
where $1/\alpha+1/\beta=1$, $\alpha,\beta\in(1,\infty)$ and an arbitrary constant $c\in\mathbb{R}$. Since $\mathsf{F}_{\alpha}[\rho(\theta)]=\max_{\{E_x\}}\mathsf{f}_{\alpha}[p(\theta)]\geq \mathsf{f}_{\alpha}[p(\theta)]$ [recall Eqs.~(\ref{eq:newvalpha}) and~(\ref{eq:Valpha})] the bound also applies to the quantum statistical speed~(\ref{eq:SchattenSpeed}).  This is in analogy to the bounds on the generalized quantum Fisher information presented in Sec.~\ref{sec:lbounds}. Notice, however, that in contrast to Eq.~(\ref{eq:moments}), the denominator on the right-hand side does not depend on the probability distribution, and, in order to obtain the tightest possible bound, the operator norm $\|M-c\mathbb{I}\|_{\beta}$ can be minimized by a suitable choice of $c$ which only depends on $M$. Hence, a bound for $\mathsf{F}_{\alpha}[\rho(\theta)]$ can be found by observing the sensitivity of $\langle M\rangle_{\rho(\theta)}$ to changes of $\theta$.

\section{Quantum statistical speed limits for separable states}\label{sec:separable}
We have discussed different measures for the quantum statistical speed of arbitrary quantum states under any quantum evolution. There exist upper bounds for the maximal attainable statistical speed by any quantum state for a fixed evolution. As was seen in Sec.~\ref{sec:HL}, for the quantum Fisher information and the trace speed, these bounds are given in terms of the superoperator norm of the generator of the evolution, induced by the trace norm [Eq.~(\ref{eq:HL})]. Similarly, the maximal attainable quantum statistical speed for the Schatten norm quantifiers~(\ref{eq:SchattenSpeed}) is given by their respective induced superoperator norms
\begin{align}
\mathsf{F}_{\alpha}[\rho(\theta)]&\leq \sup_{\rho(\theta)}\mathsf{F}_{\alpha}[\rho(\theta)]=\|\mathcal{L}\|_{\alpha},
\end{align}
where
\begin{align}
\|\mathcal{L}\|_{\alpha}:=\sup_{\substack{\rho\geq 0\\\mathrm{Tr}\rho=1}}\|\mathcal{L}[\rho]\|_{\alpha}=\sup_{|\Psi\rangle}\|\mathcal{L}[|\Psi\rangle\langle\Psi|]\|_{\alpha}
\end{align}
generalizes Eq.~(\ref{eq:SOPnorm1}). All quantifiers of quantum statistical speed discussed in this paper, i.e., the quantum Fisher information~(\ref{eq:QFI}) and the Schatten norms~(\ref{eq:SchattenSpeed}), which include the trace speed as special case, are convex functions of the quantum states. They will therefore assume their maximal values for pure states. Notice that by~(\ref{eq:SchattenHierarchy}) the limits for $\alpha>1$ represent bounds for the case $\alpha=1$. 

These upper bounds can usually only be reached by entangled states and a stricter limit on the quantum statistical speed exists for separable states \cite{PhysRevLett.102.100401,PhysRevA.85.022321,PhysRevA.85.022322,PezzePNAS2016,Gessner2016}. Determining these bounds can be extremely useful to assess the potential of quantum correlations, and to provide means to unambiguously detect them.

\subsection{Upper bounds on the quantum statistical speed for separable states}\label{sec:seplim}
A quantum state of an $N$-partite quantum system is fully separable when it can be represented as a convex combination of product states. If the phase shift is imprinted only locally into the subsystems, i.e.,
\begin{align}
\rho_{\mathrm{sep}}(\theta)=\sum_ip_i\rho^{(i)}_1(\theta)\otimes\cdots\otimes\rho^{(i)}_N(\theta),
\end{align}
the quantum statistical speed limits for separable states are given by
\begin{align}\label{eq:sepbound}
\sup_{\rho_{\mathrm{sep}}}\mathsf{F}_{\alpha}[\rho_{\mathrm{sep}}(\theta)]=\sup_{\Psi_1\otimes\cdots\otimes\Psi_N}\left\|[\mathcal{L}[\Psi_1\otimes\cdots\otimes\Psi_N]\right\|_{\alpha},
\end{align}
where we used again the convexity property. Similarly, one can determine speed limits for states that are $k$-separable or separable in a specific partition \cite{PhysRevA.85.022321,PhysRevA.85.022322,BosonicSqueezing}. One may go further and investigate bounds for more general classes of correlated quantum states \cite{Adesso2016}.

Interestingly, for a pure state under Hamiltonian evolution all quantifiers of quantum statistical speed considered here reduce to simple functions of the variance of $H$, recall Eqs.~(\ref{eq:F2pure}),~(\ref{eq:F1pure}) and~(\ref{eq:Valphapure}). The bounds~(\ref{eq:sepbound}) for separable states under local Hamiltonians thus reduce to the corresponding bounds on the variance which are well known. For example, $N$-qubit states $\rho_k$ that are $k$-separable, i.e., states in which not more than $k$ particles are entangled, obey the bound \cite{PhysRevA.71.052302,PhysRevA.85.022321,PhysRevA.85.022322}
\begin{align}\label{eq:sepbound1}
\max_{\rho_k}(\Delta J_{\mathbf{n}})^2_{\rho_k}=\frac{sk^2+r^2}{4},
\end{align}
where $s=\lfloor \frac{N}{k}\rfloor$, $r=N-sk$, and the generating Hamiltonian $J_{\mathbf{n}}$ is a sum of local qubit Hamiltonians. Combining this with Eqs.~(\ref{eq:Valpha}) and~(\ref{eq:Valphapure}), we obtain the bounds
\begin{align}\label{eq:SchattenkSep}
\mathsf{F}_{\alpha}[\rho_k,J_{\mathbf{n}}]\leq \sup_{\rho_k}\mathsf{F}_{\alpha}[\rho_k,J_{\mathbf{n}}]=2^{\frac{1-\alpha}{\alpha}}\sqrt{sk^2+r^2}
\end{align}
for the Schatten speed of order $\alpha$ of $k$-separable states. For instance, with $\alpha=1$ and $k=1$ we find that states whose trace speed exceeds $F_1[\rho,J_{\mathbf{n}}]=\sqrt{N}$ are necessarily entangled, i.e., they cannot be fully separable. These results can be easily generalized beyond qubit systems.

These bounds represent necessary conditions for $k$-separability, however, they are not violated by all entangled states, not even all pure states \cite{PhysRevA.82.012337}. One can further sharpen these bounds and provide additional information on the microscopic distribution of quantum correlations among the subsystems by using state-dependent bounds. To be more specific, let us define by $\mathcal{A}=\{\mathcal{A}_1,\cdots,\mathcal{A}_l\}$ a partition where the $\mathcal{A}_i$ describe families of subsystems. We call states $\mathcal{A}$-separable if they do not exhibit quantum correlations among the subsystems in $\mathcal{A}$, whereas quantum correlations can be present within each of the $\mathcal{A}_i$. For a Hamiltonian $H_{\mathcal{A}}=\sum_{k=1}^lH_k$ that is local in the partition $\mathcal{A}$, the variance of $\mathcal{A}$-separable states $\rho_{\mathcal{A}}$ is bounded by \cite{Gessner2016,ResolutionEnhanced,BosonicSqueezing}
\begin{align}\label{eq:sepbound2}
(\Delta H_{\mathcal{A}})^2_{\rho_{\mathcal{A}}}\leq \sum_{k=1}^l(\Delta H_k)^2_{\rho_{\mathcal{A}}}.
\end{align}
This condition is in fact necessary and sufficient for separability of all pure states, i.e., for each entangled pure state there exists at least one Hamiltonian $H_{\mathcal{A}}$ for which the bound is violated \cite{Gessner2016,BosonicSqueezing}. Again, in combination with Eqs.~(\ref{eq:Valpha}) and~(\ref{eq:Valphapure}), we find the following bound for $\mathcal{A}$-separable states:
\begin{align}\label{eq:SchattenASep}
\mathsf{F}_{\alpha}[\rho_{\mathcal{A}},H_{\mathcal{A}}]\leq 2^{\frac{1}{\alpha}}\sqrt{\sum_{k=1}^l(\Delta H_k)^2_{\rho_{\mathcal{A}}}}.
\end{align}
If the operators $H_k$ have a bounded spectrum, the local variances on the right-hand side may be further bounded using Eq.~(\ref{eq:BD}).

For pure states under Hamiltonian evolutions all quantum statistical speed measures are equivalent from the point of view of entanglement detection, since in these cases all separability bounds are equivalent to the variance. For mixed states, the hierarchies~(\ref{eq:F1F2bound}) and~(\ref{eq:SchattenHierarchy}) show that the quantum Fisher information is the most effective of all entanglement witnesses considered. Among all Schatten measures of quantum statistical speed, the trace speed detects the largest class of entangled states.

The expression~(\ref{eq:sepbound}) enables us to explicitly determine the separability bounds also for more general, possibly nonunitary evolutions. These methods are expected to be especially convenient for the Hilbert-Schmidt case $\alpha=2$, which can be combined effectively with vectorization techniques.

Alternatively, one may combine the additivity and convexity properties of the quantum Fisher information \cite{Varenna} with Eq.~(\ref{eq:F1F2pure}) to derive separability bounds for a general local generator $\mathcal{L}=\sum_{i=1}^N\mathcal{L}_i$ as
\begin{align}
F_2[\rho_{\mathrm{sep}}(\theta)]&\leq\sup_{\rho_{\mathrm{sep}}}F_2[\rho_{\mathrm{sep}}(\theta)]\notag\\
&\leq \sum_{i=1}^N\sup_{|\Psi_i(\theta)\rangle}F_2[\Psi_i(\theta)]\notag\\
&= \sum_{i=1}^N\sup_{|\Psi_i(\theta)\rangle}F_1[\Psi_i(\theta)]^2= \sum_{i=1}^N\|\mathcal{L}_i\|_1^2.
\end{align}
In the case of local Hamiltonian evolutions we recover the well-known bounds \cite{PhysRevLett.96.010401,PhysRevLett.102.100401,Varenna} with Eq.~(\ref{eq:HLHF1}).

Using Eq.~(\ref{eq:nonHboundF1}) we obtain that separable states under a local non-Hermitian Hamiltonian evolution with $H_{\mathrm{eff}}=\sum_{i=1}^NH_{i,\mathrm{eff}}$ have a maximal quantum Fisher information of
\begin{align}
F_2[\rho_{\mathrm{sep}},H_{\mathrm{eff}}]\leq 4\sum_{i=1}^N\min_{r_i\in\mathbb{R}}\|H_{i,\mathrm{eff}}-r_i\mathbb{I} \|_{\infty}^2.
\end{align}

\subsection{Relation to quantum speed limits}
Quantum speed limits have been investigated extensively in the literature, originally motivated by searches for a quantitative energy-time uncertainty relation \cite{Mandelstam1945,Margolus1998,PhysRevLett.65.1697,PhysRevA.67.052109,Braunstein1996}. Recently, these approaches have been generalized to nonunitary quantum evolutions and open systems \cite{PhysRevLett.110.050402,PhysRevLett.110.050403,Taddei2014,PhysRevX.6.021031,Deffner2017}. Following a general geometric approach \cite{PhysRevD.23.357,PhysRevLett.110.050402,PhysRevX.6.021031}, the length of a curve between two quantum states $\rho(0),\rho(\tau)$, parametrized by the dynamical evolution $[0,\tau]\ni t\longmapsto\rho(t)$, is given in terms of the quantum statistical speed $S$ as
\begin{align}\label{eq:geometricspeedlimit}
l[\rho(0),\rho(\tau)] =\int_0^{\tau} dtS[\rho(t)].
\end{align}
The shortest possible length $l_{\mathrm{gd}}$ is achieved for a geodesic evolution $\rho(t)$. The inequality $l_{\mathrm{gd}}[\rho(0),\rho(\tau)]\leq l[\rho(0),\rho(\tau)]$ is sometimes referred to as a geometric quantum speed limit \cite{PhysRevLett.110.050402,PhysRevX.6.021031,Deffner2017}. If the geodesic distance $l_{\mathrm{gd}}$ associated with the metric defined by $S$ is known, as is the case, e.g., for the quantum Fisher information $F_2$ \cite{PhysRevD.23.357}, this can be used to derive fundamental bounds on the required time to evolve from one state to another \cite{PhysRevLett.110.050402,PhysRevX.6.021031,Deffner2017}.

We have identified limits for the quantum statistical speed (recall also Sec.~\ref{sec:HL}) of two families of speed measures ($S=F_{\alpha}$ and $S=\mathsf{F}_{\alpha}$), which lead to upper bounds on the right-hand side of Eq.~(\ref{eq:geometricspeedlimit}). In multipartite systems, separable quantum states are unable to saturate these bounds and we further derived speed limits for separable quantum states (Sec.~\ref{sec:seplim}). The determination of the geodesic distance for the metrics induced by these measures of quantum statistical speed, however, remains open for future investigations.

\section{Applications and interpretations of the trace speed}\label{sec:Applications}
Entanglement is considered a key resource for quantum information theory. However, in many cases, a clear identification of the technological advantage provided by entangled states is far from obvious. Usually, not all entangled states can lead to the desired quantum gain and the set of useful states depends on the task at hand. For example, not all entangled states are recognized as such by the quantum Fisher information; it only recognizes those states that are useful for quantum metrology \cite{PhysRevLett.102.100401}. The trace speed~(\ref{eq:F1}), in turn, being a lower bound to the quantum Fisher information [Eq.~(\ref{eq:F1F2bound})], while exhibiting an equivalent separability bound (Sec.~\ref{sec:separable}), recognizes only a subset of the states that are useful for quantum metrology. This raises the question if there exists a quantum technology for which the trace speed emerges as a natural measure and therefore identifies those states that are useful for this specific task. In this section we discuss two possible technologies for which entangled states characterized by the trace speed are useful.

\subsection{Pairwise distinguishability of quantum states}
The trace distance allows for a natural interpretation in terms of the distinguishability of two quantum states by a single optimal measurement \cite{Helstrom1976,Hayashi2006,NielsenChuang}. In this scheme, Alice prepares one of two states $\rho$ or $\sigma$ with equal probability and sends it to Bob, who tries to determine which of the two states he received with an optimal measurement. It can be shown that the probability for Bob to successfully identify the correct state is given by \cite{Hayashi2006}
\begin{align}\label{eq:Bob}
P=\frac{1+D_1(\rho,\sigma)}{2},
\end{align}
where $D_1(\rho,\sigma)$ was defined in Eq.~(\ref{eq:Trace}). The result can be generalized to include a bias towards one of the two states (see, e.g., Ref.~\cite{PhysRevA.92.042108}). In any case, this scheme relies on the prior information about the two possible states and their probabilities. Its intrinsic Bayesian nature and its restriction to two states mark the difference from other tasks related to the distinguishability of quantum states, such as the one presented in Ref.~\cite{PhysRevD.23.357}, whose interpretation is linked to the Bures distance rather than the trace distance.

Based on this interpretation we can imagine a quantum game whose figure of merit directly links to the trace speed. Picture, for instance, that Alice tries to send a binary message to Bob, where each bit is encoded in one of two quantum states. Alice prepares the two states by manipulating one of two identical copies of the same initial state during an evolution time $\theta$. She then sends the unchanged initial state $\rho(0)$ to transmit the information 0 and the evolved states $\rho(\theta)$ for 1. Ideally, she would try to make $\rho(\theta)$ orthogonal to $\rho(0)$, in which case Bob's probability~(\ref{eq:Bob}) to read the message correctly reaches one. If, however, for some reason, Alice's preparation device only allows for small evolution times $\theta$, a high trace speed of $\rho(\theta)$ becomes crucial to minimize Bob's error. In this case, \textit{only} the entanglement recognized by $F_1[\rho(\theta)]$ helps Alice to generate pairs of states that are more easily distinguishable than any pair that could be created from an initially separable state.

\subsection{Median-unbiased quantum phase estimation}
The main objective of the quantum theory of phase estimation is the minimization of the variance between the estimated and actual value of a parameter of interest. The variance is a natural cost function for mean-unbiased estimators and Gaussian distributions. In most practical cases these concepts apply, enforced by the central limit theorem. Mean-unbiased phase estimation is, however, just one of the possible strategies to point estimation and in some situations it can be beneficial to consider different concepts of unbiasedness \cite{Brown1947,Lehmann1951,THS3}. Furthermore, the variance may not exist, in which case the central limit theorem does not apply; an example of particular relevance in physics is the Lorentz (Cauchy) distribution. In these cases, alternative measures of statistical dispersion besides the variance are more natural. These alternative cost functions, in turn, lead to different analogs of the Cram\'{e}r-Rao bound \cite{Barankin1949,Alamo1964,Stangenhaus1977,Sung1988}; recall also the Barankin bounds which were discussed in Sec.~\ref{sec:Barankin}. In this section we discuss the example of median-unbiased phase estimation and use our results on the trace speed to introduce a corresponding quantum bound which can be saturated by an optimal quantum measurement.

Let us consider an estimator $\theta_{\mathrm{est}}(x)$ for $\theta$ as a function of a random variable $x$ with the continuous probability distribution $p_x(\theta)$. We denote by $g_{\theta_{\mathrm{est}}}(y|\theta)$ the probability for the estimator $\theta_{\mathrm{est}}$ to take on the value $y$. An estimator $\theta_{\mathrm{est}}$ is median unbiased if $\theta$ coincides with the median of $g_{\theta_{\mathrm{est}}}$, i.e.,
\begin{align}\label{eq:medianunbiased}
\frac{1}{2}=\int_{-\infty}^\theta dyg_{\theta_{\mathrm{est}}}(y|\theta)=\int\limits_{\{x:\:\theta_{\mathrm{est}}(x)\leq \theta\}}dxp_x(\theta).
\end{align}
A median-unbiased estimator thus balances the frequency of over- and underestimation. In contrast to mean-unbiased estimators, median-unbiased estimators are invariant under one-to-one transformations: If $\theta_{\mathrm{est}}$ is median unbiased for $\theta$ then $f(\theta_{\mathrm{est}})$ is median unbiased for $f(\theta)$ for injective functions $f$. It is easy to verify that median-unbiased estimators minimize the mean absolute deviation $\int dx|\theta_{\mathrm{est}}(x)-\theta|p_x(\theta)$, whereas mean-unbiased estimators minimize the variance. The mean absolute deviation therefore offers a natural cost function for median-unbiased estimators. However, a general Cram\'{e}r-Rao-type bound cannot be derived for this function \cite{Alamo1964,Sung1988}; recall also that the Barankin bounds~(\ref{eq:Barankin}) do not apply for $\beta=1$. Instead, a suitable bound can be obtained based on an analog of the central limit theorem for the sample median.

Let us therefore first recall the standard form of the central limit theorem (see, e.g., Ref.~\cite{THS3}): Let $x$ be a random variable with probability density $p_x(\theta)$ with mean $\theta$ and variance $\sigma^2$. The distribution of sample-mean values obtained from random samples of fixed size $m$ asymptotically (for many repeated samples of fixed size $m$) approaches a normal distribution with center $\theta$ and variance $\sigma^2/m$. From the standard deviation of the sample-mean distribution, we identify $\sigma$ as a natural cost function. Now let us consider the following variation of the above theorem for the sample median (see, e.g., Theorem 11.2.8 in Ref.~\cite{THS3}): Let $x$ be a random variable with probability density $p_x(\theta)$ with median $\theta$. The distribution of sample-median values obtained from random samples of fixed size $m$ asymptotically approaches a normal distribution with center $\theta$ and variance $1/4mp_{\theta}(\theta)^2$. Here the standard deviation of the sample median leads to $1/2p_{\theta}(\theta)$, which has been suggested as a natural quantifier of statistical dispersion in this context \cite{Alamo1964,Sung1988}.

Based on the above reasoning, we employ the quantity $1/2g_{\theta_{\mathrm{est}}}(\theta|\theta)$ as a measure of dispersion for median-unbiased estimators. For any median-unbiased estimator $\theta_{\mathrm{est}}$ the following holds \cite{Stangenhaus1977} (see Refs.~\cite{Sung1988,So1994} for further generalizations):
\begin{align}\label{eq:Stangenhaus}
\frac{1}{2g_{\theta_{\mathrm{est}}}(\theta|\theta)}\geq \frac{1}{f_1[p(\theta)]},
\end{align}
where $f_1[p(\theta)]$ is the generalized Fisher information~(\ref{eq:generalizedFisher}) for $\alpha=1$. 

To summarize, if $\theta_{\mathrm{est}}(x_1,\dots,x_m)$ is a median-unbiased estimator for the parameter $\theta$, obtained from a sample of $m$ events, we obtain a statistical distribution of $\theta_{\mathrm{est}}$ described by the probability density $g_{\theta_{\mathrm{est}}}(y|\theta)$ whose median is $\theta$. Repeating the estimation $n\gg 1$ times (each time with a fixed sample size of $m$) leads to a statistical distribution of the medians, which is normal and has a variance of $1/4mg_{\theta_{\mathrm{est}}}(\theta|\theta)^2$. The quantity $\frac{1}{2g_{\theta_{\mathrm{est}}}(\theta|\theta)}$ thus quantifies the uncertainty of the estimation. According to Eq.~(\ref{eq:Stangenhaus}), the ultimate precision bound for median-unbiased estimators is given by $\frac{1}{f_1[p(\theta)]}$.

Going further, we may assume that the parameter $\theta$ is imprinted in a quantum state $\rho(\theta)$ and thus the obtained probability distributions depend on the choice of quantum measurement. Using Eq.~(\ref{eq:F1}), we can now define the quantum bound associated with the classical bound~(\ref{eq:Stangenhaus}) as
\begin{align}
\frac{1}{2g_{\theta_{\mathrm{est}}}(\theta|\theta)}\geq \frac{1}{\max_{\{E_x\}}f_1[p(\theta)]}=\frac{1}{F_1[\rho(\theta)]}.
\end{align}
We emphasize again that this bound is saturable by an optimal measurement. This relation reveals the fundamental relevance of the trace speed for median-unbiased quantum phase estimation protocols: The trace speed determines the precision bounds for median-unbiased quantum phase estimation in analogy to the role of the quantum Fisher information in the standard quantum Cram\'{e}r-Rao bound~(\ref{eq:QCR}).

We can further deduce that there exist limits for the precision of median-unbiased quantum phase estimation when only classically correlated states are available as resources. The subset of entangled states which is recognized as such by the separability criterion of the trace speed [states that violate Eq.~(\ref{eq:SchattenkSep}) for $\alpha=1$] defines the set of useful entangled states for this specific technological task.

\section{Conclusions}
We have shown that the family of Schatten norm distance and speed is accessible by an optimal quantum measurement from the classical probability distributions which are observable in experiments. This family includes the trace speed~(\ref{eq:F1}) as a case of particular interest, which can be rigorously linked to the quantum Fisher information. This link is provided by introducing the generalized Fisher information which produces the trace distance and the quantum Fisher information as special cases. We have further shown that the generalized quantum Fisher information for all $1\leq\alpha\leq 2$ reduces to the variance for pure states and we identified an optimal measurement strategy which involves only two projectors and does not depend on the value of the phase $\theta$. The Schatten speed quantifiers further allow for a determination of upper bounds for the statistical speed under arbitrary evolutions using induced superoperator norms. This can be used to derive bounds on the statistical speed of separable states leading to experimentally accessible entanglement witnesses, following methods similar to those reported in Ref.~\cite{Strobel2014}. Moreover, the Schatten speed provides computable bounds for the quantum Fisher information.

Besides the well-known aspect of parameter sensitivity, the concept of statistical speed is also intimately related to the concept of asymmetry. To assess asymmetry, the variations of a given state under a generator of transformations are quantified \cite{Marvian2014}. The Schatten speed introduced here thus provides a family of observable asymmetry quantifiers, providing direct links to multipartite entanglement. This has been pointed out in Refs.~\cite{Girolami2017,Zhang2016} for the case $\alpha=2$, where the Hilbert-Schmidt norm was measured directly by letting two identical copies of the system interfere. Our results suggest that a potentially more powerful entanglement witness can be obtained, e.g., by extracting the trace speed from the measurement statistics, involving only a single copy of the system.

In the recent literature a series of quantum speed limits have been derived \cite{PhysRevA.67.052109,PhysRevLett.110.050403,PhysRevLett.110.050402,PhysRevX.6.021031}. These are bounds on the required time to evolve from one state to another and are expressed as an integral over the statistical speed along a trajectory in state space. The results presented in this article can be used in this context, giving rise to a family of experimentally accessible quantum speed measures with a geometric interpretation.

An important unsettled question in quantum information theory concerns the precise role of entanglement for quantum technologies. It is well known that entanglement does not provide a universal resource for all tasks of quantum information science. By identifying specific witnesses of entanglement as the figure of merit for specific technological tasks, it becomes possible to characterize the set of entangled quantum states that are useful for this particular task, thereby shedding light onto this question. The quantum Fisher information achieves this for the case of quantum metrology \cite{PhysRevLett.102.100401}. The trace speed recognizes a smaller set of entangled states. We have discussed two possible technological interpretations for the trace speed, related to the pairwise distinguishability of parametrically prepared quantum states and to median-unbiased quantum phase estimation, respectively. A trace speed above the classical limit therefore defines a necessary and sufficient criterion for useful entanglement for these tasks.

\section*{Acknowledgments}
M.G. acknowledges support from the Alexander von Humboldt foundation.

\appendix
\section{Proof of Eq.~(\ref{eq:ordering})}\label{sec:ordering}
In this appendix we prove that the distance~(\ref{eq:dalpha}) obeys the following ordering $(d_{\alpha}(p,q))^{\alpha}\leq (d_{\beta}(p,q))^{\beta}$ for $\alpha\geq \beta$.

We first show that for $a,b\geq 0$, $n\in\mathbb{R}$, and $n\geq 1$
\begin{align}\label{eq:abn}
|a-b|^n\leq |a^n-b^n|.
\end{align}
To this end, we demonstrate that
\begin{align}\label{eq:lemma1}
(1-p)^n\leq 1-p^n
\end{align}
holds for arbitrary $0\leq p\leq 1$, $n\in\mathbb{R}$, and $n\geq 1$. For $n=1$ the statement is trivial. For $n>1$, the function $f(p)=1-p^n-(1-p)^n$ with domain $p\in[0,1]$ has a unique maximum at $p=1/2$. This can be seen from the derivative $f'(p)=n[(1-p)^{n-1}-p^{n-1}]$, which is positive for $p\in[0,\frac{1}{2})$, zero at $p=\frac{1}{2}$, and negative for $p\in(\frac{1}{2},1]$. The second derivative $f''(p)=-n(n-1)[(1-p)^{n-2}+p^{n-2}]<0$ confirms the maximum. Since $f(p)$ is monotonically increasing in the interval $p\in[0,\frac{1}{2})$ and symmetric about the maximum, $f(p)=f(1-p)$, we have $f(p)\geq f(0)=0$ for all $p\in[0,1]$ and Eq.~(\ref{eq:lemma1}) follows.

Next we consider $a\geq b>0$ [in the cases where one or both variables are zero, Eq.~(\ref{eq:abn}) holds trivially] and use Eq.~(\ref{eq:lemma1}) with $p=b/a$. We obtain
\begin{align}
\left(1-\frac{b}{a}\right)^n\leq 1-\left(\frac{b}{a}\right)^n.
\end{align}
Multiplying both sides by $a^n$ yields $\left(a-b\right)^n\leq a^n-b^n$. In the case of $b\geq a$ we employ $p=a/b$ and obtain $\left(b-a\right)^n\leq b^n-a^n$. This completes the proof of Eq.~(\ref{eq:abn}). By replacing $a\rightarrow a^{\frac{1}{n}}$ and $b\rightarrow b^{\frac{1}{n}}$ in Eq.~(\ref{eq:abn}), we find that the reverse inequality holds for $n\leq 1$.

We now apply Eq.~(\ref{eq:abn}) with $a=p_x^{\frac{1}{\alpha}}$, $b=q_x^{\frac{1}{\alpha}}$ and $n=\alpha/\beta$, where $\alpha\geq \beta\geq 1$. We obtain
\begin{align}
\left|p_x^{\frac{1}{\alpha}}-q_x^{\frac{1}{\alpha}}\right|^{\alpha}\leq\left|p_x^{\frac{1}{\beta}}-q_x^{\frac{1}{\beta}}\right|^{\beta}.
\end{align}
Summation over $x$ yields the ordering relation~(\ref{eq:ordering}) of the $(d_{\alpha}(p,q))^{\alpha}$. This generalizes the relation between Hellinger and Kolmogorov distances \cite{NielsenChuang}, which is recovered for $\alpha=2$ and $\beta=1$.

\section{Proof of Eq.~(\ref{eq:monotonicity})}\label{app:monotonicity}
We first recall a general ordering relation for absolute moments. For $\beta\geq \alpha\geq 0$, a probability distribution $p_x$, and an arbitrary function $\varphi_x$, the following holds:
\begin{align}\label{eq:generalmeanorder}
\left(\sum_xp_x|\varphi_x|^{\alpha}\right)^{\frac{1}{\alpha}}\leq \left(\sum_xp_x|\varphi_x|^{\beta}\right)^{\frac{1}{\beta}}.
\end{align}
This inequality follows from Jensen's inequality, which states that $h(\sum_xp_xg_x)\leq \sum_xp_xh(g_x)$ for an arbitrary function $g_x$ and a convex function $h$. Equation~(\ref{eq:generalmeanorder}) can be obtained by applying Jensen's inequality for the mean value of $g_x=|\varphi_x|^{\alpha}$ with the convex function $h(y)=y^{\frac{\beta}{\alpha}}$, where $\beta>\alpha$.

The ordering relation~(\ref{eq:monotonicity}) for the generalized Fisher information now follows from Eq.~(\ref{eq:generalmeanorder}) for $p_x=p_x(\theta)$ and $\varphi_x=\frac{\partial}{\partial \theta}\log p_x(\theta)$ \cite{Boekee1977}.

\section{Proof of Eq.~(\ref{eq:moments})}\label{app:moments}
H\"{o}lder's inequality states that for $p,q\in(1,\infty)$ and $1/p+1/q=1$, the following holds:
\begin{align}\label{eq:holder}
\sum_x|f_xg_x|\leq\left(\sum_x|f_x|^p\right)^{\frac{1}{p}}\left(\sum_x|g_x|^q\right)^{\frac{1}{q}}.
\end{align}

To prove Eq.~(\ref{eq:moments}), we use $|\sum_xf_xg_x|\leq \sum_x|f_xg_x|$ and then apply H\"older's inequality~(\ref{eq:holder}) with $f_x=(m_x-g(\theta))p_x(\theta)^{\frac{1}{p}}$ and $g_x=p_x(\theta)^{-\frac{1}{p}}\frac{dp_x(\theta)}{d\theta}$ for $p=\beta$ and $q=\alpha$. The claim follows after noticing that $\sum_xf_xg_x=\sum_x(m_x-g(\theta))\frac{dp_x(\theta)}{d\theta}=\frac{d\langle M\rangle_{\rho(\theta)}}{d\theta}$, due to $\sum_xg(\theta)\frac{dp_x(\theta)}{d\theta}=g(\theta)\frac{d}{d\theta}\sum_xp_x(\theta)=0$ \cite{Varenna}.

The moments in the denominator on the right-hand side of Eq.~(\ref{eq:moments}) are furthermore ordered according to Eq.~(\ref{eq:generalmeanorder}).

\section{Proof of Eq.~(\ref{eq:F1})}\label{app:F1}
In this appendix, we prove Eq.~(\ref{eq:F1}), which demonstrates that the diagram in Fig.~\ref{fig:diagram} commutes for the trace distance. We first notice that for any Hermitian operator $X=X^{\dagger}$,
\begin{align}\label{eq:optmeas}
\max_{\{E_x\}}\sum_x|\mathrm{Tr}\{E_xX\}|=\mathrm{Tr}|X|.
\end{align}
To prove this, we use the spectral decomposition of $X$, separating positive and negative eigenvalues as
\begin{align}\label{eq:JordanHahn}
X&=\underbrace{\sum_{\lambda_i>0}\lambda_i|\lambda_i\rangle\langle \lambda_i|}_{X_+}+\underbrace{\sum_{\lambda_i<0}\lambda_i|\lambda_i\rangle\langle \lambda_i|}_{X_-}\notag\\
&=X_++X_-,
\end{align}
where $X_+$ and $-X_-$ are positive operators \cite{NielsenChuang}. This decomposition is also known as the Jordan-Hahn decomposition. Notice that $|X|=X_+-X_-$. We now have
\begin{align}
|\mathrm{Tr}\{E_xX\}|&=|\underbrace{\mathrm{Tr}\{E_xX_+\}}_{\geq 0}+\underbrace{\mathrm{Tr}\{E_xX_-\}}_{\leq 0}|\notag\\&\leq |\underbrace{\mathrm{Tr}\{E_xX_+\}}_{\geq 0}\underbrace{-\mathrm{Tr}\{E_xX_-\}}_{\geq 0}|\notag\\&=|\mathrm{Tr}\{E_x|X|\}|\notag\\&=\mathrm{Tr}\{E_x|X|\}.
\end{align}
Carrying out the sum over $x$, we obtain, using $\sum_xE_x=\mathbb{I}$,
\begin{align}\label{eq:optmeas1}
\sum_x|\mathrm{Tr}\{E_xX\}|\leq \mathrm{Tr}|X|.
\end{align}
Choosing a POVM with projectors $E_+=\sum_{\lambda_i>0}|\lambda_i\rangle\langle \lambda_i|$ and $E_-=\sum_{\lambda_i<0}|\lambda_i\rangle\langle \lambda_i|$, we obtain
\begin{align}
\sum_x|\mathrm{Tr}\{E_xX\}|=|\mathrm{Tr}\{E_+X\}|+|\mathrm{Tr}\{E_-X\}|=\mathrm{Tr}|X|.
\end{align}
We would obtain the same result for a measurement employing the rank-1 projectors $E_i=|\lambda_i\rangle\langle\lambda_i|$. Thus, the upper bound is achievable by an optimal projective measurement. This proves the statement~(\ref{eq:optmeas}). This result can be generalized to include the full class of Schatten norms, as will be done in Appendix~\ref{app:Dalpha}.

Optimizing now Eq.~(\ref{eq:clFisher1}) with $p_x(\theta)=\mathrm{Tr}\{E_x\rho(\theta)\}$, we find
\begin{align}
\max_{\{E_x\}}f_1[p(\theta)]&=\max_{\{E_x\}}\sum_x\left|\frac{\partial p_x(\theta)}{\partial\theta}\right|\notag\\
&=\max_{\{E_x\}}\sum_x\left|\mathrm{Tr}\{E_x\frac{d \rho(\theta)}{d\theta}\}\right|\notag\\
&=\mathrm{Tr}\left|\frac{d\rho(\theta)}{d\theta}\right|.
\end{align}
In the last step we used Eq.~(\ref{eq:optmeas}), based on the fact that the operator $d \rho(\theta)/d\theta$ is Hermitian.

\section{Proof of Eq.~(\ref{eq:F1F2pure})}\label{app:F1F2pure}
We now show that the bound~(\ref{eq:F1F2bound}) for the quantum Fisher information by the trace speed is saturated by arbitrary pure states for generic evolutions [Eq.~(\ref{eq:F1F2pure})]. This can be seen by using the correspondence of trace distance and fidelity for pure states. The inequality \cite{Fuchs1999}
\begin{align}\label{eq:FvdG}
D_1(\rho,\sigma)\leq \sqrt{1-\mathcal{F}(\rho,\sigma)^2}
\end{align}
becomes an equality when $\rho$ and $\sigma$ are pure states \cite{Kitaev1997}. In this case, we can express the trace distance as a function of the Bures distance. We insert [recall the definition~(\ref{eq:Bures})]
\begin{align}
\mathcal{F}(\Psi,\phi)^2&=(1-D^2_2(\Psi,\phi))^2
\end{align}
into Eq.~(\ref{eq:FvdG}) and obtain
\begin{align}
D_1(\Psi,\phi)=D_2(\Psi,\phi)\sqrt{2-D^2_2(\Psi,\phi)},
\end{align}
where $\Psi=|\Psi\rangle\langle\Psi|$ and $\phi=|\phi\rangle\langle\phi|$ are pure states. Introducing a parametric family of pure states $\Psi(\theta)=|\Psi(\theta)\rangle\langle\Psi(\theta)|$ and using
\begin{align}
D_2(\Psi(\theta_0+\theta),\Psi(\theta_0))=S_2[\Psi(\theta_0)]\theta+\mathcal{O}(\theta^2),
\end{align}
we obtain
\begin{align}
D_1(\Psi(\theta_0+\theta),\Psi(\theta_0))&=S_2[\Psi(\theta_0)]\theta\sqrt{2-S^2_2[\Psi(\theta_0)]\theta^2}+\mathcal{O}(\theta^2)\notag\\
&=\sqrt{2}S_2[\Psi(\theta_0)]\theta+\mathcal{O}(\theta^2).
\end{align}
By comparison with
\begin{align}
D_1(\Psi(\theta_0+\theta),\Psi(\theta_0))=S_1[\Psi(\theta_0)]\theta+\mathcal{O}(\theta^2),
\end{align}
we find that $S_1[\Psi(\theta_0)]=\sqrt{2}S_2[\Psi(\theta_0)]$, which, using Eqs.~(\ref{eq:V1}) and~(\ref{eq:V2}) finally yields Eq.~(\ref{eq:F1F2pure}).

We further show that the two coinciding quantum values~(\ref{eq:F1F2pure}) are obtained by the same optimal measurement, i.e., that the maxima in Eqs.~(\ref{eq:F1}) and~(\ref{eq:QFI}), respectively, are achieved by the same POVM. The first hint that this must be the case is given by the inequality~(\ref{eq:F1F2bound}), which can only be saturated if the maximal POVMs coincide. To confirm this explicitly, recall that the maximum in Eq.~(\ref{eq:F1}) is obtained by projectors onto the eigenstates of $\frac{d\Psi(\theta)}{d\theta}$. Generally, the optimal measurement to achieve the maximum~(\ref{eq:QFI}) is given by the projectors onto the eigenstates of the symmetric logarithmic derivative $L_{\theta}$ \cite{PhysRevLett.72.3439}, defined in Eq.~(\ref{eq:SLD}). For a pure state we may write $\Psi(\theta)^2=\Psi(\theta)$, leading to 
\begin{align}
\frac{d\Psi(\theta)}{d\theta}=\frac{d\Psi(\theta)^2}{d\theta}=\frac{d\Psi(\theta)}{d\theta}\Psi(\theta)+\Psi(\theta)\frac{d\Psi(\theta)}{d\theta}.
\end{align}
In comparison with the definition~(\ref{eq:SLD}), one observes that $L_{\theta}=2\frac{d\Psi(\theta)}{d\theta}$ for pure states. This implies that the eigenstates of $L_{\theta}$ coincide with those of $\frac{d\Psi(\theta)}{d\theta}$, and indeed the same measurement is optimal for both cases.

\section{Proof of Eq.~(\ref{eq:F1nonH})}\label{app:F1nonH}
Here we derive the trace speed for a pure state subject to a non-Hermitian time evolution~(\ref{eq:nonHevo}). In the first steps of the following derivation we employ methods similar to those used in Ref.~\cite{PhysRevX.6.041044} for a proof of Eq.~(\ref{eq:F1pure}). For a pure state, the trace speed reads
\begin{align}
F_{1}[\Psi,H-i\Gamma]=\left\|\frac{\partial|\Psi\rangle\langle\Psi|}{\partial \theta}\right\|_1&=\left\|H_{\mathrm{eff}}|\Psi\rangle\langle\Psi|-|\Psi\rangle\langle\Psi|H_{\mathrm{eff}}^{\dagger}\right\|_1\notag\\
&=\||\Phi'\rangle\langle\Psi|-|\Psi\rangle\langle\Phi'|\|_1,
\end{align}
where $|\Phi'\rangle=H_{\mathrm{eff}}|\Psi\rangle$. We have $\langle\Phi'|\Phi'\rangle=\langle\Psi|H_{\mathrm{eff}}^{\dagger}H_{\mathrm{eff}}|\Psi\rangle$, and the unit vector
\begin{align}
|\Phi\rangle=\frac{|\Phi'\rangle}{\sqrt{\langle\Phi'|\Phi'\rangle}}.
\end{align}
Hence,
\begin{align}\label{eq:dpsidtheta1norm}
\left\|\frac{\partial|\Psi\rangle\langle\Psi|}{\partial \theta}\right\|_1&=\sqrt{\langle\Psi|H_{\mathrm{eff}}^{\dagger}H_{\mathrm{eff}}|\Psi\rangle}\||\Phi\rangle\langle\Psi|-|\Psi\rangle\langle\Phi|\|_1.
\end{align}
The matrix under the norm is rank 2 and anti-Hermitian and can thus be diagonalized analytically. To do this, let us expand the unit vector $|\Phi\rangle$ in terms of $|\Psi\rangle$ and some orthonormal vector $|\Psi_{\perp}\rangle$, as
\begin{align}\label{eq:phi}
|\Phi\rangle=c_1|\Psi\rangle+c_2|\Psi_{\perp}\rangle.
\end{align}
It holds $\langle\Phi|\Phi\rangle=|c_1|^2+|c_2|^2=1$ and
\begin{align}\label{eq:c1}
c_1=\langle\Psi|\Phi\rangle=\frac{\langle\Psi|\Phi'\rangle}{\sqrt{\langle\Phi'|\Phi'\rangle}}=\frac{\langle\Psi|H_{\mathrm{eff}}|\Psi\rangle}{\sqrt{\langle\Psi|H_{\mathrm{eff}}^{\dagger}H_{\mathrm{eff}}|\Psi\rangle}}.
\end{align}
We thus have
\begin{align}\label{eq:dpsidtheta}
|\Phi\rangle\langle\Psi|-|\Psi\rangle\langle\Phi|=(c_1-c_1^*)|\Psi\rangle\langle\Psi|+c_2|\Psi_{\perp}\rangle\langle\Psi|-c_2^*|\Psi\rangle\langle\Psi_{\perp}|.
\end{align}
Determining the trace norm of the operator corresponds to finding the singular values of the matrix
\begin{align}
X=\begin{pmatrix}
2i \mathrm{Im}(c_1) & -c_2^*\\
c_2 & 0
\end{pmatrix}.
\end{align}
These, in turn, correspond to the square roots of the eigenvalues of the matrix
\begin{align}
X^{\dagger}X=-XX=\begin{pmatrix}
4 \mathrm{Im}(c_1)^2 + |c_2|^2 & 2i c_2^* \mathrm{Im}(c_1)\\
2i c_2 \mathrm{Im}(c_1) & |c_2|^2.
\end{pmatrix}
\end{align}
Straightforward diagonalization yields
\begin{align}
\|X\|_1&=\mathrm{Tr}\sqrt{X^{\dagger}X}\notag\\
&=\sqrt{|c_2|^2 + 2\mathrm{Im}(c_1)^2- 2\mathrm{Im}(c_1)\sqrt{|c_2|^2 + \mathrm{Im}(c_1)^2}}\notag\\&\quad + \sqrt{|c_2|^2 + 2\mathrm{Im}(c_1)^2+ 2\mathrm{Im}(c_1)\sqrt{|c_2|^2 + \mathrm{Im}(c_1)^2}}.
\end{align}
Inserting
\begin{align}
|c_2|^2=1-|c_1|^2=1-\mathrm{Im}(c_1)^2-\mathrm{Re}(c_1)^2
\end{align}
into the preceding line yields
\begin{align}
\|X\|_1
&=\left|\sqrt{1-\mathrm{Re}(c_1)^2} - \mathrm{Im}(c_1)\right| + \left|\sqrt{1-\mathrm{Re}(c_1)^2} + \mathrm{Im}(c_1)\right|.
\end{align}
We further notice that
\begin{align}
\mathrm{Re}(c_1)=\frac{\langle\Psi|H|\Psi\rangle}{\sqrt{\langle\Psi|H_{\mathrm{eff}}^{\dagger}H_{\mathrm{eff}}|\Psi\rangle}}
\end{align}
and
\begin{align}
\mathrm{Im}(c_1)=-\frac{\langle\Psi|\Gamma|\Psi\rangle}{\sqrt{\langle\Psi|H_{\mathrm{eff}}^{\dagger}H_{\mathrm{eff}}|\Psi\rangle}}.
\end{align}
This allows us to write the trace speed as
\begin{align}\label{eq:int1}
F_{1}[\Psi,H-i\Gamma]&=\sqrt{\langle\Psi|H_{\mathrm{eff}}^{\dagger}H_{\mathrm{eff}}|\Psi\rangle}\|X\|_1\\
&=\left|\sqrt{\langle\Psi|H_{\mathrm{eff}}^{\dagger}H_{\mathrm{eff}}|\Psi\rangle-\langle\Psi|H|\Psi\rangle^2} +\langle\Psi|\Gamma|\Psi\rangle\right| \notag\\&\quad+ \left|\sqrt{\langle\Psi|H_{\mathrm{eff}}^{\dagger}H_{\mathrm{eff}}|\Psi\rangle-\langle\Psi|H|\Psi\rangle^2} -\langle\Psi|\Gamma|\Psi\rangle\right|.\notag
\end{align}
Next consider that
\begin{align}
0&\leq \langle\Psi|(H_{\mathrm{eff}}-\langle\Psi|H_{\mathrm{eff}}|\Psi\rangle)^{\dagger}(H_{\mathrm{eff}}-\langle\Psi|H_{\mathrm{eff}}|\Psi\rangle)|\Psi\rangle\notag\\
&=\langle\Psi|H_{\mathrm{eff}}^{\dagger}H_{\mathrm{eff}}|\Psi\rangle-\langle\Psi|H_{\mathrm{eff}}|\Psi\rangle^*\langle\Psi|H_{\mathrm{eff}}|\Psi\rangle\notag\\&\quad-\langle\Psi|H_{\mathrm{eff}}|\Psi\rangle\langle\Psi|H_{\mathrm{eff}}|\Psi\rangle^*+\langle\Psi|H_{\mathrm{eff}}|\Psi\rangle^*\langle\Psi|H_{\mathrm{eff}}|\Psi\rangle\notag\\
&=\langle\Psi|H_{\mathrm{eff}}^{\dagger}H_{\mathrm{eff}}|\Psi\rangle-|\langle\Psi|H_{\mathrm{eff}}|\Psi\rangle|^2.
\end{align}
Furthermore,
\begin{align}
|\langle\Psi|H_{\mathrm{eff}}|\Psi\rangle|^2=|\langle\Psi|H|\Psi\rangle|^2+|\langle\Psi|\Gamma|\Psi\rangle|^2.
\end{align}
Hence, we find
\begin{align}
&\quad\sqrt{\langle\Psi|H_{\mathrm{eff}}^{\dagger}H_{\mathrm{eff}}|\Psi\rangle-\langle\Psi|H|\Psi\rangle^2}\notag\\
&=\sqrt{\langle\Psi|H_{\mathrm{eff}}^{\dagger}H_{\mathrm{eff}}|\Psi\rangle-|\langle\Psi|H_{\mathrm{eff}}|\Psi\rangle|^2+\langle\Psi|\Gamma|\Psi\rangle^2}\notag\\&\geq\sqrt{\langle\Psi|\Gamma|\Psi\rangle^2}=|\langle\Psi|\Gamma|\Psi\rangle|\geq\pm\langle\Psi|\Gamma|\Psi\rangle.
\end{align}
This allows us to remove the absolute values in Eq.~(\ref{eq:int1}), leading to
\begin{align}
F_{1}[\Psi,H-i\Gamma]
&=2\sqrt{\langle\Psi|H_{\mathrm{eff}}^{\dagger}H_{\mathrm{eff}}|\Psi\rangle-\langle\Psi|H|\Psi\rangle^2}.
\end{align}

\section{Proof of Eq.~(\ref{eq:nonHboundF1})}\label{app:nonHboundF1}
In this appendix we determine the maximum trace speed for non-Hermitian evolutions~(\ref{eq:nonHboundF1}). Due to convexity, the maximum will be attained by a pure state. Hence
\begin{align}
\sup_{\rho}F_{1}[\rho,H-i\Gamma]&=2\sup_{|\Psi\rangle}\sqrt{\langle\Psi|H_{\mathrm{eff}}^{\dagger}H_{\mathrm{eff}}|\Psi\rangle-\langle\Psi|H|\Psi\rangle^2}.
\end{align}
Thus, we write using~(\ref{eq:ineq1})
\begin{align}
\sup_{\rho}F_{1}[\rho,H-i\Gamma]&\leq 2 \sup_{|\Psi\rangle}\sqrt{\langle\Psi|(H_{\mathrm{eff}}-r\mathbb{I})^{\dagger}(H_{\mathrm{eff}}-r\mathbb{I})|\Psi\rangle}\notag\\
&= 2 \sup_{|\Psi\rangle}\|(H_{\mathrm{eff}}-r\mathbb{I})|\Psi\rangle \|_{l_2}\notag\\
&= 2\|H_{\mathrm{eff}}-r\mathbb{I} \|_{\infty},
\end{align}
where $\||\varphi\rangle\|_{l_2}=\sqrt{\langle\varphi|\varphi\rangle}$ is the (Euclidean) $l_2$-norm and $\|X\|_{\infty}=\sigma_{\max}(X)$ is the operator norm, given by the largest singular value. Since the above result holds for arbitrary $r$, we can further minimize the expression with respect to $r$, leading to Eq.~(\ref{eq:nonHboundF1}).

Determining the $r$ which minimizes the expression can, however, be a nontrivial task. The operator norm $\|H_{\mathrm{eff}}-r\mathbb{I} \|_{\infty}$ is given by the maximum singular value of $H_{\mathrm{eff}}-r\mathbb{I}$, i.e., the square root of the maximum eigenvalue of
\begin{align}
(H_{\mathrm{eff}}-r\mathbb{I})^{\dagger}(H_{\mathrm{eff}}-r\mathbb{I})=(H-r\mathbb{I})^2+\Gamma^2. 
\end{align}
Let us first consider the case $H=0$. In this case we obtain
\begin{align}
\min_r\|H_{\mathrm{eff}}-r\mathbb{I} \|_{\infty}=\min_r\sqrt{\lambda_{\max}(\Gamma)^2+r^2}=|\lambda_{\max}(\Gamma)|.
\end{align}

Even for $H\neq 0$, we could invoke the triangle inequality to derive the simple general upper bound
\begin{align}
\min_r\|H_{\mathrm{eff}}-r\mathbb{I} \|_{\infty}&=\min_r\|H-i\Gamma-r\mathbb{I} \|_{\infty}\notag\\
&\leq \|H\|_{\infty}+\min_r\|\Gamma-r\mathbb{I} \|_{\infty}\notag\\
&=|\lambda_{\max}(H)|+|\lambda_{\max}(\Gamma)|.
\end{align}
This bound can usually be improved. In fact, for $\Gamma=0$ we know that Eq.~(\ref{eq:HLHF1}) provides a sharper bound for Eq.~(\ref{eq:nonHboundF1}) than the above expression.

Let us illustrate a more general scenario with the $2\times 2$ case for $[H,\Gamma]=0$, which corresponds to a qubit subject to dephasing. The two eigenvalues of $(H-r\mathbb{I})^2+\Gamma^2$ are given by
\begin{align}
(\lambda_{i}(H)-r)^2+\lambda_i(\Gamma)^2
\end{align}
for $i\in\{\max,\min\}$. As a function of $r$ they describe two positive parabolas, with minima at $\lambda_{\max}(H)$ and $\lambda_{\min}(H)$, respectively. If the larger of the two minima, i.e., $\lambda_{\max}(\Gamma)^2$, is larger than the entire smaller parabola, i.e., $\lambda_{\max}(\Gamma)^2\geq (\lambda_{\min}(H)-r)^2+\lambda_{\min}(\Gamma)^2$ for all $r$, we can identity $|\lambda_{\max}(\Gamma)|$ as the minimal operator norm. If this is not the case, the largest eigenvalue after minimizing over $r$ is given by the value of the two parabolas at their crossing point at
\begin{align}
r_0=\frac{\lambda_{\max}(H) + \lambda_{\min}(H)}{2} + \frac{\lambda_{\max}(\Gamma)^2 - \lambda_{\min}(\Gamma)^2}{2 (\lambda_{\max}(H) - \lambda_{\min}(H))}.
\end{align}
This value is given by
\begin{align}
y_{0}&=\frac{\lambda_{\max}(\Gamma)^2+\lambda_{\min}(\Gamma)^2}{2}+\frac{(\lambda_{\max}(\Gamma)^2-\lambda_{\min}(\Gamma)^2)^2}{4(\lambda_{\max}(H)-\lambda_{\min}(H))^2}\notag\\&\quad+\frac{(\lambda_{\max}(H)-\lambda_{\min}(H))^2}{4}.
\end{align}
Since $y_{0}\geq \lambda_{\max}(\Gamma)^2$, we have
\begin{align}
\min_r\|H_{\mathrm{eff}}-r\mathbb{I} \|_{\infty}\leq \sqrt{y_{0}}.
\end{align}
For $\Gamma=0$ we recover Eq.~(\ref{eq:HLHF1}) by inserting this result into Eq.~(\ref{eq:nonHboundF1}).

\section{Optimal quantum measurement for pure states}\label{app:optmeas}
For pure states $|\Psi(\theta)\rangle$ under unitary evolution and projective measurements $\{E(x)=|x\rangle\langle x|\}_x$, we obtain $p_x(\theta)=|\langle x|\Psi(\theta)\rangle|^2$. This leads to
\begin{align}
\frac{\partial p_x(\theta)}{\partial\theta}&=\langle x|\frac{d\Psi(\theta)}{d\theta}|x\rangle\notag\\&=-i\langle x|H|\Psi(\theta)\rangle\langle\Psi(\theta)|x\rangle+i\langle x|\Psi(\theta)\rangle\langle\Psi(\theta)|H|x\rangle\notag\\
&=2\mathrm{Im}(\langle x|H|\Psi(\theta)\rangle\langle\Psi(\theta)|x\rangle).
\end{align}
Inserting these results into Eq.~(\ref{eq:generalizedFisher}), the generalized Fisher information can be written as
\begin{align}\label{eq:falphaweak}
f_{\alpha}[p(\theta)]&=\sum_x p_x(\theta)\left|\frac{1}{p_x(\theta)}\frac{\partial p_x(\theta)}{\partial\theta}\right|^{\alpha}\notag\\
&=2^{\alpha}\sum_x |\langle x|\Psi(\theta)\rangle|^2\left|\mathrm{Im}\left(\frac{\langle x|H|\Psi(\theta)\rangle}{\langle x|\Psi(\theta)\rangle}\right)\right|^{\alpha}.
\end{align}
We point out that this corresponds to the absolute moments of order $\alpha$ of the imaginary parts of the weak values for the observable $H$ with initial state $|\Psi(\theta)\rangle$ and final states $|x\rangle$ \cite{PhysRevLett.60.1351}. Consequently, the generalized quantum Fisher information~(\ref{eq:Falpha}) can be interpreted as the maximum of this quantity over all sets of final states.

An optimal quantum measurement which maximizes Eq.~(\ref{eq:Falpha}) for $1\leq \alpha\leq 2$ is achieved by the projectors onto $\frac{d\Psi(\theta)}{d\theta}$. We now explicitly construct such a measurement and demonstrate that it achieves the value~(\ref{eq:FalphaVar}) for all $\alpha$. Recall from Eq.~(\ref{eq:dpsidtheta1norm}) that this operator is of rank 2; thus only two projectors will be needed. Here we consider unitary evolution, generated by the Hamiltonian $H$. From Eq.~(\ref{eq:c1}) it follows, for $H_{\mathrm{eff}}=H$, that $c_1-c_1^*=0$ and hence, with Eq.~(\ref{eq:dpsidtheta}),
\begin{align}
\left.\frac{d\Psi(\theta)}{d\theta}\right|_{\theta=\theta_0}&=-iH|\Psi\rangle\langle\Psi|+i|\Psi\rangle\langle\Psi|H\notag\\&=-i\sqrt{\langle\Psi|H^2|\Psi\rangle}(c_2|\Psi_{\perp}\rangle\langle\Psi|-c_2^*|\Psi\rangle\langle\Psi_{\perp}|),
\end{align}
with $|\Psi\rangle=|\Psi(\theta_0)\rangle$. The eigenstates of $\frac{d\Psi(\theta)}{d\theta}$ are given by
\begin{align}
|\varphi_{\pm}\rangle=\frac{1}{\sqrt{2}}(|\Psi\rangle\pm ie^{i\phi}|\Psi_{\perp}\rangle),
\end{align}
where $c_2=|c_2|e^{i\phi}$. For the specific choice of projectors $\{E_+=|\varphi_+\rangle\langle\varphi_+|,E_-=|\varphi_-\rangle\langle\varphi_-|\}$, we obtain $\langle\varphi_{\pm}|\Psi\rangle=\frac{1}{\sqrt{2}}$ and
\begin{align}\label{eq:weak}
\frac{\langle\varphi_{\pm}|H|\Psi\rangle}{\langle\varphi_{\pm}|\Psi\rangle}=\langle\Psi|H|\Psi\rangle\mp ie^{-i\phi}\langle\Psi_{\perp}|H|\Psi\rangle.
\end{align}
Using
\begin{align}
c_2=\frac{\langle\Psi_{\perp}|H|\Psi\rangle}{\sqrt{\langle\Psi|H^2|\Psi\rangle}}
\end{align}
and $|c_2|=\sqrt{1-|c_1|^2}=\sqrt{1-\frac{\langle\Psi|H|\Psi\rangle^2}{\langle\Psi|H^2|\Psi\rangle}}$ [recall Eqs.~(\ref{eq:phi}) and~(\ref{eq:c1})], we obtain
\begin{align}
e^{-i\phi}\sqrt{\langle\Psi_{\perp}|H|\Psi\rangle}&=e^{-i\phi}c_2\sqrt{\langle\Psi|H^2|\Psi\rangle}=|c_2|\sqrt{\langle\Psi|H^2|\Psi\rangle}\notag\\
&=\sqrt{\langle\Psi|H^2|\Psi\rangle-\langle\Psi|H|\Psi\rangle^2}=(\Delta H)_{|\Psi\rangle}.
\end{align}
Inserting this into Eq.~(\ref{eq:weak}) yields
\begin{align}
\left|\mathrm{Im}\left(\frac{\langle\varphi_{\pm}|H|\Psi\rangle}{\langle\varphi_{\pm}|\Psi\rangle}\right)\right|=(\Delta H)_{|\Psi\rangle}.
\end{align}
From Eq.~(\ref{eq:falphaweak}) we now obtain the generalized Fisher information at $\theta=\theta_0$ as
\begin{align}\label{eq:falpha2DH}
f_{\alpha}[p(\theta)]&=\left(2(\Delta H)_{|\Psi\rangle}\right)^{\alpha}.
\end{align}
Comparison with Eq.~(\ref{eq:FalphaVar}) and the definition~(\ref{eq:Falpha}) shows that the measurement is optimal for $1\leq \alpha\leq 2$. This measurement still achieves Eq.~(\ref{eq:FalphaVar}) for other values of $\alpha$, but in these cases its optimality is not proven. The states $|\varphi_{\pm}\rangle$ can be written in terms of $H$ and $|\Psi\rangle$ by introducing the state $|\widetilde{\Psi}\rangle=e^{i\phi}|\Psi_{\perp}\rangle$ [see Eq.~(\ref{eq:psitilde})].

\section{Proof of Eq.~(\ref{eq:Dalpha})}\label{app:Dalpha}
First we generalize the result~(\ref{eq:optmeas}) to arbitrary Schatten norms with $\alpha\geq 1$. For an arbitrary Hermitian operator $X$, we have
\begin{align}\label{eq:DalphaG}
\max_{\{E_x\}}\sum_x|\mathrm{Tr}\{E_xX\}|^{\alpha}=\mathrm{Tr}|X|^{\alpha}.
\end{align}
The proof is similar to the one presented in Appendix~\ref{app:F1}, with an additional step involving the H\"{o}lder inequality. We make use of the Jordan-Hahn decomposition~(\ref{eq:JordanHahn}) of $X=\sum_{i}\lambda_i|\varphi_i\rangle\langle\varphi_i|=X_++X_-$ \cite{NielsenChuang}, obtaining
\begin{align}
\sum_x|\mathrm{Tr}\{E_xX\}|^{\alpha}&\leq\sum_x|\mathrm{Tr}\{E_x|X|\}|^{\alpha}\notag\\
&=\sum_x(\mathrm{Tr}\{E_x(X_+-X_-)\})^{\alpha}\notag\\
&=\sum_x\left(\sum_i|\lambda_i|\langle\varphi_i|E_x|\varphi_i\rangle\right)^{\alpha}.\label{eq:lastline}
\end{align}
Using the H\"{o}lder inequality~(\ref{eq:holder}) with $f_i=|\lambda_i|\langle\varphi_i|E_x|\varphi_i\rangle^{\frac{1}{\alpha}}$, $g_i=\langle\varphi_i|E_x|\varphi_i\rangle^{1-\frac{1}{\alpha}}$ and $q=\alpha$ [thus $p=\alpha/(\alpha-1)$] yields
\begin{align}
\sum_i|\lambda_i|\langle\varphi_i|E_x|\varphi_i\rangle\leq\left(\sum_i|\lambda_i|^{\alpha}\langle\varphi_i|E_x|\varphi_i\rangle\right)^{\frac{1}{\alpha}}\underbrace{\left(\sum_i\langle\varphi_i|E_x|\varphi_i\rangle\right)^{1-\frac{1}{\alpha}}}_{=1},
\end{align}
for $\alpha>1$, whereas for $\alpha=1$ equality holds trivially. Inserting this result into Eq.~(\ref{eq:lastline}), we find
\begin{align}
\sum_x|\mathrm{Tr}\{E_xX\}|^{\alpha}&\leq\sum_x\sum_i|\lambda_i|^{\alpha}\langle\varphi_i|E_x|\varphi_i\rangle\\
&=\sum_x\mathrm{Tr}\{E_x|X|^{\alpha}\}\\
&=\mathrm{Tr}|X|^{\alpha}
\end{align}
for arbitrary POVMs. 

By choosing, e.g., a POVM which contains the projectors $E_x=|\varphi_x\rangle\langle\varphi_x|$, we obtain $\mathrm{Tr}\{E_xX\}=\lambda_x$ and hence $\sum_x|\mathrm{Tr}\{E_xX\}|^{\alpha}=\sum_x|\lambda_x|^{\alpha}=\mathrm{Tr}|X|^{\alpha}$. This shows that the upper bound can be reached by an optimal POVM and proves Eq.~(\ref{eq:DalphaG}).

Using Eq.~(\ref{eq:DalphaG}) with $X=\rho-\sigma$ yields Eq.~(\ref{eq:Dalpha}). Similarly, we can use $X=d \rho(\theta)/d\theta$ to prove Eq.~(\ref{eq:Valpha}). This verifies that the diagram in Fig.~\ref{fig:diagram} commutes for the Schatten norms of all $\alpha$.

\section{Proof of Eq.~(\ref{eq:SchattenBound})}\label{app:SchattenBound}
The proof for Eq.~(\ref{eq:SchattenBound}) is similar to the one presented in Appendix~\ref{app:moments}. The statement follows analogously using H\"older's inequality~(\ref{eq:holder}) with $f_x=\frac{dp_x(\theta)}{d\theta}$ and $g_x=m_x-c$ for $p=\alpha$, $q=\beta$, where the $m_x$ are the eigenvalues of the observable $M$. The constant $c\in\mathbb{R}$ may be a function of $\theta$ but must be independent of $x$.


\begin{thebibliography}{109}%
\makeatletter
\providecommand \@ifxundefined [1]{%
 \@ifx{#1\undefined}
}%
\providecommand \@ifnum [1]{%
 \ifnum #1\expandafter \@firstoftwo
 \else \expandafter \@secondoftwo
 \fi
}%
\providecommand \@ifx [1]{%
 \ifx #1\expandafter \@firstoftwo
 \else \expandafter \@secondoftwo
 \fi
}%
\providecommand \natexlab [1]{#1}%
\providecommand \enquote  [1]{``#1''}%
\providecommand \bibnamefont  [1]{#1}%
\providecommand \bibfnamefont [1]{#1}%
\providecommand \citenamefont [1]{#1}%
\providecommand \href@noop [0]{\@secondoftwo}%
\providecommand \href [0]{\begingroup \@sanitize@url \@href}%
\providecommand \@href[1]{\@@startlink{#1}\@@href}%
\providecommand \@@href[1]{\endgroup#1\@@endlink}%
\providecommand \@sanitize@url [0]{\catcode `\\12\catcode `\$12\catcode
  `\&12\catcode `\#12\catcode `\^12\catcode `\_12\catcode `\%12\relax}%
\providecommand \@@startlink[1]{}%
\providecommand \@@endlink[0]{}%
\providecommand \url  [0]{\begingroup\@sanitize@url \@url }%
\providecommand \@url [1]{\endgroup\@href {#1}{\urlprefix }}%
\providecommand \urlprefix  [0]{URL }%
\providecommand \Eprint [0]{\href }%
\providecommand \doibase [0]{http://dx.doi.org/}%
\providecommand \selectlanguage [0]{\@gobble}%
\providecommand \bibinfo  [0]{\@secondoftwo}%
\providecommand \bibfield  [0]{\@secondoftwo}%
\providecommand \translation [1]{[#1]}%
\providecommand \BibitemOpen [0]{}%
\providecommand \bibitemStop [0]{}%
\providecommand \bibitemNoStop [0]{.\EOS\space}%
\providecommand \EOS [0]{\spacefactor3000\relax}%
\providecommand \BibitemShut  [1]{\csname bibitem#1\endcsname}%
\let\auto@bib@innerbib\@empty
\bibitem [{\citenamefont {Braunstein}\ and\ \citenamefont
  {Caves}(1994)}]{PhysRevLett.72.3439}%
  \BibitemOpen
  \bibfield  {author} {\bibinfo {author} {\bibfnamefont {S.~L.}\ \bibnamefont
  {Braunstein}}\ and\ \bibinfo {author} {\bibfnamefont {C.~M.}\ \bibnamefont
  {Caves}},\ }\href {\doibase 10.1103/PhysRevLett.72.3439} {\bibfield
  {journal} {\bibinfo  {journal} {Phys. Rev. Lett.}\ }\textbf {\bibinfo
  {volume} {72}},\ \bibinfo {pages} {3439} (\bibinfo {year}
  {1994})}\BibitemShut {NoStop}%
\bibitem [{\citenamefont {Giovannetti}\ \emph {et~al.}(2006)\citenamefont
  {Giovannetti}, \citenamefont {Lloyd},\ and\ \citenamefont
  {Maccone}}]{PhysRevLett.96.010401}%
  \BibitemOpen
  \bibfield  {author} {\bibinfo {author} {\bibfnamefont {V.}~\bibnamefont
  {Giovannetti}}, \bibinfo {author} {\bibfnamefont {S.}~\bibnamefont {Lloyd}},
  \ and\ \bibinfo {author} {\bibfnamefont {L.}~\bibnamefont {Maccone}},\ }\href
  {\doibase 10.1103/PhysRevLett.96.010401} {\bibfield  {journal} {\bibinfo
  {journal} {Phys. Rev. Lett.}\ }\textbf {\bibinfo {volume} {96}},\ \bibinfo
  {pages} {010401} (\bibinfo {year} {2006})}\BibitemShut {NoStop}%
\bibitem [{\citenamefont {Pezz\'e}\ and\ \citenamefont
  {Smerzi}(2014)}]{Varenna}%
  \BibitemOpen
  \bibfield  {author} {\bibinfo {author} {\bibfnamefont {L.}~\bibnamefont
  {Pezz\'e}}\ and\ \bibinfo {author} {\bibfnamefont {A.}~\bibnamefont
  {Smerzi}},\ }in\ \href@noop {} {\emph {\bibinfo {booktitle} {Atom
  Interferometry, Proceedings of the International School of Physics "Enrico
  Fermi", Course 188, Varenna}}},\ \bibinfo {editor} {edited by\ \bibinfo
  {editor} {\bibfnamefont {G.}~\bibnamefont {Tino}}\ and\ \bibinfo {editor}
  {\bibfnamefont {M.}~\bibnamefont {Kasevich}}}\ (\bibinfo  {publisher} {IOS
  Press},\ \bibinfo {address} {Amsterdam, Netherlands},\ \bibinfo {year}
  {2014})\BibitemShut {NoStop}%
\bibitem [{\citenamefont {Pezz\'e}\ and\ \citenamefont
  {Smerzi}(2009)}]{PhysRevLett.102.100401}%
  \BibitemOpen
  \bibfield  {author} {\bibinfo {author} {\bibfnamefont {L.}~\bibnamefont
  {Pezz\'e}}\ and\ \bibinfo {author} {\bibfnamefont {A.}~\bibnamefont
  {Smerzi}},\ }\href {\doibase 10.1103/PhysRevLett.102.100401} {\bibfield
  {journal} {\bibinfo  {journal} {Phys. Rev. Lett.}\ }\textbf {\bibinfo
  {volume} {102}},\ \bibinfo {pages} {100401} (\bibinfo {year}
  {2009})}\BibitemShut {NoStop}%
\bibitem [{\citenamefont {Pezz\`{e}}\ \emph {et~al.}(2016)\citenamefont
  {Pezz\`{e}}, \citenamefont {Smerzi}, \citenamefont {Oberthaler},
  \citenamefont {Schmied},\ and\ \citenamefont {Treutlein}}]{LucaRMP}%
  \BibitemOpen
  \bibfield  {author} {\bibinfo {author} {\bibfnamefont {L.}~\bibnamefont
  {Pezz\`{e}}}, \bibinfo {author} {\bibfnamefont {A.}~\bibnamefont {Smerzi}},
  \bibinfo {author} {\bibfnamefont {M.~K.}\ \bibnamefont {Oberthaler}},
  \bibinfo {author} {\bibfnamefont {R.}~\bibnamefont {Schmied}}, \ and\
  \bibinfo {author} {\bibfnamefont {P.}~\bibnamefont {Treutlein}},\ }\href
  {http://arxiv.org/abs/1609.01609} {\enquote {\bibinfo {title} {Non-classical
  states of atomic ensembles: fundamentals and applications in quantum
  metrology},}\ }\bibinfo {howpublished} {arXiv:1609.01609 [quant-ph]}
  (\bibinfo {year} {2016})\BibitemShut {NoStop}%
\bibitem [{\citenamefont {Strobel}\ \emph {et~al.}(2014)\citenamefont
  {Strobel}, \citenamefont {Muessel}, \citenamefont {Linnemann}, \citenamefont
  {Zibold}, \citenamefont {Hume}, \citenamefont {Pezz{\`e}}, \citenamefont
  {Smerzi},\ and\ \citenamefont {Oberthaler}}]{Strobel2014}%
  \BibitemOpen
  \bibfield  {author} {\bibinfo {author} {\bibfnamefont {H.}~\bibnamefont
  {Strobel}}, \bibinfo {author} {\bibfnamefont {W.}~\bibnamefont {Muessel}},
  \bibinfo {author} {\bibfnamefont {D.}~\bibnamefont {Linnemann}}, \bibinfo
  {author} {\bibfnamefont {T.}~\bibnamefont {Zibold}}, \bibinfo {author}
  {\bibfnamefont {D.~B.}\ \bibnamefont {Hume}}, \bibinfo {author}
  {\bibfnamefont {L.}~\bibnamefont {Pezz{\`e}}}, \bibinfo {author}
  {\bibfnamefont {A.}~\bibnamefont {Smerzi}}, \ and\ \bibinfo {author}
  {\bibfnamefont {M.~K.}\ \bibnamefont {Oberthaler}},\ }\href {\doibase
  10.1126/science.1250147} {\bibfield  {journal} {\bibinfo  {journal}
  {Science}\ }\textbf {\bibinfo {volume} {345}},\ \bibinfo {pages} {424}
  (\bibinfo {year} {2014})}\BibitemShut {NoStop}%
\bibitem [{\citenamefont {Pezz\`e}\ \emph {et~al.}(2016)\citenamefont
  {Pezz\`e}, \citenamefont {Li}, \citenamefont {Li},\ and\ \citenamefont
  {Smerzi}}]{PezzePNAS2016}%
  \BibitemOpen
  \bibfield  {author} {\bibinfo {author} {\bibfnamefont {L.}~\bibnamefont
  {Pezz\`e}}, \bibinfo {author} {\bibfnamefont {Y.}~\bibnamefont {Li}},
  \bibinfo {author} {\bibfnamefont {W.}~\bibnamefont {Li}}, \ and\ \bibinfo
  {author} {\bibfnamefont {A.}~\bibnamefont {Smerzi}},\ }\href {\doibase
  10.1073/pnas.1603346113} {\bibfield  {journal} {\bibinfo  {journal} {Proc.
  Natl. Acad. Sci.}\ }\textbf {\bibinfo {volume} {113}},\ \bibinfo {pages}
  {11459} (\bibinfo {year} {2016})}\BibitemShut {NoStop}%
\bibitem [{\citenamefont {Zhang}\ \emph {et~al.}(2017)\citenamefont {Zhang},
  \citenamefont {Yadin}, \citenamefont {Hou}, \citenamefont {Cao},
  \citenamefont {Liu}, \citenamefont {Huang}, \citenamefont {Maity},
  \citenamefont {Vedral}, \citenamefont {Li}, \citenamefont {Guo},\ and\
  \citenamefont {Girolami}}]{Zhang2016}%
  \BibitemOpen
  \bibfield  {author} {\bibinfo {author} {\bibfnamefont {C.}~\bibnamefont
  {Zhang}}, \bibinfo {author} {\bibfnamefont {B.}~\bibnamefont {Yadin}},
  \bibinfo {author} {\bibfnamefont {Z.-B.}\ \bibnamefont {Hou}}, \bibinfo
  {author} {\bibfnamefont {H.}~\bibnamefont {Cao}}, \bibinfo {author}
  {\bibfnamefont {B.-H.}\ \bibnamefont {Liu}}, \bibinfo {author} {\bibfnamefont
  {Y.-F.}\ \bibnamefont {Huang}}, \bibinfo {author} {\bibfnamefont
  {R.}~\bibnamefont {Maity}}, \bibinfo {author} {\bibfnamefont
  {V.}~\bibnamefont {Vedral}}, \bibinfo {author} {\bibfnamefont {C.-F.}\
  \bibnamefont {Li}}, \bibinfo {author} {\bibfnamefont {G.-C.}\ \bibnamefont
  {Guo}}, \ and\ \bibinfo {author} {\bibfnamefont {D.}~\bibnamefont
  {Girolami}},\ }\href {\doibase 10.1103/PhysRevA.96.042327} {\bibfield
  {journal} {\bibinfo  {journal} {Phys. Rev. A}\ }\textbf {\bibinfo {volume}
  {96}},\ \bibinfo {pages} {042327} (\bibinfo {year} {2017})}\BibitemShut
  {NoStop}%
\bibitem [{\citenamefont {Fuchs}\ and\ \citenamefont {van~de
  Graaf}(1999)}]{Fuchs1999}%
  \BibitemOpen
  \bibfield  {author} {\bibinfo {author} {\bibfnamefont {C.~A.}\ \bibnamefont
  {Fuchs}}\ and\ \bibinfo {author} {\bibfnamefont {J.}~\bibnamefont {van~de
  Graaf}},\ }\href {\doibase 10.1109/18.761271} {\bibfield  {journal} {\bibinfo
   {journal} {IEEE Trans. Inf. Theo.}\ }\textbf {\bibinfo {volume} {45}},\
  \bibinfo {pages} {1216} (\bibinfo {year} {1999})}\BibitemShut {NoStop}%
\bibitem [{\citenamefont {Nielsen}\ and\ \citenamefont
  {Chuang}(2000)}]{NielsenChuang}%
  \BibitemOpen
  \bibfield  {author} {\bibinfo {author} {\bibfnamefont {M.~A.}\ \bibnamefont
  {Nielsen}}\ and\ \bibinfo {author} {\bibfnamefont {I.~L.}\ \bibnamefont
  {Chuang}},\ }\href@noop {} {\emph {\bibinfo {title} {Quantum Computation and
  Quantum Information}}}\ (\bibinfo  {publisher} {Cambridge University Press},\
  \bibinfo {address} {New York, NY},\ \bibinfo {year} {2000})\BibitemShut
  {NoStop}%
\bibitem [{\citenamefont {Hayashi}(2006)}]{Hayashi2006}%
  \BibitemOpen
  \bibfield  {author} {\bibinfo {author} {\bibfnamefont {M.}~\bibnamefont
  {Hayashi}},\ }\href@noop {} {\emph {\bibinfo {title} {Quantum Information}}}\
  (\bibinfo  {publisher} {Springer},\ \bibinfo {address} {Berlin, Germany},\
  \bibinfo {year} {2006})\BibitemShut {NoStop}%
\bibitem [{\citenamefont {Bengtsson}\ and\ \citenamefont
  {\.{Z}yczkowski}(2006)}]{Zyczkowski2006}%
  \BibitemOpen
  \bibfield  {author} {\bibinfo {author} {\bibfnamefont {I.}~\bibnamefont
  {Bengtsson}}\ and\ \bibinfo {author} {\bibfnamefont {K.}~\bibnamefont
  {\.{Z}yczkowski}},\ }\href@noop {} {\emph {\bibinfo {title} {Geometry of
  Quantum States}}}\ (\bibinfo  {publisher} {Cambridge University Press},\
  \bibinfo {address} {Cambridge, UK},\ \bibinfo {year} {2006})\BibitemShut
  {NoStop}%
\bibitem [{\citenamefont {Spehner}(2014)}]{Spehner2014}%
  \BibitemOpen
  \bibfield  {author} {\bibinfo {author} {\bibfnamefont {D.}~\bibnamefont
  {Spehner}},\ }\href {\doibase 10.1063/1.4885832} {\bibfield  {journal}
  {\bibinfo  {journal} {J. Math. Phys.}\ }\textbf {\bibinfo {volume} {55}},\
  \bibinfo {pages} {075211} (\bibinfo {year} {2014})}\BibitemShut {NoStop}%
\bibitem [{\citenamefont {Wigner}\ and\ \citenamefont
  {Yanase}(1963)}]{WignerYanase}%
  \BibitemOpen
  \bibfield  {author} {\bibinfo {author} {\bibfnamefont {E.~P.}\ \bibnamefont
  {Wigner}}\ and\ \bibinfo {author} {\bibfnamefont {M.~M.}\ \bibnamefont
  {Yanase}},\ }\href@noop {} {\bibfield  {journal} {\bibinfo  {journal} {Proc.
  Natl. Acad. Sci.}\ }\textbf {\bibinfo {volume} {49}},\ \bibinfo {pages} {910}
  (\bibinfo {year} {1963})}\BibitemShut {NoStop}%
\bibitem [{\citenamefont {Petz}(1996)}]{Petz1996}%
  \BibitemOpen
  \bibfield  {author} {\bibinfo {author} {\bibfnamefont {D.}~\bibnamefont
  {Petz}},\ }\href {\doibase 10.1016/0024-3795(94)00211-8} {\bibfield
  {journal} {\bibinfo  {journal} {Linear Algebra and its Applications}\
  }\textbf {\bibinfo {volume} {244}},\ \bibinfo {pages} {81 } (\bibinfo {year}
  {1996})}\BibitemShut {NoStop}%
\bibitem [{\citenamefont {Wootters}(1981)}]{PhysRevD.23.357}%
  \BibitemOpen
  \bibfield  {author} {\bibinfo {author} {\bibfnamefont {W.~K.}\ \bibnamefont
  {Wootters}},\ }\href {\doibase 10.1103/PhysRevD.23.357} {\bibfield  {journal}
  {\bibinfo  {journal} {Phys. Rev. D}\ }\textbf {\bibinfo {volume} {23}},\
  \bibinfo {pages} {357} (\bibinfo {year} {1981})}\BibitemShut {NoStop}%
\bibitem [{\citenamefont {Gibilisco}\ and\ \citenamefont
  {Isola}(2003)}]{Gibilisco2003}%
  \BibitemOpen
  \bibfield  {author} {\bibinfo {author} {\bibfnamefont {P.}~\bibnamefont
  {Gibilisco}}\ and\ \bibinfo {author} {\bibfnamefont {T.}~\bibnamefont
  {Isola}},\ }\href {\doibase 10.1063/1.1598279} {\bibfield  {journal}
  {\bibinfo  {journal} {J. Math. Phys.}\ }\textbf {\bibinfo {volume} {44}},\
  \bibinfo {pages} {3752} (\bibinfo {year} {2003})}\BibitemShut {NoStop}%
\bibitem [{\citenamefont {Zanardi}\ \emph {et~al.}(2007)\citenamefont
  {Zanardi}, \citenamefont {Giorda},\ and\ \citenamefont
  {Cozzini}}]{PhysRevLett.99.100603}%
  \BibitemOpen
  \bibfield  {author} {\bibinfo {author} {\bibfnamefont {P.}~\bibnamefont
  {Zanardi}}, \bibinfo {author} {\bibfnamefont {P.}~\bibnamefont {Giorda}}, \
  and\ \bibinfo {author} {\bibfnamefont {M.}~\bibnamefont {Cozzini}},\ }\href
  {\doibase 10.1103/PhysRevLett.99.100603} {\bibfield  {journal} {\bibinfo
  {journal} {Phys. Rev. Lett.}\ }\textbf {\bibinfo {volume} {99}},\ \bibinfo
  {pages} {100603} (\bibinfo {year} {2007})}\BibitemShut {NoStop}%
\bibitem [{\citenamefont {Petz}\ and\ \citenamefont {Ghinea}(2011)}]{Petz2011}%
  \BibitemOpen
  \bibfield  {author} {\bibinfo {author} {\bibfnamefont {D.}~\bibnamefont
  {Petz}}\ and\ \bibinfo {author} {\bibfnamefont {C.}~\bibnamefont {Ghinea}},\
  }in\ \href {\doibase 10.1142/9789814338745_0015} {\emph {\bibinfo {booktitle}
  {QP-PQ: Quantum Probability and White Noise Analysis. Vol. 27}}},\ \bibinfo
  {editor} {edited by\ \bibinfo {editor} {\bibfnamefont {R.}~\bibnamefont
  {Rebolledo}}\ and\ \bibinfo {editor} {\bibfnamefont {M.}~\bibnamefont
  {Orszag}}}\ (\bibinfo  {publisher} {World Scientific},\ \bibinfo {address}
  {Singapore},\ \bibinfo {year} {2011})\ pp.\ \bibinfo {pages}
  {261--281}\BibitemShut {NoStop}%
\bibitem [{\citenamefont {Brody}\ and\ \citenamefont
  {Graefe}(2012)}]{PhysRevLett.109.230405}%
  \BibitemOpen
  \bibfield  {author} {\bibinfo {author} {\bibfnamefont {D.~C.}\ \bibnamefont
  {Brody}}\ and\ \bibinfo {author} {\bibfnamefont {E.-M.}\ \bibnamefont
  {Graefe}},\ }\href {\doibase 10.1103/PhysRevLett.109.230405} {\bibfield
  {journal} {\bibinfo  {journal} {Phys. Rev. Lett.}\ }\textbf {\bibinfo
  {volume} {109}},\ \bibinfo {pages} {230405} (\bibinfo {year}
  {2012})}\BibitemShut {NoStop}%
\bibitem [{\citenamefont {T\'oth}\ and\ \citenamefont
  {Petz}(2013)}]{PhysRevA.87.032324}%
  \BibitemOpen
  \bibfield  {author} {\bibinfo {author} {\bibfnamefont {G.}~\bibnamefont
  {T\'oth}}\ and\ \bibinfo {author} {\bibfnamefont {D.}~\bibnamefont {Petz}},\
  }\href {\doibase 10.1103/PhysRevA.87.032324} {\bibfield  {journal} {\bibinfo
  {journal} {Phys. Rev. A}\ }\textbf {\bibinfo {volume} {87}},\ \bibinfo
  {pages} {032324} (\bibinfo {year} {2013})}\BibitemShut {NoStop}%
\bibitem [{\citenamefont {Hauke}\ \emph {et~al.}(2016)\citenamefont {Hauke},
  \citenamefont {Heyl}, \citenamefont {Tagliacozzo},\ and\ \citenamefont
  {Zoller}}]{Hauke2016}%
  \BibitemOpen
  \bibfield  {author} {\bibinfo {author} {\bibfnamefont {P.}~\bibnamefont
  {Hauke}}, \bibinfo {author} {\bibfnamefont {M.}~\bibnamefont {Heyl}},
  \bibinfo {author} {\bibfnamefont {L.}~\bibnamefont {Tagliacozzo}}, \ and\
  \bibinfo {author} {\bibfnamefont {P.}~\bibnamefont {Zoller}},\ }\href
  {http://dx.doi.org/10.1038/nphys3700} {\bibfield  {journal} {\bibinfo
  {journal} {Nat. Phys.}\ }\textbf {\bibinfo {volume} {12}},\ \bibinfo {pages}
  {778} (\bibinfo {year} {2016})}\BibitemShut {NoStop}%
\bibitem [{\citenamefont {Pezz\`e}\ \emph {et~al.}(2017)\citenamefont
  {Pezz\`e}, \citenamefont {Gabbrielli}, \citenamefont {Lepori},\ and\
  \citenamefont {Smerzi}}]{PhysRevLett.119.250401}%
  \BibitemOpen
  \bibfield  {author} {\bibinfo {author} {\bibfnamefont {L.}~\bibnamefont
  {Pezz\`e}}, \bibinfo {author} {\bibfnamefont {M.}~\bibnamefont {Gabbrielli}},
  \bibinfo {author} {\bibfnamefont {L.}~\bibnamefont {Lepori}}, \ and\ \bibinfo
  {author} {\bibfnamefont {A.}~\bibnamefont {Smerzi}},\ }\href {\doibase
  10.1103/PhysRevLett.119.250401} {\bibfield  {journal} {\bibinfo  {journal}
  {Phys. Rev. Lett.}\ }\textbf {\bibinfo {volume} {119}},\ \bibinfo {pages}
  {250401} (\bibinfo {year} {2017})}\BibitemShut {NoStop}%
\bibitem [{\citenamefont {Laine}\ \emph {et~al.}(2010)\citenamefont {Laine},
  \citenamefont {Piilo},\ and\ \citenamefont {Breuer}}]{Laine2010}%
  \BibitemOpen
  \bibfield  {author} {\bibinfo {author} {\bibfnamefont {E.-M.}\ \bibnamefont
  {Laine}}, \bibinfo {author} {\bibfnamefont {J.}~\bibnamefont {Piilo}}, \ and\
  \bibinfo {author} {\bibfnamefont {H.-P.}\ \bibnamefont {Breuer}},\ }\href
  {http://stacks.iop.org/0295-5075/92/i=6/a=60010} {\bibfield  {journal}
  {\bibinfo  {journal} {Europhys. Lett.}\ }\textbf {\bibinfo {volume} {92}},\
  \bibinfo {pages} {60010} (\bibinfo {year} {2010})}\BibitemShut {NoStop}%
\bibitem [{\citenamefont {Breuer}\ \emph {et~al.}(2016)\citenamefont {Breuer},
  \citenamefont {Laine}, \citenamefont {Piilo},\ and\ \citenamefont
  {Vacchini}}]{RevModPhys.88.021002}%
  \BibitemOpen
  \bibfield  {author} {\bibinfo {author} {\bibfnamefont {H.-P.}\ \bibnamefont
  {Breuer}}, \bibinfo {author} {\bibfnamefont {E.-M.}\ \bibnamefont {Laine}},
  \bibinfo {author} {\bibfnamefont {J.}~\bibnamefont {Piilo}}, \ and\ \bibinfo
  {author} {\bibfnamefont {B.}~\bibnamefont {Vacchini}},\ }\href {\doibase
  10.1103/RevModPhys.88.021002} {\bibfield  {journal} {\bibinfo  {journal}
  {Rev. Mod. Phys.}\ }\textbf {\bibinfo {volume} {88}},\ \bibinfo {pages}
  {021002} (\bibinfo {year} {2016})}\BibitemShut {NoStop}%
\bibitem [{\citenamefont {Gessner}(2017)}]{GessnerDiss}%
  \BibitemOpen
  \bibfield  {author} {\bibinfo {author} {\bibfnamefont {M.}~\bibnamefont
  {Gessner}},\ }\href {\doibase 10.1007/978-3-319-44459-8} {\emph {\bibinfo
  {title} {Dynamics and Characterization of Composite Quantum Systems}}}\
  (\bibinfo  {publisher} {Springer International Publishing},\ \bibinfo {year}
  {2017})\BibitemShut {NoStop}%
\bibitem [{\citenamefont {Cianciaruso}\ \emph {et~al.}(2017)\citenamefont
  {Cianciaruso}, \citenamefont {Maniscalco},\ and\ \citenamefont
  {Adesso}}]{PhysRevA.96.012105}%
  \BibitemOpen
  \bibfield  {author} {\bibinfo {author} {\bibfnamefont {M.}~\bibnamefont
  {Cianciaruso}}, \bibinfo {author} {\bibfnamefont {S.}~\bibnamefont
  {Maniscalco}}, \ and\ \bibinfo {author} {\bibfnamefont {G.}~\bibnamefont
  {Adesso}},\ }\href {\doibase 10.1103/PhysRevA.96.012105} {\bibfield
  {journal} {\bibinfo  {journal} {Phys. Rev. A}\ }\textbf {\bibinfo {volume}
  {96}},\ \bibinfo {pages} {012105} (\bibinfo {year} {2017})}\BibitemShut
  {NoStop}%
\bibitem [{\citenamefont {T\'{o}th}\ and\ \citenamefont
  {Apellaniz}(2014)}]{Toth2014}%
  \BibitemOpen
  \bibfield  {author} {\bibinfo {author} {\bibfnamefont {G.}~\bibnamefont
  {T\'{o}th}}\ and\ \bibinfo {author} {\bibfnamefont {I.}~\bibnamefont
  {Apellaniz}},\ }\href {\doibase 10.1088/1751-8113/47/42/424006} {\bibfield
  {journal} {\bibinfo  {journal} {J. Phys. A}\ }\textbf {\bibinfo {volume}
  {47}},\ \bibinfo {pages} {424006} (\bibinfo {year} {2014})}\BibitemShut
  {NoStop}%
\bibitem [{\citenamefont {Giovannetti}\ \emph {et~al.}(2003)\citenamefont
  {Giovannetti}, \citenamefont {Lloyd},\ and\ \citenamefont
  {Maccone}}]{PhysRevA.67.052109}%
  \BibitemOpen
  \bibfield  {author} {\bibinfo {author} {\bibfnamefont {V.}~\bibnamefont
  {Giovannetti}}, \bibinfo {author} {\bibfnamefont {S.}~\bibnamefont {Lloyd}},
  \ and\ \bibinfo {author} {\bibfnamefont {L.}~\bibnamefont {Maccone}},\ }\href
  {\doibase 10.1103/PhysRevA.67.052109} {\bibfield  {journal} {\bibinfo
  {journal} {Phys. Rev. A}\ }\textbf {\bibinfo {volume} {67}},\ \bibinfo
  {pages} {052109} (\bibinfo {year} {2003})}\BibitemShut {NoStop}%
\bibitem [{\citenamefont {Taddei}\ \emph {et~al.}(2013)\citenamefont {Taddei},
  \citenamefont {Escher}, \citenamefont {Davidovich},\ and\ \citenamefont
  {de~Matos~Filho}}]{PhysRevLett.110.050402}%
  \BibitemOpen
  \bibfield  {author} {\bibinfo {author} {\bibfnamefont {M.~M.}\ \bibnamefont
  {Taddei}}, \bibinfo {author} {\bibfnamefont {B.~M.}\ \bibnamefont {Escher}},
  \bibinfo {author} {\bibfnamefont {L.}~\bibnamefont {Davidovich}}, \ and\
  \bibinfo {author} {\bibfnamefont {R.~L.}\ \bibnamefont {de~Matos~Filho}},\
  }\href {\doibase 10.1103/PhysRevLett.110.050402} {\bibfield  {journal}
  {\bibinfo  {journal} {Phys. Rev. Lett.}\ }\textbf {\bibinfo {volume} {110}},\
  \bibinfo {pages} {050402} (\bibinfo {year} {2013})}\BibitemShut {NoStop}%
\bibitem [{\citenamefont {del Campo}\ \emph {et~al.}(2013)\citenamefont {del
  Campo}, \citenamefont {Egusquiza}, \citenamefont {Plenio},\ and\
  \citenamefont {Huelga}}]{PhysRevLett.110.050403}%
  \BibitemOpen
  \bibfield  {author} {\bibinfo {author} {\bibfnamefont {A.}~\bibnamefont {del
  Campo}}, \bibinfo {author} {\bibfnamefont {I.~L.}\ \bibnamefont {Egusquiza}},
  \bibinfo {author} {\bibfnamefont {M.~B.}\ \bibnamefont {Plenio}}, \ and\
  \bibinfo {author} {\bibfnamefont {S.~F.}\ \bibnamefont {Huelga}},\ }\href
  {\doibase 10.1103/PhysRevLett.110.050403} {\bibfield  {journal} {\bibinfo
  {journal} {Phys. Rev. Lett.}\ }\textbf {\bibinfo {volume} {110}},\ \bibinfo
  {pages} {050403} (\bibinfo {year} {2013})}\BibitemShut {NoStop}%
\bibitem [{\citenamefont {Pires}\ \emph {et~al.}(2016)\citenamefont {Pires},
  \citenamefont {Cianciaruso}, \citenamefont {C\'eleri}, \citenamefont
  {Adesso},\ and\ \citenamefont {Soares-Pinto}}]{PhysRevX.6.021031}%
  \BibitemOpen
  \bibfield  {author} {\bibinfo {author} {\bibfnamefont {D.~P.}\ \bibnamefont
  {Pires}}, \bibinfo {author} {\bibfnamefont {M.}~\bibnamefont {Cianciaruso}},
  \bibinfo {author} {\bibfnamefont {L.~C.}\ \bibnamefont {C\'eleri}}, \bibinfo
  {author} {\bibfnamefont {G.}~\bibnamefont {Adesso}}, \ and\ \bibinfo {author}
  {\bibfnamefont {D.~O.}\ \bibnamefont {Soares-Pinto}},\ }\href {\doibase
  10.1103/PhysRevX.6.021031} {\bibfield  {journal} {\bibinfo  {journal} {Phys.
  Rev. X}\ }\textbf {\bibinfo {volume} {6}},\ \bibinfo {pages} {021031}
  (\bibinfo {year} {2016})}\BibitemShut {NoStop}%
\bibitem [{\citenamefont {Smerzi}(2012)}]{PhysRevLett.109.150410}%
  \BibitemOpen
  \bibfield  {author} {\bibinfo {author} {\bibfnamefont {A.}~\bibnamefont
  {Smerzi}},\ }\href {\doibase 10.1103/PhysRevLett.109.150410} {\bibfield
  {journal} {\bibinfo  {journal} {Phys. Rev. Lett.}\ }\textbf {\bibinfo
  {volume} {109}},\ \bibinfo {pages} {150410} (\bibinfo {year}
  {2012})}\BibitemShut {NoStop}%
\bibitem [{\citenamefont {Streltsov}\ \emph {et~al.}(2017)\citenamefont
  {Streltsov}, \citenamefont {Adesso},\ and\ \citenamefont
  {Plenio}}]{Streltsov2016}%
  \BibitemOpen
  \bibfield  {author} {\bibinfo {author} {\bibfnamefont {A.}~\bibnamefont
  {Streltsov}}, \bibinfo {author} {\bibfnamefont {G.}~\bibnamefont {Adesso}}, \
  and\ \bibinfo {author} {\bibfnamefont {M.~B.}\ \bibnamefont {Plenio}},\
  }\href {\doibase 10.1103/RevModPhys.89.041003} {\bibfield  {journal}
  {\bibinfo  {journal} {Rev. Mod. Phys.}\ }\textbf {\bibinfo {volume} {89}},\
  \bibinfo {pages} {041003} (\bibinfo {year} {2017})}\BibitemShut {NoStop}%
\bibitem [{\citenamefont {Horodecki}\ \emph {et~al.}(2009)\citenamefont
  {Horodecki}, \citenamefont {Horodecki}, \citenamefont {Horodecki},\ and\
  \citenamefont {Horodecki}}]{Horodecki2009}%
  \BibitemOpen
  \bibfield  {author} {\bibinfo {author} {\bibfnamefont {R.}~\bibnamefont
  {Horodecki}}, \bibinfo {author} {\bibfnamefont {P.}~\bibnamefont
  {Horodecki}}, \bibinfo {author} {\bibfnamefont {M.}~\bibnamefont
  {Horodecki}}, \ and\ \bibinfo {author} {\bibfnamefont {K.}~\bibnamefont
  {Horodecki}},\ }\href {http://dx.doi.org/10.1103/RevModPhys.81.865}
  {\bibfield  {journal} {\bibinfo  {journal} {Rev. Mod. Phys.}\ }\textbf
  {\bibinfo {volume} {81}},\ \bibinfo {pages} {865} (\bibinfo {year}
  {2009})}\BibitemShut {NoStop}%
\bibitem [{\citenamefont {Modi}\ \emph {et~al.}(2012)\citenamefont {Modi},
  \citenamefont {Brodutch}, \citenamefont {Cable}, \citenamefont {Paterek},\
  and\ \citenamefont {Vedral}}]{Modi2012}%
  \BibitemOpen
  \bibfield  {author} {\bibinfo {author} {\bibfnamefont {K.}~\bibnamefont
  {Modi}}, \bibinfo {author} {\bibfnamefont {A.}~\bibnamefont {Brodutch}},
  \bibinfo {author} {\bibfnamefont {H.}~\bibnamefont {Cable}}, \bibinfo
  {author} {\bibfnamefont {T.}~\bibnamefont {Paterek}}, \ and\ \bibinfo
  {author} {\bibfnamefont {V.}~\bibnamefont {Vedral}},\ }\href {\doibase
  10.1103/RevModPhys.84.1655} {\bibfield  {journal} {\bibinfo  {journal} {Rev.
  Mod. Phys.}\ }\textbf {\bibinfo {volume} {84}},\ \bibinfo {pages} {1655}
  (\bibinfo {year} {2012})}\BibitemShut {NoStop}%
\bibitem [{\citenamefont {Adesso}\ \emph {et~al.}(2016)\citenamefont {Adesso},
  \citenamefont {Bromley},\ and\ \citenamefont {Cianciaruso}}]{Adesso2016}%
  \BibitemOpen
  \bibfield  {author} {\bibinfo {author} {\bibfnamefont {G.}~\bibnamefont
  {Adesso}}, \bibinfo {author} {\bibfnamefont {T.~R.}\ \bibnamefont {Bromley}},
  \ and\ \bibinfo {author} {\bibfnamefont {M.}~\bibnamefont {Cianciaruso}},\
  }\href {\doibase 10.1088/1751-8113/49/47/473001} {\bibfield  {journal}
  {\bibinfo  {journal} {J. Phys. A}\ }\textbf {\bibinfo {volume} {49}},\
  \bibinfo {pages} {473001} (\bibinfo {year} {2016})}\BibitemShut {NoStop}%
\bibitem [{\citenamefont {Gessner}\ \emph
  {et~al.}(2017{\natexlab{a}})\citenamefont {Gessner}, \citenamefont {Breuer},\
  and\ \citenamefont {Buchleitner}}]{Gessner2017}%
  \BibitemOpen
  \bibfield  {author} {\bibinfo {author} {\bibfnamefont {M.}~\bibnamefont
  {Gessner}}, \bibinfo {author} {\bibfnamefont {H.-P.}\ \bibnamefont {Breuer}},
  \ and\ \bibinfo {author} {\bibfnamefont {A.}~\bibnamefont {Buchleitner}},\
  }in\ \href {\doibase 10.1007/978-3-319-53412-1_14} {\emph {\bibinfo
  {booktitle} {Lectures on General Quantum Correlations and their
  Applications}}},\ \bibinfo {editor} {edited by\ \bibinfo {editor}
  {\bibfnamefont {F.~F.}\ \bibnamefont {Fanchini}}, \bibinfo {editor}
  {\bibfnamefont {D.~d.~O.}\ \bibnamefont {Soares~Pinto}}, \ and\ \bibinfo
  {editor} {\bibfnamefont {G.}~\bibnamefont {Adesso}}}\ (\bibinfo  {publisher}
  {Springer International Publishing},\ \bibinfo {address} {Cham,
  Switzerland},\ \bibinfo {year} {2017})\ pp.\ \bibinfo {pages}
  {275--307}\BibitemShut {NoStop}%
\bibitem [{\citenamefont {Luo}(2003)}]{Luo2003}%
  \BibitemOpen
  \bibfield  {author} {\bibinfo {author} {\bibfnamefont {S.}~\bibnamefont
  {Luo}},\ }\href {\doibase 10.1103/PhysRevLett.91.180403} {\bibfield
  {journal} {\bibinfo  {journal} {Phys. Rev. Lett.}\ }\textbf {\bibinfo
  {volume} {91}},\ \bibinfo {pages} {180403} (\bibinfo {year}
  {2003})}\BibitemShut {NoStop}%
\bibitem [{\citenamefont {Marvian}\ and\ \citenamefont
  {Spekkens}(2014)}]{Marvian2014}%
  \BibitemOpen
  \bibfield  {author} {\bibinfo {author} {\bibfnamefont {I.}~\bibnamefont
  {Marvian}}\ and\ \bibinfo {author} {\bibfnamefont {R.~W.}\ \bibnamefont
  {Spekkens}},\ }\href {\doibase 10.1038/ncomms4821} {\bibfield  {journal}
  {\bibinfo  {journal} {Nat. Comm.}\ }\textbf {\bibinfo {volume} {5}},\
  \bibinfo {pages} {3821} (\bibinfo {year} {2014})}\BibitemShut {NoStop}%
\bibitem [{\citenamefont {Horodecki}\ \emph {et~al.}(2003)\citenamefont
  {Horodecki}, \citenamefont {Horodecki},\ and\ \citenamefont
  {Oppenheim}}]{PhysRevA.67.062104}%
  \BibitemOpen
  \bibfield  {author} {\bibinfo {author} {\bibfnamefont {M.}~\bibnamefont
  {Horodecki}}, \bibinfo {author} {\bibfnamefont {P.}~\bibnamefont
  {Horodecki}}, \ and\ \bibinfo {author} {\bibfnamefont {J.}~\bibnamefont
  {Oppenheim}},\ }\href {\doibase 10.1103/PhysRevA.67.062104} {\bibfield
  {journal} {\bibinfo  {journal} {Phys. Rev. A}\ }\textbf {\bibinfo {volume}
  {67}},\ \bibinfo {pages} {062104} (\bibinfo {year} {2003})}\BibitemShut
  {NoStop}%
\bibitem [{\citenamefont {Streltsov}\ \emph {et~al.}(2016)\citenamefont
  {Streltsov}, \citenamefont {Kampermann}, \citenamefont {W\"{o}lk},
  \citenamefont {Gessner},\ and\ \citenamefont {Bru\ss}}]{Streltsov2016a}%
  \BibitemOpen
  \bibfield  {author} {\bibinfo {author} {\bibfnamefont {A.}~\bibnamefont
  {Streltsov}}, \bibinfo {author} {\bibfnamefont {H.}~\bibnamefont
  {Kampermann}}, \bibinfo {author} {\bibfnamefont {S.}~\bibnamefont
  {W\"{o}lk}}, \bibinfo {author} {\bibfnamefont {M.}~\bibnamefont {Gessner}}, \
  and\ \bibinfo {author} {\bibfnamefont {D.}~\bibnamefont {Bru\ss}},\ }\href
  {https://arxiv.org/abs/1612.07570} {\enquote {\bibinfo {title} {Maximal
  coherence and the resource theory of purity},}\ }\bibinfo {howpublished}
  {arXiv:1612.07570 [quant-ph]} (\bibinfo {year} {2016})\BibitemShut {NoStop}%
\bibitem [{\citenamefont {Hellinger}(1909)}]{Hellinger1909}%
  \BibitemOpen
  \bibfield  {author} {\bibinfo {author} {\bibfnamefont {E.}~\bibnamefont
  {Hellinger}},\ }\href {\doibase 10.1515/crll.1909.136.210} {\bibfield
  {journal} {\bibinfo  {journal} {Journal f\"{u}r die reine und angewandte
  {M}athematik}\ }\textbf {\bibinfo {volume} {136}},\ \bibinfo {pages} {210}
  (\bibinfo {year} {1909})}\BibitemShut {NoStop}%
\bibitem [{\citenamefont {Bhattacharyya}(1946)}]{Bhattacharyya1946}%
  \BibitemOpen
  \bibfield  {author} {\bibinfo {author} {\bibfnamefont {A.}~\bibnamefont
  {Bhattacharyya}},\ }\href {http://www.jstor.org/stable/25047882} {\bibfield
  {journal} {\bibinfo  {journal} {Sankhya}\ }\textbf {\bibinfo {volume} {7}},\
  \bibinfo {pages} {401} (\bibinfo {year} {1946})}\BibitemShut {NoStop}%
\bibitem [{\citenamefont {Jeffreys}(1946)}]{Jeffreys1946}%
  \BibitemOpen
  \bibfield  {author} {\bibinfo {author} {\bibfnamefont {H.}~\bibnamefont
  {Jeffreys}},\ }\href {\doibase 10.1098/rspa.1946.0056} {\bibfield  {journal}
  {\bibinfo  {journal} {Proc. R. Soc. A}\ }\textbf {\bibinfo {volume} {186}},\
  \bibinfo {pages} {453} (\bibinfo {year} {1946})}\BibitemShut {NoStop}%
\bibitem [{\citenamefont {Matusita}(1967)}]{Matusita1967}%
  \BibitemOpen
  \bibfield  {author} {\bibinfo {author} {\bibfnamefont {K.}~\bibnamefont
  {Matusita}},\ }\href {\doibase 10.1007/BF02911675} {\bibfield  {journal}
  {\bibinfo  {journal} {Ann. Inst. Stat. Math.}\ }\textbf {\bibinfo {volume}
  {19}},\ \bibinfo {pages} {181} (\bibinfo {year} {1967})}\BibitemShut
  {NoStop}%
\bibitem [{\citenamefont {Fuchs}(1995)}]{Fuchs1995}%
  \BibitemOpen
  \bibfield  {author} {\bibinfo {author} {\bibfnamefont {C.~A.}\ \bibnamefont
  {Fuchs}},\ }\emph {\bibinfo {title} {Distinguishability and Accessible
  Information in Quantum Theory}},\ \href@noop {} {Ph.D. thesis},\ \bibinfo
  {school} {University of New Mexico, Albuquerque} (\bibinfo {year}
  {1995})\BibitemShut {NoStop}%
\bibitem [{\citenamefont {Bures}(1969)}]{Bures1969}%
  \BibitemOpen
  \bibfield  {author} {\bibinfo {author} {\bibfnamefont {D.}~\bibnamefont
  {Bures}},\ }\href {\doibase 10.1090/S0002-9947-1969-0236719-2} {\bibfield
  {journal} {\bibinfo  {journal} {Trans. Amer. Math. Soc.}\ }\textbf {\bibinfo
  {volume} {135}},\ \bibinfo {pages} {199} (\bibinfo {year}
  {1969})}\BibitemShut {NoStop}%
\bibitem [{\citenamefont {H\"{u}bner}(1992)}]{Huebner1992}%
  \BibitemOpen
  \bibfield  {author} {\bibinfo {author} {\bibfnamefont {M.}~\bibnamefont
  {H\"{u}bner}},\ }\href {\doibase 10.1016/0375-9601(92)91004-B} {\bibfield
  {journal} {\bibinfo  {journal} {Phys. Lett. A}\ }\textbf {\bibinfo {volume}
  {163}},\ \bibinfo {pages} {239} (\bibinfo {year} {1992})}\BibitemShut
  {NoStop}%
\bibitem [{\citenamefont {Uhlmann}(1976)}]{Uhlmann1976}%
  \BibitemOpen
  \bibfield  {author} {\bibinfo {author} {\bibfnamefont {A.}~\bibnamefont
  {Uhlmann}},\ }\href {\doibase 10.1016/0034-4877(76)90060-4} {\bibfield
  {journal} {\bibinfo  {journal} {Rep. Math. Phys.}\ }\textbf {\bibinfo
  {volume} {9}},\ \bibinfo {pages} {273} (\bibinfo {year} {1976})}\BibitemShut
  {NoStop}%
\bibitem [{\citenamefont {Josza}(1994)}]{Josza1994}%
  \BibitemOpen
  \bibfield  {author} {\bibinfo {author} {\bibfnamefont {R.}~\bibnamefont
  {Josza}},\ }\href {\doibase 10.1080/09500349414552171} {\bibfield  {journal}
  {\bibinfo  {journal} {J. Mod. Opt.}\ }\textbf {\bibinfo {volume} {41}},\
  \bibinfo {pages} {2315} (\bibinfo {year} {1994})}\BibitemShut {NoStop}%
\bibitem [{\citenamefont {Fr\'{e}chet}(1943)}]{Frechet1943}%
  \BibitemOpen
  \bibfield  {author} {\bibinfo {author} {\bibfnamefont {M.}~\bibnamefont
  {Fr\'{e}chet}},\ }\href {http://www.jstor.org/stable/1401114} {\bibfield
  {journal} {\bibinfo  {journal} {Rev. Intl. Stat. Inst.}\ }\textbf {\bibinfo
  {volume} {11}},\ \bibinfo {pages} {182} (\bibinfo {year} {1943})}\BibitemShut
  {NoStop}%
\bibitem [{\citenamefont {Rao}(1945)}]{Rao1945}%
  \BibitemOpen
  \bibfield  {author} {\bibinfo {author} {\bibfnamefont {C.~R.}\ \bibnamefont
  {Rao}},\ }\href@noop {} {\bibfield  {journal} {\bibinfo  {journal} {Bull.
  Calcutta Math. Soc.}\ }\textbf {\bibinfo {volume} {37}},\ \bibinfo {pages}
  {81} (\bibinfo {year} {1945})}\BibitemShut {NoStop}%
\bibitem [{\citenamefont {Cram\'{e}r}(1946)}]{Cramer1946}%
  \BibitemOpen
  \bibfield  {author} {\bibinfo {author} {\bibfnamefont {H.}~\bibnamefont
  {Cram\'{e}r}},\ }\href@noop {} {\emph {\bibinfo {title} {Mathematical Methods
  of Statistics}}}\ (\bibinfo  {publisher} {Princeton University Press},\
  \bibinfo {address} {Princeton, NJ},\ \bibinfo {year} {1946})\BibitemShut
  {NoStop}%
\bibitem [{\citenamefont {Paris}(2009)}]{paris2009}%
  \BibitemOpen
  \bibfield  {author} {\bibinfo {author} {\bibfnamefont {M.~G.}\ \bibnamefont
  {Paris}},\ }\href {http://dx.doi.org/10.1142/S0219749909004839} {\bibfield
  {journal} {\bibinfo  {journal} {Intl. J. Quant. Inf.}\ }\textbf {\bibinfo
  {volume} {7}},\ \bibinfo {pages} {125} (\bibinfo {year} {2009})}\BibitemShut
  {NoStop}%
\bibitem [{\citenamefont {Giovannetti}\ \emph {et~al.}(2011)\citenamefont
  {Giovannetti}, \citenamefont {Lloyd},\ and\ \citenamefont
  {Maccone}}]{Giovannetti2011}%
  \BibitemOpen
  \bibfield  {author} {\bibinfo {author} {\bibfnamefont {V.}~\bibnamefont
  {Giovannetti}}, \bibinfo {author} {\bibfnamefont {S.}~\bibnamefont {Lloyd}},
  \ and\ \bibinfo {author} {\bibfnamefont {L.}~\bibnamefont {Maccone}},\ }\href
  {\doibase 10.1038/nphoton.2011.35} {\bibfield  {journal} {\bibinfo  {journal}
  {Nat. Photon.}\ }\textbf {\bibinfo {volume} {5}},\ \bibinfo {pages} {222}
  (\bibinfo {year} {2011})}\BibitemShut {NoStop}%
\bibitem [{\citenamefont {Audenaert}\ \emph {et~al.}(2007)\citenamefont
  {Audenaert}, \citenamefont {Calsamiglia}, \citenamefont {Mu\~noz Tapia},
  \citenamefont {Bagan}, \citenamefont {Masanes}, \citenamefont {Acin},\ and\
  \citenamefont {Verstraete}}]{PhysRevLett.98.160501}%
  \BibitemOpen
  \bibfield  {author} {\bibinfo {author} {\bibfnamefont {K.~M.~R.}\
  \bibnamefont {Audenaert}}, \bibinfo {author} {\bibfnamefont {J.}~\bibnamefont
  {Calsamiglia}}, \bibinfo {author} {\bibfnamefont {R.}~\bibnamefont {Mu\~noz
  Tapia}}, \bibinfo {author} {\bibfnamefont {E.}~\bibnamefont {Bagan}},
  \bibinfo {author} {\bibfnamefont {L.}~\bibnamefont {Masanes}}, \bibinfo
  {author} {\bibfnamefont {A.}~\bibnamefont {Acin}}, \ and\ \bibinfo {author}
  {\bibfnamefont {F.}~\bibnamefont {Verstraete}},\ }\href {\doibase
  10.1103/PhysRevLett.98.160501} {\bibfield  {journal} {\bibinfo  {journal}
  {Phys. Rev. Lett.}\ }\textbf {\bibinfo {volume} {98}},\ \bibinfo {pages}
  {160501} (\bibinfo {year} {2007})}\BibitemShut {NoStop}%
\bibitem [{\citenamefont {Temme}\ and\ \citenamefont
  {Verstraete}(2015)}]{Temme2015}%
  \BibitemOpen
  \bibfield  {author} {\bibinfo {author} {\bibfnamefont {K.}~\bibnamefont
  {Temme}}\ and\ \bibinfo {author} {\bibfnamefont {F.}~\bibnamefont
  {Verstraete}},\ }\href {\doibase 10.1063/1.4905843} {\bibfield  {journal}
  {\bibinfo  {journal} {J. Math. Phys.}\ }\textbf {\bibinfo {volume} {56}},\
  \bibinfo {pages} {012202} (\bibinfo {year} {2015})}\BibitemShut {NoStop}%
\bibitem [{foo()}]{footnotePseudoDistance}%
  \BibitemOpen
  \href@noop {} {}\bibinfo {note} {A distance measure satisfies the properties
  of non-negativity (yielding zero if and only if both elements are identical),
  symmetry, and the triangle inequality. A pseudodistance may not satisfy the
  latter two but still conveys a useful sense of discrepancy between two
  objects, a prominent example being the relative entropy.}\BibitemShut {Stop}%
\bibitem [{\citenamefont {Helstrom}(1976)}]{Helstrom1976}%
  \BibitemOpen
  \bibfield  {author} {\bibinfo {author} {\bibfnamefont {C.~W.}\ \bibnamefont
  {Helstrom}},\ }\href@noop {} {\emph {\bibinfo {title} {Quantum Detection and
  Estimation Theory. Mathematics in Science and Engineering, vol. 123}}}\
  (\bibinfo  {publisher} {Academic Press},\ \bibinfo {year} {1976})\BibitemShut
  {NoStop}%
\bibitem [{\citenamefont {Boekee}(1977{\natexlab{a}})}]{Boekee1977}%
  \BibitemOpen
  \bibfield  {author} {\bibinfo {author} {\bibfnamefont {D.~E.}\ \bibnamefont
  {Boekee}},\ }\emph {\bibinfo {title} {A Generalization of the {F}isher
  Information Measure}},\ \href@noop {} {Ph.D. thesis},\ \bibinfo  {school}
  {Technische Hogeschool Delft} (\bibinfo {year}
  {1977}{\natexlab{a}})\BibitemShut {NoStop}%
\bibitem [{\citenamefont {Barankin}(1949)}]{Barankin1949}%
  \BibitemOpen
  \bibfield  {author} {\bibinfo {author} {\bibfnamefont {E.~W.}\ \bibnamefont
  {Barankin}},\ }\href {\doibase 10.1214/aoms/1177729943} {\bibfield  {journal}
  {\bibinfo  {journal} {Ann. Math. Stat.}\ }\textbf {\bibinfo {volume} {20}},\
  \bibinfo {pages} {477} (\bibinfo {year} {1949})}\BibitemShut {NoStop}%
\bibitem [{\citenamefont {Stangenhaus}(1977)}]{Stangenhaus1977}%
  \BibitemOpen
  \bibfield  {author} {\bibinfo {author} {\bibfnamefont {G.}~\bibnamefont
  {Stangenhaus}},\ }\emph {\bibinfo {title} {Optimum estimation under
  generalized unbiasedness}},\ \href@noop {} {Ph.D. thesis},\ \bibinfo
  {school} {Iowa State University} (\bibinfo {year} {1977})\BibitemShut
  {NoStop}%
\bibitem [{\citenamefont {Ferentinos}\ and\ \citenamefont
  {Papaioannou}(1981)}]{Ferentinos1981}%
  \BibitemOpen
  \bibfield  {author} {\bibinfo {author} {\bibfnamefont {K.}~\bibnamefont
  {Ferentinos}}\ and\ \bibinfo {author} {\bibfnamefont {T.}~\bibnamefont
  {Papaioannou}},\ }\href {\doibase 10.1016/S0019-9958(81)90263-1} {\bibfield
  {journal} {\bibinfo  {journal} {Information and Control}\ }\textbf {\bibinfo
  {volume} {51}},\ \bibinfo {pages} {193 } (\bibinfo {year}
  {1981})}\BibitemShut {NoStop}%
\bibitem [{\citenamefont {Lutwak}\ \emph {et~al.}(2005)\citenamefont {Lutwak},
  \citenamefont {Yang},\ and\ \citenamefont {Zhang}}]{Lutwak2005}%
  \BibitemOpen
  \bibfield  {author} {\bibinfo {author} {\bibfnamefont {E.}~\bibnamefont
  {Lutwak}}, \bibinfo {author} {\bibfnamefont {D.}~\bibnamefont {Yang}}, \ and\
  \bibinfo {author} {\bibfnamefont {G.}~\bibnamefont {Zhang}},\ }\href
  {\doibase 10.1109/TIT.2004.840871} {\bibfield  {journal} {\bibinfo  {journal}
  {IEEE Trans. Inf. Theo.}\ }\textbf {\bibinfo {volume} {51}},\ \bibinfo
  {pages} {473} (\bibinfo {year} {2005})}\BibitemShut {NoStop}%
\bibitem [{\citenamefont {Bercher}(2012)}]{Bercher2012}%
  \BibitemOpen
  \bibfield  {author} {\bibinfo {author} {\bibfnamefont {J.-F.}\ \bibnamefont
  {Bercher}},\ }\href {\doibase 10.1088/1751-8113/45/25/255303} {\bibfield
  {journal} {\bibinfo  {journal} {J. Phys. A}\ }\textbf {\bibinfo {volume}
  {45}},\ \bibinfo {pages} {255303} (\bibinfo {year} {2012})}\BibitemShut
  {NoStop}%
\bibitem [{\citenamefont {Luis}\ and\ \citenamefont
  {Rodil}(2013)}]{PhysRevA.87.034101}%
  \BibitemOpen
  \bibfield  {author} {\bibinfo {author} {\bibfnamefont {A.}~\bibnamefont
  {Luis}}\ and\ \bibinfo {author} {\bibfnamefont {A.}~\bibnamefont {Rodil}},\
  }\href {\doibase 10.1103/PhysRevA.87.034101} {\bibfield  {journal} {\bibinfo
  {journal} {Phys. Rev. A}\ }\textbf {\bibinfo {volume} {87}},\ \bibinfo
  {pages} {034101} (\bibinfo {year} {2013})}\BibitemShut {NoStop}%
\bibitem [{\citenamefont {R\'{e}nyi}(1961)}]{Renyi1961}%
  \BibitemOpen
  \bibfield  {author} {\bibinfo {author} {\bibfnamefont {A.}~\bibnamefont
  {R\'{e}nyi}},\ }in\ \href@noop {} {\emph {\bibinfo {booktitle} {Proceedings
  of the fourth {B}erkeley Symposium on Mathematics, Statistics and
  Probability}}}\ (\bibinfo {year} {1961})\ pp.\ \bibinfo {pages}
  {547--561}\BibitemShut {NoStop}%
\bibitem [{\citenamefont {Boekee}(1977{\natexlab{b}})}]{Boekee1977a}%
  \BibitemOpen
  \bibfield  {author} {\bibinfo {author} {\bibfnamefont {D.}~\bibnamefont
  {Boekee}},\ }in\ \href@noop {} {\emph {\bibinfo {booktitle} {Topics in
  Information Theory, Vol. 16}}},\ \bibinfo {editor} {edited by\ \bibinfo
  {editor} {\bibfnamefont {I.}~\bibnamefont {Csisz\'{a}r}}\ and\ \bibinfo
  {editor} {\bibfnamefont {P.}~\bibnamefont {Elias}}}\ (\bibinfo  {publisher}
  {J\'{a}nos Bolyai Mathematical Society and North-Holland},\ \bibinfo
  {address} {Keszthely, Hungary},\ \bibinfo {year} {1977})\ pp.\ \bibinfo
  {pages} {113--123}\BibitemShut {NoStop}%
\bibitem [{\citenamefont {Bercher}(2013)}]{Bercher2013}%
  \BibitemOpen
  \bibfield  {author} {\bibinfo {author} {\bibfnamefont {J.-F.}\ \bibnamefont
  {Bercher}},\ }\href {\doibase http://dx.doi.org/10.1016/j.physa.2013.03.062}
  {\bibfield  {journal} {\bibinfo  {journal} {Physica A}\ }\textbf {\bibinfo
  {volume} {392}},\ \bibinfo {pages} {3140 } (\bibinfo {year}
  {2013})}\BibitemShut {NoStop}%
\bibitem [{\citenamefont {Cianchi}\ \emph {et~al.}(2014)\citenamefont
  {Cianchi}, \citenamefont {Lutwak}, \citenamefont {Yang},\ and\ \citenamefont
  {Zhang}}]{Cianchi2014}%
  \BibitemOpen
  \bibfield  {author} {\bibinfo {author} {\bibfnamefont {A.}~\bibnamefont
  {Cianchi}}, \bibinfo {author} {\bibfnamefont {E.}~\bibnamefont {Lutwak}},
  \bibinfo {author} {\bibfnamefont {D.}~\bibnamefont {Yang}}, \ and\ \bibinfo
  {author} {\bibfnamefont {G.}~\bibnamefont {Zhang}},\ }\href {\doibase
  10.1109/TIT.2013.2284498} {\bibfield  {journal} {\bibinfo  {journal} {IEEE
  Trans. Inf. Theo.}\ }\textbf {\bibinfo {volume} {60}},\ \bibinfo {pages}
  {643} (\bibinfo {year} {2014})}\BibitemShut {NoStop}%
\bibitem [{\citenamefont {Gessner}\ \emph
  {et~al.}(2017{\natexlab{b}})\citenamefont {Gessner}, \citenamefont
  {Pezz\`e},\ and\ \citenamefont {Smerzi}}]{ResolutionEnhanced}%
  \BibitemOpen
  \bibfield  {author} {\bibinfo {author} {\bibfnamefont {M.}~\bibnamefont
  {Gessner}}, \bibinfo {author} {\bibfnamefont {L.}~\bibnamefont {Pezz\`e}}, \
  and\ \bibinfo {author} {\bibfnamefont {A.}~\bibnamefont {Smerzi}},\ }\href
  {\doibase 10.1103/PhysRevA.95.032326} {\bibfield  {journal} {\bibinfo
  {journal} {Phys. Rev. A}\ }\textbf {\bibinfo {volume} {95}},\ \bibinfo
  {pages} {032326} (\bibinfo {year} {2017}{\natexlab{b}})}\BibitemShut
  {NoStop}%
\bibitem [{\citenamefont {Oszmaniec}\ \emph {et~al.}(2016)\citenamefont
  {Oszmaniec}, \citenamefont {Augusiak}, \citenamefont {Gogolin}, \citenamefont
  {Ko\l{}ody\ifmmode~\acute{n}\else \'{n}\fi{}ski}, \citenamefont {Ac\'{\i}n},\
  and\ \citenamefont {Lewenstein}}]{PhysRevX.6.041044}%
  \BibitemOpen
  \bibfield  {author} {\bibinfo {author} {\bibfnamefont {M.}~\bibnamefont
  {Oszmaniec}}, \bibinfo {author} {\bibfnamefont {R.}~\bibnamefont {Augusiak}},
  \bibinfo {author} {\bibfnamefont {C.}~\bibnamefont {Gogolin}}, \bibinfo
  {author} {\bibfnamefont {J.}~\bibnamefont {Ko\l{}ody\ifmmode~\acute{n}\else
  \'{n}\fi{}ski}}, \bibinfo {author} {\bibfnamefont {A.}~\bibnamefont
  {Ac\'{\i}n}}, \ and\ \bibinfo {author} {\bibfnamefont {M.}~\bibnamefont
  {Lewenstein}},\ }\href {\doibase 10.1103/PhysRevX.6.041044} {\bibfield
  {journal} {\bibinfo  {journal} {Phys. Rev. X}\ }\textbf {\bibinfo {volume}
  {6}},\ \bibinfo {pages} {041044} (\bibinfo {year} {2016})}\BibitemShut
  {NoStop}%
\bibitem [{\citenamefont {Moiseyev}(2011)}]{Moiseyev2011}%
  \BibitemOpen
  \bibfield  {author} {\bibinfo {author} {\bibfnamefont {N.}~\bibnamefont
  {Moiseyev}},\ }\href@noop {} {\emph {\bibinfo {title} {Non-Hermitian Quantum
  Mechanics}}}\ (\bibinfo  {publisher} {Cambridge University Press},\ \bibinfo
  {address} {Cambridge, UK},\ \bibinfo {year} {2011})\BibitemShut {NoStop}%
\bibitem [{\citenamefont {Breuer}\ and\ \citenamefont
  {Petruccione}(2002)}]{BreuerPetruccione2006}%
  \BibitemOpen
  \bibfield  {author} {\bibinfo {author} {\bibfnamefont {H.-P.}\ \bibnamefont
  {Breuer}}\ and\ \bibinfo {author} {\bibfnamefont {F.}~\bibnamefont
  {Petruccione}},\ }\href@noop {} {\emph {\bibinfo {title} {The Theory of Open
  Quantum Systems}}}\ (\bibinfo  {publisher} {Oxford University Press},\
  \bibinfo {address} {Oxford, UK},\ \bibinfo {year} {2002})\BibitemShut
  {NoStop}%
\bibitem [{\citenamefont {Kitaev}(1997)}]{Kitaev1997}%
  \BibitemOpen
  \bibfield  {author} {\bibinfo {author} {\bibfnamefont {A.~Y.}\ \bibnamefont
  {Kitaev}},\ }\href {\doibase 10.1070/RM1997v052n06ABEH002155} {\bibfield
  {journal} {\bibinfo  {journal} {Russian Mathematical Surveys}\ }\textbf
  {\bibinfo {volume} {52}},\ \bibinfo {pages} {1191} (\bibinfo {year}
  {1997})}\BibitemShut {NoStop}%
\bibitem [{\citenamefont {Amosov}\ \emph {et~al.}(2000)\citenamefont {Amosov},
  \citenamefont {Holevo},\ and\ \citenamefont {Werner}}]{Amosov2000}%
  \BibitemOpen
  \bibfield  {author} {\bibinfo {author} {\bibfnamefont {G.~G.}\ \bibnamefont
  {Amosov}}, \bibinfo {author} {\bibfnamefont {A.~S.}\ \bibnamefont {Holevo}},
  \ and\ \bibinfo {author} {\bibfnamefont {R.~F.}\ \bibnamefont {Werner}},\
  }\href@noop {} {\bibfield  {journal} {\bibinfo  {journal} {Probl. Inform.
  Transm.}\ }\textbf {\bibinfo {volume} {36}},\ \bibinfo {pages} {305}
  (\bibinfo {year} {2000})}\BibitemShut {NoStop}%
\bibitem [{\citenamefont {Watrous}(2005)}]{Watrous2005}%
  \BibitemOpen
  \bibfield  {author} {\bibinfo {author} {\bibfnamefont {J.}~\bibnamefont
  {Watrous}},\ }\href {http://dl.acm.org/citation.cfm?id=2011608.2011614}
  {\bibfield  {journal} {\bibinfo  {journal} {Quantum Info. Comput.}\ }\textbf
  {\bibinfo {volume} {5}},\ \bibinfo {pages} {58} (\bibinfo {year}
  {2005})}\BibitemShut {NoStop}%
\bibitem [{\citenamefont {Devetak}\ \emph {et~al.}(2006)\citenamefont
  {Devetak}, \citenamefont {Junge}, \citenamefont {King},\ and\ \citenamefont
  {Ruskai}}]{Devetak2006}%
  \BibitemOpen
  \bibfield  {author} {\bibinfo {author} {\bibfnamefont {I.}~\bibnamefont
  {Devetak}}, \bibinfo {author} {\bibfnamefont {M.}~\bibnamefont {Junge}},
  \bibinfo {author} {\bibfnamefont {C.}~\bibnamefont {King}}, \ and\ \bibinfo
  {author} {\bibfnamefont {M.~B.}\ \bibnamefont {Ruskai}},\ }\href {\doibase
  10.1007/s00220-006-0034-0} {\bibfield  {journal} {\bibinfo  {journal} {Comm.
  Math. Phys.}\ }\textbf {\bibinfo {volume} {266}},\ \bibinfo {pages} {37}
  (\bibinfo {year} {2006})}\BibitemShut {NoStop}%
\bibitem [{\citenamefont {Gilchrist}\ \emph {et~al.}(2005)\citenamefont
  {Gilchrist}, \citenamefont {Langford},\ and\ \citenamefont
  {Nielsen}}]{PhysRevA.71.062310}%
  \BibitemOpen
  \bibfield  {author} {\bibinfo {author} {\bibfnamefont {A.}~\bibnamefont
  {Gilchrist}}, \bibinfo {author} {\bibfnamefont {N.~K.}\ \bibnamefont
  {Langford}}, \ and\ \bibinfo {author} {\bibfnamefont {M.~A.}\ \bibnamefont
  {Nielsen}},\ }\href {\doibase 10.1103/PhysRevA.71.062310} {\bibfield
  {journal} {\bibinfo  {journal} {Phys. Rev. A}\ }\textbf {\bibinfo {volume}
  {71}},\ \bibinfo {pages} {062310} (\bibinfo {year} {2005})}\BibitemShut
  {NoStop}%
\bibitem [{\citenamefont {Bhatia}\ and\ \citenamefont
  {Davis}(2000)}]{Bhatia2000}%
  \BibitemOpen
  \bibfield  {author} {\bibinfo {author} {\bibfnamefont {R.}~\bibnamefont
  {Bhatia}}\ and\ \bibinfo {author} {\bibfnamefont {C.}~\bibnamefont {Davis}},\
  }\href {http://www.jstor.org/stable/2589180} {\bibfield  {journal} {\bibinfo
  {journal} {The American Mathematical Monthly}\ }\textbf {\bibinfo {volume}
  {107}},\ \bibinfo {pages} {353} (\bibinfo {year} {2000})}\BibitemShut
  {NoStop}%
\bibitem [{\citenamefont {Escher}\ \emph {et~al.}(2011)\citenamefont {Escher},
  \citenamefont {de~Matos~Filho},\ and\ \citenamefont
  {Davidovich}}]{Escher2011}%
  \BibitemOpen
  \bibfield  {author} {\bibinfo {author} {\bibfnamefont {B.~M.}\ \bibnamefont
  {Escher}}, \bibinfo {author} {\bibfnamefont {R.~L.}\ \bibnamefont
  {de~Matos~Filho}}, \ and\ \bibinfo {author} {\bibfnamefont {L.}~\bibnamefont
  {Davidovich}},\ }\href {\doibase 10.1038/nphys1958} {\bibfield  {journal}
  {\bibinfo  {journal} {Nat. Phys.}\ }\textbf {\bibinfo {volume} {7}},\
  \bibinfo {pages} {406} (\bibinfo {year} {2011})}\BibitemShut {NoStop}%
\bibitem [{\citenamefont {Alipour}\ \emph {et~al.}(2014)\citenamefont
  {Alipour}, \citenamefont {Mehboudi},\ and\ \citenamefont
  {Rezakhani}}]{PhysRevLett.112.120405}%
  \BibitemOpen
  \bibfield  {author} {\bibinfo {author} {\bibfnamefont {S.}~\bibnamefont
  {Alipour}}, \bibinfo {author} {\bibfnamefont {M.}~\bibnamefont {Mehboudi}}, \
  and\ \bibinfo {author} {\bibfnamefont {A.~T.}\ \bibnamefont {Rezakhani}},\
  }\href {\doibase 10.1103/PhysRevLett.112.120405} {\bibfield  {journal}
  {\bibinfo  {journal} {Phys. Rev. Lett.}\ }\textbf {\bibinfo {volume} {112}},\
  \bibinfo {pages} {120405} (\bibinfo {year} {2014})}\BibitemShut {NoStop}%
\bibitem [{\citenamefont {Demkowicz-Dobrza\ifmmode~\acute{n}\else
  \'{n}\fi{}ski}\ \emph {et~al.}(2012)\citenamefont
  {Demkowicz-Dobrza\ifmmode~\acute{n}\else \'{n}\fi{}ski}, \citenamefont
  {Ko\l{}ody\ifmmode~\acute{n}\else \'{n}\fi{}ski},\ and\ \citenamefont
  {Gu\ifmmode \mbox{\c{t}}\else \c{t}\fi{}\ifmmode~\u{a}\else
  \u{a}\fi{}}}]{Rafal2012}%
  \BibitemOpen
  \bibfield  {author} {\bibinfo {author} {\bibfnamefont {R.}~\bibnamefont
  {Demkowicz-Dobrza\ifmmode~\acute{n}\else \'{n}\fi{}ski}}, \bibinfo {author}
  {\bibfnamefont {J.}~\bibnamefont {Ko\l{}ody\ifmmode~\acute{n}\else
  \'{n}\fi{}ski}}, \ and\ \bibinfo {author} {\bibfnamefont {M.}~\bibnamefont
  {Gu\ifmmode \mbox{\c{t}}\else \c{t}\fi{}\ifmmode~\u{a}\else \u{a}\fi{}}},\
  }\href {\doibase 10.1038/ncomms2067} {\bibfield  {journal} {\bibinfo
  {journal} {Nat. Comm.}\ }\textbf {\bibinfo {volume} {3}},\ \bibinfo {pages}
  {1063} (\bibinfo {year} {2012})}\BibitemShut {NoStop}%
\bibitem [{\citenamefont {Smirne}\ \emph {et~al.}(2016)\citenamefont {Smirne},
  \citenamefont {Ko\l{}ody\ifmmode~\acute{n}\else \'{n}\fi{}ski}, \citenamefont
  {Huelga},\ and\ \citenamefont {Demkowicz-Dobrza\ifmmode~\acute{n}\else
  \'{n}\fi{}ski}}]{PhysRevLett.116.120801}%
  \BibitemOpen
  \bibfield  {author} {\bibinfo {author} {\bibfnamefont {A.}~\bibnamefont
  {Smirne}}, \bibinfo {author} {\bibfnamefont {J.}~\bibnamefont
  {Ko\l{}ody\ifmmode~\acute{n}\else \'{n}\fi{}ski}}, \bibinfo {author}
  {\bibfnamefont {S.~F.}\ \bibnamefont {Huelga}}, \ and\ \bibinfo {author}
  {\bibfnamefont {R.}~\bibnamefont {Demkowicz-Dobrza\ifmmode~\acute{n}\else
  \'{n}\fi{}ski}},\ }\href {\doibase 10.1103/PhysRevLett.116.120801} {\bibfield
   {journal} {\bibinfo  {journal} {Phys. Rev. Lett.}\ }\textbf {\bibinfo
  {volume} {116}},\ \bibinfo {pages} {120801} (\bibinfo {year}
  {2016})}\BibitemShut {NoStop}%
\bibitem [{\citenamefont {Yu}(2013)}]{Yu2013}%
  \BibitemOpen
  \bibfield  {author} {\bibinfo {author} {\bibfnamefont {S.}~\bibnamefont
  {Yu}},\ }\href {https://arxiv.org/abs/1302.5311} {\enquote {\bibinfo {title}
  {{Q}uantum {F}isher {I}nformation as the {C}onvex {R}oof of {V}ariance},}\
  }\bibinfo {howpublished} {arXiv:1302.5311 [quant-ph]} (\bibinfo {year}
  {2013})\BibitemShut {NoStop}%
\bibitem [{\citenamefont {Ma}\ \emph {et~al.}(2011)\citenamefont {Ma},
  \citenamefont {Wang}, \citenamefont {Sun},\ and\ \citenamefont
  {Nori}}]{SpinSqueezing}%
  \BibitemOpen
  \bibfield  {author} {\bibinfo {author} {\bibfnamefont {J.}~\bibnamefont
  {Ma}}, \bibinfo {author} {\bibfnamefont {X.}~\bibnamefont {Wang}}, \bibinfo
  {author} {\bibfnamefont {C.}~\bibnamefont {Sun}}, \ and\ \bibinfo {author}
  {\bibfnamefont {F.}~\bibnamefont {Nori}},\ }\href {\doibase
  http://dx.doi.org/10.1016/j.physrep.2011.08.003} {\bibfield  {journal}
  {\bibinfo  {journal} {Phys. Rep.}\ }\textbf {\bibinfo {volume} {509}},\
  \bibinfo {pages} {89 } (\bibinfo {year} {2011})}\BibitemShut {NoStop}%
\bibitem [{\citenamefont {Wineland}\ \emph {et~al.}(1992)\citenamefont
  {Wineland}, \citenamefont {Bollinger}, \citenamefont {Itano}, \citenamefont
  {Moore},\ and\ \citenamefont {Heinzen}}]{PhysRevA.46.R6797}%
  \BibitemOpen
  \bibfield  {author} {\bibinfo {author} {\bibfnamefont {D.~J.}\ \bibnamefont
  {Wineland}}, \bibinfo {author} {\bibfnamefont {J.~J.}\ \bibnamefont
  {Bollinger}}, \bibinfo {author} {\bibfnamefont {W.~M.}\ \bibnamefont
  {Itano}}, \bibinfo {author} {\bibfnamefont {F.~L.}\ \bibnamefont {Moore}}, \
  and\ \bibinfo {author} {\bibfnamefont {D.~J.}\ \bibnamefont {Heinzen}},\
  }\href {\doibase 10.1103/PhysRevA.46.R6797} {\bibfield  {journal} {\bibinfo
  {journal} {Phys. Rev. A}\ }\textbf {\bibinfo {volume} {46}},\ \bibinfo
  {pages} {R6797} (\bibinfo {year} {1992})}\BibitemShut {NoStop}%
\bibitem [{\citenamefont {Girolami}\ and\ \citenamefont
  {Yadin}(2017)}]{Girolami2017}%
  \BibitemOpen
  \bibfield  {author} {\bibinfo {author} {\bibfnamefont {D.}~\bibnamefont
  {Girolami}}\ and\ \bibinfo {author} {\bibfnamefont {B.}~\bibnamefont
  {Yadin}},\ }\href {\doibase 10.3390/e19030124} {\bibfield  {journal}
  {\bibinfo  {journal} {Entropy}\ }\textbf {\bibinfo {volume} {19}},\ \bibinfo
  {pages} {124} (\bibinfo {year} {2017})}\BibitemShut {NoStop}%
\bibitem [{\citenamefont {Hyllus}\ \emph {et~al.}(2012)\citenamefont {Hyllus},
  \citenamefont {Laskowski}, \citenamefont {Krischek}, \citenamefont
  {Schwemmer}, \citenamefont {Wieczorek}, \citenamefont {Weinfurter},
  \citenamefont {Pezz\'e},\ and\ \citenamefont {Smerzi}}]{PhysRevA.85.022321}%
  \BibitemOpen
  \bibfield  {author} {\bibinfo {author} {\bibfnamefont {P.}~\bibnamefont
  {Hyllus}}, \bibinfo {author} {\bibfnamefont {W.}~\bibnamefont {Laskowski}},
  \bibinfo {author} {\bibfnamefont {R.}~\bibnamefont {Krischek}}, \bibinfo
  {author} {\bibfnamefont {C.}~\bibnamefont {Schwemmer}}, \bibinfo {author}
  {\bibfnamefont {W.}~\bibnamefont {Wieczorek}}, \bibinfo {author}
  {\bibfnamefont {H.}~\bibnamefont {Weinfurter}}, \bibinfo {author}
  {\bibfnamefont {L.}~\bibnamefont {Pezz\'e}}, \ and\ \bibinfo {author}
  {\bibfnamefont {A.}~\bibnamefont {Smerzi}},\ }\href {\doibase
  10.1103/PhysRevA.85.022321} {\bibfield  {journal} {\bibinfo  {journal} {Phys.
  Rev. A}\ }\textbf {\bibinfo {volume} {85}},\ \bibinfo {pages} {022321}
  (\bibinfo {year} {2012})}\BibitemShut {NoStop}%
\bibitem [{\citenamefont {T\'oth}(2012)}]{PhysRevA.85.022322}%
  \BibitemOpen
  \bibfield  {author} {\bibinfo {author} {\bibfnamefont {G.}~\bibnamefont
  {T\'oth}},\ }\href {\doibase 10.1103/PhysRevA.85.022322} {\bibfield
  {journal} {\bibinfo  {journal} {Phys. Rev. A}\ }\textbf {\bibinfo {volume}
  {85}},\ \bibinfo {pages} {022322} (\bibinfo {year} {2012})}\BibitemShut
  {NoStop}%
\bibitem [{\citenamefont {Gessner}\ \emph {et~al.}(2016)\citenamefont
  {Gessner}, \citenamefont {Pezz\`e},\ and\ \citenamefont
  {Smerzi}}]{Gessner2016}%
  \BibitemOpen
  \bibfield  {author} {\bibinfo {author} {\bibfnamefont {M.}~\bibnamefont
  {Gessner}}, \bibinfo {author} {\bibfnamefont {L.}~\bibnamefont {Pezz\`e}}, \
  and\ \bibinfo {author} {\bibfnamefont {A.}~\bibnamefont {Smerzi}},\ }\href
  {\doibase 10.1103/PhysRevA.94.020101} {\bibfield  {journal} {\bibinfo
  {journal} {Phys. Rev. A}\ }\textbf {\bibinfo {volume} {94}},\ \bibinfo
  {pages} {020101(R)} (\bibinfo {year} {2016})}\BibitemShut {NoStop}%
\bibitem [{\citenamefont {Gessner}\ \emph
  {et~al.}(2017{\natexlab{c}})\citenamefont {Gessner}, \citenamefont
  {Pezz\`e},\ and\ \citenamefont {Smerzi}}]{BosonicSqueezing}%
  \BibitemOpen
  \bibfield  {author} {\bibinfo {author} {\bibfnamefont {M.}~\bibnamefont
  {Gessner}}, \bibinfo {author} {\bibfnamefont {L.}~\bibnamefont {Pezz\`e}}, \
  and\ \bibinfo {author} {\bibfnamefont {A.}~\bibnamefont {Smerzi}},\ }\href
  {\doibase 10.22331/q-2017-07-14-17} {\bibfield  {journal} {\bibinfo
  {journal} {Quantum}\ }\textbf {\bibinfo {volume} {1}},\ \bibinfo {pages} {17}
  (\bibinfo {year} {2017}{\natexlab{c}})}\BibitemShut {NoStop}%
\bibitem [{\citenamefont {Chen}(2005)}]{PhysRevA.71.052302}%
  \BibitemOpen
  \bibfield  {author} {\bibinfo {author} {\bibfnamefont {Z.}~\bibnamefont
  {Chen}},\ }\href {\doibase 10.1103/PhysRevA.71.052302} {\bibfield  {journal}
  {\bibinfo  {journal} {Phys. Rev. A}\ }\textbf {\bibinfo {volume} {71}},\
  \bibinfo {pages} {052302} (\bibinfo {year} {2005})}\BibitemShut {NoStop}%
\bibitem [{\citenamefont {Hyllus}\ \emph {et~al.}(2010)\citenamefont {Hyllus},
  \citenamefont {G\"uhne},\ and\ \citenamefont {Smerzi}}]{PhysRevA.82.012337}%
  \BibitemOpen
  \bibfield  {author} {\bibinfo {author} {\bibfnamefont {P.}~\bibnamefont
  {Hyllus}}, \bibinfo {author} {\bibfnamefont {O.}~\bibnamefont {G\"uhne}}, \
  and\ \bibinfo {author} {\bibfnamefont {A.}~\bibnamefont {Smerzi}},\ }\href
  {\doibase 10.1103/PhysRevA.82.012337} {\bibfield  {journal} {\bibinfo
  {journal} {Phys. Rev. A}\ }\textbf {\bibinfo {volume} {82}},\ \bibinfo
  {pages} {012337} (\bibinfo {year} {2010})}\BibitemShut {NoStop}%
\bibitem [{\citenamefont {Mandelstam}\ and\ \citenamefont
  {Tamm}(1945)}]{Mandelstam1945}%
  \BibitemOpen
  \bibfield  {author} {\bibinfo {author} {\bibfnamefont {L.}~\bibnamefont
  {Mandelstam}}\ and\ \bibinfo {author} {\bibfnamefont {I.}~\bibnamefont
  {Tamm}},\ }\href@noop {} {\bibfield  {journal} {\bibinfo  {journal} {J. Phys.
  USSR}\ }\textbf {\bibinfo {volume} {9}},\ \bibinfo {pages} {249} (\bibinfo
  {year} {1945})}\BibitemShut {NoStop}%
\bibitem [{\citenamefont {Margolus}\ and\ \citenamefont
  {Levitin}(1998)}]{Margolus1998}%
  \BibitemOpen
  \bibfield  {author} {\bibinfo {author} {\bibfnamefont {N.}~\bibnamefont
  {Margolus}}\ and\ \bibinfo {author} {\bibfnamefont {L.~B.}\ \bibnamefont
  {Levitin}},\ }\href {\doibase https://doi.org/10.1016/S0167-2789(98)00054-2}
  {\bibfield  {journal} {\bibinfo  {journal} {Physica D: Nonlinear Phenomena}\
  }\textbf {\bibinfo {volume} {120}},\ \bibinfo {pages} {188 } (\bibinfo {year}
  {1998})}\BibitemShut {NoStop}%
\bibitem [{\citenamefont {Anandan}\ and\ \citenamefont
  {Aharonov}(1990)}]{PhysRevLett.65.1697}%
  \BibitemOpen
  \bibfield  {author} {\bibinfo {author} {\bibfnamefont {J.}~\bibnamefont
  {Anandan}}\ and\ \bibinfo {author} {\bibfnamefont {Y.}~\bibnamefont
  {Aharonov}},\ }\href {\doibase 10.1103/PhysRevLett.65.1697} {\bibfield
  {journal} {\bibinfo  {journal} {Phys. Rev. Lett.}\ }\textbf {\bibinfo
  {volume} {65}},\ \bibinfo {pages} {1697} (\bibinfo {year}
  {1990})}\BibitemShut {NoStop}%
\bibitem [{\citenamefont {Braunstein}\ \emph {et~al.}(1996)\citenamefont
  {Braunstein}, \citenamefont {Caves},\ and\ \citenamefont
  {Milburn}}]{Braunstein1996}%
  \BibitemOpen
  \bibfield  {author} {\bibinfo {author} {\bibfnamefont {S.~L.}\ \bibnamefont
  {Braunstein}}, \bibinfo {author} {\bibfnamefont {C.~M.}\ \bibnamefont
  {Caves}}, \ and\ \bibinfo {author} {\bibfnamefont {G.}~\bibnamefont
  {Milburn}},\ }\href {\doibase https://doi.org/10.1006/aphy.1996.0040}
  {\bibfield  {journal} {\bibinfo  {journal} {Annals of Physics}\ }\textbf
  {\bibinfo {volume} {247}},\ \bibinfo {pages} {135 } (\bibinfo {year}
  {1996})}\BibitemShut {NoStop}%
\bibitem [{\citenamefont {Taddei}(2014)}]{Taddei2014}%
  \BibitemOpen
  \bibfield  {author} {\bibinfo {author} {\bibfnamefont {M.~M.}\ \bibnamefont
  {Taddei}},\ }\emph {\bibinfo {title} {Quantum Speed Limits for General
  Physical Processes}},\ \href@noop {} {Ph.D. thesis},\ \bibinfo  {school}
  {Universidade Federal do Rio de Janeiro} (\bibinfo {year} {2014}),\ \Eprint
  {http://arxiv.org/abs/1407.4343} {1407.4343} \BibitemShut {NoStop}%
\bibitem [{\citenamefont {Deffner}\ and\ \citenamefont
  {Campbell}(2017)}]{Deffner2017}%
  \BibitemOpen
  \bibfield  {author} {\bibinfo {author} {\bibfnamefont {S.}~\bibnamefont
  {Deffner}}\ and\ \bibinfo {author} {\bibfnamefont {S.}~\bibnamefont
  {Campbell}},\ }\href {\doibase 10.1088/1751-8121/aa86c6} {\bibfield
  {journal} {\bibinfo  {journal} {Journal of Physics A: Mathematical and
  Theoretical}\ }\textbf {\bibinfo {volume} {50}},\ \bibinfo {pages} {453001}
  (\bibinfo {year} {2017})}\BibitemShut {NoStop}%
\bibitem [{\citenamefont {Wi\ss{}mann}\ \emph {et~al.}(2015)\citenamefont
  {Wi\ss{}mann}, \citenamefont {Breuer},\ and\ \citenamefont
  {Vacchini}}]{PhysRevA.92.042108}%
  \BibitemOpen
  \bibfield  {author} {\bibinfo {author} {\bibfnamefont {S.}~\bibnamefont
  {Wi\ss{}mann}}, \bibinfo {author} {\bibfnamefont {H.-P.}\ \bibnamefont
  {Breuer}}, \ and\ \bibinfo {author} {\bibfnamefont {B.}~\bibnamefont
  {Vacchini}},\ }\href {\doibase 10.1103/PhysRevA.92.042108} {\bibfield
  {journal} {\bibinfo  {journal} {Phys. Rev. A}\ }\textbf {\bibinfo {volume}
  {92}},\ \bibinfo {pages} {042108} (\bibinfo {year} {2015})}\BibitemShut
  {NoStop}%
\bibitem [{\citenamefont {Brown}(1947)}]{Brown1947}%
  \BibitemOpen
  \bibfield  {author} {\bibinfo {author} {\bibfnamefont {G.~W.}\ \bibnamefont
  {Brown}},\ }\href {http://www.jstor.org/stable/2236236} {\bibfield  {journal}
  {\bibinfo  {journal} {The Annals of Mathematical Statistics}\ }\textbf
  {\bibinfo {volume} {18}},\ \bibinfo {pages} {582} (\bibinfo {year}
  {1947})}\BibitemShut {NoStop}%
\bibitem [{\citenamefont {Lehmann}(1951)}]{Lehmann1951}%
  \BibitemOpen
  \bibfield  {author} {\bibinfo {author} {\bibfnamefont {E.~L.}\ \bibnamefont
  {Lehmann}},\ }\href {http://www.jstor.org/stable/2236928} {\bibfield
  {journal} {\bibinfo  {journal} {The Annals of Mathematical Statistics}\
  }\textbf {\bibinfo {volume} {22}},\ \bibinfo {pages} {587} (\bibinfo {year}
  {1951})}\BibitemShut {NoStop}%
\bibitem [{\citenamefont {Lehmann}\ and\ \citenamefont {Romano}(2005)}]{THS3}%
  \BibitemOpen
  \bibfield  {author} {\bibinfo {author} {\bibfnamefont {E.}~\bibnamefont
  {Lehmann}}\ and\ \bibinfo {author} {\bibfnamefont {J.~P.}\ \bibnamefont
  {Romano}},\ }\href@noop {} {\emph {\bibinfo {title} {Testing Statistical
  Hypotheses}}}\ (\bibinfo  {publisher} {Springer},\ \bibinfo {address} {New
  York, NY},\ \bibinfo {year} {2005})\BibitemShut {NoStop}%
\bibitem [{\citenamefont {Alamo}(1964)}]{Alamo1964}%
  \BibitemOpen
  \bibfield  {author} {\bibinfo {author} {\bibfnamefont {J.~B.}\ \bibnamefont
  {Alamo}},\ }\href@noop {} {\bibfield  {journal} {\bibinfo  {journal}
  {Trabajos de estad\'{i}stica e investigaci\'{o}n operativa}\ }\textbf
  {\bibinfo {volume} {15}},\ \bibinfo {pages} {93} (\bibinfo {year}
  {1964})}\BibitemShut {NoStop}%
\bibitem [{\citenamefont {Sung}(1988)}]{Sung1988}%
  \BibitemOpen
  \bibfield  {author} {\bibinfo {author} {\bibfnamefont {N.~K.}\ \bibnamefont
  {Sung}},\ }\emph {\bibinfo {title} {Cram\'{e}r-Rao analogues for
  median-unbiased estimators}},\ \href@noop {} {Ph.D. thesis},\ \bibinfo
  {school} {Iowa State University} (\bibinfo {year} {1988})\BibitemShut
  {NoStop}%
\bibitem [{\citenamefont {So}(1994)}]{So1994}%
  \BibitemOpen
  \bibfield  {author} {\bibinfo {author} {\bibfnamefont {B.-S.}\ \bibnamefont
  {So}},\ }\href@noop {} {\bibfield  {journal} {\bibinfo  {journal} {Journal of
  the Korean Statistical Society}\ }\textbf {\bibinfo {volume} {23}},\ \bibinfo
  {pages} {187} (\bibinfo {year} {1994})}\BibitemShut {NoStop}%
\bibitem [{\citenamefont {Aharonov}\ \emph {et~al.}(1988)\citenamefont
  {Aharonov}, \citenamefont {Albert},\ and\ \citenamefont
  {Vaidman}}]{PhysRevLett.60.1351}%
  \BibitemOpen
  \bibfield  {author} {\bibinfo {author} {\bibfnamefont {Y.}~\bibnamefont
  {Aharonov}}, \bibinfo {author} {\bibfnamefont {D.~Z.}\ \bibnamefont
  {Albert}}, \ and\ \bibinfo {author} {\bibfnamefont {L.}~\bibnamefont
  {Vaidman}},\ }\href {\doibase 10.1103/PhysRevLett.60.1351} {\bibfield
  {journal} {\bibinfo  {journal} {Phys. Rev. Lett.}\ }\textbf {\bibinfo
  {volume} {60}},\ \bibinfo {pages} {1351} (\bibinfo {year}
  {1988})}\BibitemShut {NoStop}%
\end{thebibliography}
\end{document}